\documentclass[conf]{new-aiaa}
\usepackage{graphicx}
\usepackage{tabularx}
\usepackage{caption}
\usepackage{subcaption}
\usepackage{amsmath}
\usepackage{longtable,tabularx}
\usepackage{makecell}
\usepackage{color}  

\title{Data-Driven Modal Decomposition Analysis of Unsteady Flow in a Multi-Stage Turbine} 

\author{Yalu Zhu\footnote{Assistant Specialist, Department of Mechanical and Aerospace Engineering, yalu.zhu@uci.edu (Corresponding author), Member AIAA.} 
and Feng Liu\footnote{Professor, Department of Mechanical and Aerospace Engineering, Fellow AIAA.}}
\affil{University of California, Irvine, Irvine, CA, 92697-3975}
\begin{document}

\maketitle


\begin{abstract}
Two data-driven modal analysis approaches, proper orthogonal decomposition (POD) and dynamic mode decomposition (DMD), are applied to analyze the unsteady flow obtained by solving the Reynolds-averaged Navier-Stokes (RANS) equations in a 1.5-stage axial turbine. 
The reduced-order reconstructed pressure, dominant mode shapes, and dynamic features of these dominant modes in the downstream stator of the turbine are compared between POD and four DMD variants. 
It is found that the DMD methods based on the amplitude criterion, the Tissot criterion, and the sparsity-promoting DMD (SP-DMD) achieve reconstruction accuracy comparable to that of POD, while the frequency criterion proves unsuitable for the present problem. 
The second and third POD and DMD modes capture the dominant pressure fluctuation structures within the stator, and there is similarity between the corresponding POD and DMD spatial modes. The unsteady flow is primarily governed by neutral DMD modes characterized by high amplitudes and low frequencies corresponding to the basic and harmonic frequencies driven by the rotor passing frequency. While the POD analysis provides accurate reconstruction for the original snapshots, the time evolution of each POD mode does not reflect the true dynamic feature of the system. In particular, they misrepresent the fundamental frequencies of the problem.
In addition, the correlations between the dominant modes in the downstream stator and the turbine adiabatic efficiency are explored across different stator clocking configurations.
It is found that a clocking configuration with higher adiabatic efficiency at 50\% span corresponds to a larger spatial and time component of the second and third DMD mode pair, and similarly a larger second POD mode. 

\end{abstract}


\section{Introduction} \label{sec:introduction}
The Navier-Stokes (N-S) equations are considered as an accurate mathematical model for the macroscopic motion of fluids. Due to its strong nonlinearity, however, it is still a huge challenge to fast and accurately obtain the numerical solution to a complex flow, such as flows in multi-stage turbomachines. 
On the other hand, flow fields can be decomposed into a linear combination of a set of modes to approximate the original one, thereby reducing the dimension of flow data. The mode characterizes the basic flow structure in the flow field, and its corresponding coefficient represents the contribution in approximating the flow field.
There are two classes of algorithms for extracting flow field modes \cite{2011_Schmid_DMD}: (1) model-driven algorithms, such as the Arnoldi method and the Krylov subspace method. These methods require the Jacobi matrix of the N-S equations or its action on a specific flow field; (2) data-driven algorithms, such as data statistics (mean, variance, etc.) and conditional sampling techniques. This type of methods can extract flow information from flow fields without relying on the availability of a physical model. Proper orthogonal decomposition (POD) and dynamic mode decomposition (DMD) are such two powerful data-driven algorithms that can capture the dominant flow structures and the dynamic evolution of transient flows.

The concept of POD originates from the least-squares method for curve fitting. Pearson \cite{1901_Pearson_line_fit} proposed finding the best-fit line by minimizing the sum of the squared perpendicular distances from each point to the line, which is mathematically equivalent to maximizing the projection of the data set onto the fit line. By extending this idea to high-dimensional spaces, the concept of identifying the optimal mode from a set of data points forms the foundation of POD. An elaborate introduction to POD is given in Ref. \cite{2000_Chatterjee_pod_introduction}. The mathematical formulation and underlying meaning of POD will be described in Sec. \ref{sec:pod}.
In fluid dynamics, POD has been used to extract both qualitative and quantitative information about the dominant flow features, such as shock waves, boundary layers, and wakes, within a given flow field. As such it has been used as one approach for reduced-order analysis. Lumley \cite{1967_Lumley_pod_fluid} first introduced POD into fluid dynamics by using it to extract coherent structures in turbulence. Sirovich \cite{1987_Sirovich_pod_turbulence} proposed the snapshot method to select the primary flow mode in a time series of flow fields. It has also been applied to analyze unsteady flows in turbomachines \cite{2003_Cizmas_pod_ROA, 2005_Utturkar_pod_ROA, 2006_Rochuon_pod_ROA, 2016_Wang_pod_ROA, buron2023modal, shao2025spectral}. 
POD can provide a reduced-order reconstruction of the original flow field by retaining only the modes with the largest singular values \cite{luo2017flow}. The flow field reconstructed from these modes offers the most accurate approximation of the original data for a given number of modes.

The DMD method was introduced by Rowley et al. \cite{2009_Rowley_dmd} and Schmid \cite{2010_Schmid_dmd}. It extracts the evolution behavior and spatial structures of a dynamic system by performing an eigen-decomposition on a matrix constructed from a sequence of flow field snapshots. The mathematical procedure of DMD is introduced in Sec. \ref{sec:dmd}.
In recent years, a wide range of generalized and enhanced DMD variants have been proposed, such as the optimized DMD (Opt-DMD) \cite{2012_Chen_opt_dmd}, optimal mode decomposition (OMD) \cite{2013_Wynn_omd}, sparsity-promoting DMD (SP-DMD) \cite{2014_Jovanovic_SP-DMD}, exact DMD \cite{2014_Tu_exact_dmd}, multi-resolution DMD (MRDMD) \cite{2015_Kutz_Multi_dmd}, extending DMD \cite{2015_Williams_extending_dmd}, non-uniform DMD \cite{2015_Gueniat_nonuniform_dmd}, parameterized DMD \cite{2015_Sayadi_parametrized_dmd}, recursive DMD \cite{2016_Noack_recursive_dmd}, randomized DMD \cite{2016_Erichson_randomized_dmd, 2017_Bistrian_randomized_dmd}, higher-order DMD \cite{2017_Clainche_HO_dmd}, and tensor-based DMD \cite{2018_Klus_tensor-based_dmd}.

Similar to POD, DMD can be applied for reduced-order analysis and reconstruction, where the criterion for mode selection plays a crucial role. The spatial modes of POD are mutually orthogonal to each other, and the energy fraction associated with each singular value serves as the criterion for mode selection. However, DMD lacks a straightforward and universal criterion for assessing the significance of each mode.  
Simple but commonly used criteria for mode selection include mode amplitude \cite{2009_Rowley_dmd}, growth rate, and frequency. However, modes with large amplitudes may decay quickly over time, and using growth rate or frequency as selection criteria often requires prior knowledge of the underlying physical process. To address these challenges, various mode selection criteria have been proposed and applied in fluid dynamics \cite{2014_Tissot_mode_selection, 2017_Kou_DMD_criterion}. In addition, the optimization processes involved in several DMD variants, such as Opt-DMD \cite{2012_Chen_opt_dmd}, SP-DMD \cite{2014_Jovanovic_SP-DMD}, and randomized DMD \cite{2016_Erichson_randomized_dmd, 2017_Bistrian_randomized_dmd}, inherently provide mode selection criteria. The applicability of these mode selection criteria should be evaluated for a specific flow problem to determine the most suitable option.

Rowley and Dawson \cite{2017_Rowley_review} provided a comprehensive review of DMD applications in flow analysis. Based on existing studies of DMD in fluid dynamics, it can be summarized that: (1) DMD has primarily been used for unsteady flow analysis, feature extraction, and reduced-order reconstruction; (2) DMD has been applied to simple geometric configurations, such as jets \cite{2009_Rowley_dmd}, airfoil \cite{2018_Kou_dmd} and cylindrical flows \cite{min2024data}, boundary layers \cite{sayadi2014reduced}, etc. A few studies have explored its application to unsteady flows in turbomachines \cite{2016_Semlitsch_dmd, 2019_Leonard_dmd, hu2020numerical, an2022investigation, guo2023dynamic}. However, no applications in multi-stage turbomachines have been reported to date; (3) Flow data used for DMD analysis typically come from high-fidelity numerical simulations, such as direct numerical simulation (DNS) \cite{2009_Rowley_dmd, sayadi2014reduced} and large-eddy simulation (LES) \cite{2016_Semlitsch_dmd, 2019_Leonard_dmd}, or from experimental measurements, such as particle image velocimetry (PIV) \cite{hu2020numerical}. Several studies \cite{2017_Kou_DMD_criterion, 2018_Kou_dmd, hu2020numerical, an2022investigation, guo2023dynamic} have applied DMD to flow fields from solving the unsteady Reynolds-averaged N-S (URANS) equations. 
Although the unsteady fluctuations in a multi-stage turbomachine obtained from URANS simulation are mainly driven by motion of rotating blades, analyzing these unsteady flow features remains valuable for blade design and optimization by linking the aerodynamic performance of the machine to the dynamic flow characteristics.

POD and DMD are two related but distinct data-driven modal decomposition techniques. Table \ref{tab:POD_DMD_comparison} summarizes their primary features. Taira et al. \cite{2017_Taira_review} provided a comprehensive comparison of the advantages and limitations of both methods.
In addition to their inherent differences, several studies have performed comparative analyses by applying both POD and DMD to various flows, including those in turbomachinery. 
Muld et al. \cite{2012_Muld_pod_dmd} compared the flow structures around high-speed railway trains extracted by POD and DMD. 
Mariappan et al. \cite{2014_Mariappan_pod_dmd} employed both POD and DMD to analyze the dynamic stall process of airfoils. 
Yang and Ma \cite{yang2022analysis} analyzed the tip leakage flow in a transonic turbine cascade by using various POD and DMD methods.
Chen et al. \cite{chen2022modal} applied both POD and DMD to extract the flow structures in a compressor cascade and to analyze the effect of an unsteady jet on corner separation. Zhong et al. \cite{zhong2024unsteady} compared the energy scales and spatiotemporal features of the corner separation in a highly-loaded compressor cascade by using POD and DMD.
It is thus expected that a comparative evaluation of the POD and DMD methods will facilitate a deeper understanding of the unsteady flow characteristics in multi-stage turbomachines.

\begin{table}[htb!]
    \centering
    \caption{Feature comparison between POD and DMD methods}
    \begin{tabular}{lcc}
        \hline
                                    & POD          & DMD  \\
        \hline
        Mode orthogonality          & Yes          & No   \\
        Sole frequency in each mode & No           & Yes  \\
        Temporal features contained & No           & Yes  \\
        Limited to time snapshots   & No           & Yes  \\
        Sensitive to data errors    & No           & Yes  \\
        Easy to sort modes          & Easy         & Hard     \\
        Computational cost          & Small        & Large    \\
        \hline
    \end{tabular}
    \label{tab:POD_DMD_comparison}
\end{table}

In this paper, the unsteady flow fields within the downstream stator of a 1.5-stage axial turbine, obtained by solving the URANS equations, are analyzed using POD and various DMD methods. 
The remainder of this paper is organized as follows.
The two modal decomposition methods, along with various mode selection criteria in DMD, are presented in Sec. \ref{sec:mode_decomposition_method}. 
The method of generating unsteady flow fields is briefly introduced in Sec. \ref{sec:flow_generation}. 
The reduced-order reconstructed and predicted pressure, mode shapes, and dynamic features of the dominant modes in the downstream stator of the turbine are compared between POD and DMD in Secs. \ref{sec:ROR}, \ref{sec:ROA}, and \ref{sec:mode_dynamics}, respectively. 
After determining the most suitable DMD variant for the present problem, the pressure mode shapes for different stator clocking configurations are compared in Sec. \ref{sec:clocking_config}, establishing the correlation between the aerodynamic performance of the turbine and the characteristics of the modes.
The conclusion remarks are given in Sec. \ref{sec:conclusions}.

\section{Data-Driven Modal Decomposition Methods} \label{sec:mode_decomposition_method}

\subsection{Proper Orthogonal Decomposition} \label{sec:pod}
Given a set of $n$ $m$-dimensional column vectors $\mathbf{w}_l \in \mathbb{R}^m, \ l = 1, \ 2, \ ..., \ n$, where $m \gg n$, these vectors are collected into a matrix $\mathbf{W}_1^n = [\mathbf{w}_1, \ \mathbf{w}_2, \ ..., \ \mathbf{w}_n] \in \mathbb{R}^{m \times n}$. The subscript and superscript on $\mathbf{W}$ indicate the indices of the starting and ending vectors, respectively. POD, also referred to as singular value decomposition in linear algebra, is defined for the matrix $\mathbf{W}_1^n$ as
\begin{equation}
  \mathbf{W}_1^n = \mathbf{U} \mathbf{\Sigma} \mathbf{V}^T = \sum_{k=1}^{r}{\sigma_k \mathbf{u}_k \mathbf{v}_k^T}
  \label{eq:pod_svd}
\end{equation}
where $r$ is the rank of $\mathbf{W}_1^n$ with $r \le n$; both $\mathbf{U} = [\mathbf{u}_1, \ \mathbf{u}_2, \ ..., \ \mathbf{u}_r] \in \mathbb{R}^{m \times r}$ and $\mathbf{V} = [\mathbf{v}_1, \ \mathbf{v}_2, \ ..., \ \mathbf{v}_r] \in \mathbb{R}^{n \times r}$ are orthogonal matrices, i.e., satisfying $\mathbf{U}^T\mathbf{U} = \mathbf{I}$ and $\mathbf{V}^T\mathbf{V} = \mathbf{I}$; $\mathbf{\Sigma} = \mathrm{diag}(\sigma_1, \ \sigma_2, \ ..., \ \sigma_r) \in \mathbb{R}^{r \times r}$ with $\sigma_1 \geq \sigma_2 \geq ... \geq \sigma_r > 0$. $\sigma_k$ is called the $k$th singular value, and $\mathbf{u}_k$ corresponds to the $k$th POD mode shape.

The mathematical meaning of POD can be interpreted as follows. For a given set $\{ \mathbf{w}_l \}$, the intention of POD is to find a basis vector $\boldsymbol{\phi} \in \mathbb{R}^m$ which maximizes the sum of the squared projection of $\mathbf{w}_l$ onto it, i.e., to solve the following optimization problem
\begin{equation}
  \max \limits_{\boldsymbol{\phi} \in \mathbb{R}^m} \left(\| \boldsymbol{\phi}^T \mathbf{W}_1^n \|_\mathrm{F}^2 \right), \ \mathrm{s.t.} \ \|\boldsymbol{\phi}\|_\mathrm{F}^2 = 1
  \label{eq:pod_optimization_problem}
\end{equation}
where $\|\mathbf{Q}\|_\mathrm{F} = \sqrt{\mathrm{trace}(\mathbf{Q}^T\mathbf{Q})} = \sqrt{\mathrm{trace}(\mathbf{QQ}^T)}$ is the Frobenius norm of $\mathbf{Q}$.
Using the method of Lagrange multiplier, Eq. (\ref{eq:pod_optimization_problem}) is transferred into an eigenvalue problem 
\begin{equation}
  \mathbf{W}_1^n (\mathbf{W}_1^n)^T \boldsymbol{\phi} = \lambda \boldsymbol{\phi}
  \label{eq:pod_eigen_problem}
\end{equation}
Meanwhile, from Eq. (\ref{eq:pod_svd}), we obtain
\begin{equation}
  \mathbf{W}_1^n (\mathbf{W}_1^n)^T = \mathbf{U} (\mathbf{\Sigma} \mathbf{\Sigma}^T) \mathbf{U}^T
  \label{eq:pod_eigen_svd_relation_2}
\end{equation}
This equation corresponds to the eigen-decomposition of the matrix $\mathbf{W}_1^n (\mathbf{W}_1^n)^T$, where the $k$th eigenvalue and corresponding eigenvector are $\sigma_k^2$ and $\mathbf{u}_k$, respectively. Hence, a subset of solutions to Eq. (\ref{eq:pod_eigen_problem}) is $\lambda_k = \sigma_k^2$ and $\boldsymbol{\phi}_k = \mathbf{u}_k$. This implies that the POD defined by Eq. (\ref{eq:pod_svd}) provides a solution to the optimization problem defined in Eq. (\ref{eq:pod_optimization_problem}). 

The POD mode shapes and singular values offer valuable insights into the structure of the matrix $\mathbf{W}_1^n$. If $\mathbf{W}_1^n$ is composed of $n$ snapshots of unsteady flow fields, the mode shape $\mathbf{u}_k$ represents the flow structure associated with the $k$th mode, while the singular value $\sigma_k$ indicates the `energy' contained within this mode \cite{1987_Sirovich_pod_turbulence}. 
If only the first $s$ modes are retained in Eq. (\ref{eq:pod_svd}), it gives a reduced-order POD reconstruction, $\mathbf{W}^{\mathrm{POD}}$, of the original flow field $\mathbf{W}_1^n$.

\subsection{Dynamic Mode Decomposition} \label{sec:dmd}
For $n+1$ snapshots of an unsteady flow field, denoted as $\mathbf{w}_l \in \mathbb{R}^m, \ l = 1, \ 2, \ ..., \ n+1$, where $m \gg n$, it is assumed that each pair of successive snapshots satisfies a linear relation
\begin{equation}
  \mathbf{w}_{l+1} = \mathbf{A} \mathbf{w}_l, \ l = 1, \ 2, \ ..., \ n
  \label{eq:dmd_linear_relation_single}
\end{equation}
or in matrix form
\begin{equation}
  \mathbf{W}_2^{n+1} = \mathbf{A} \mathbf{W}_1^n
  \label{eq:dmd_linear_relation_matrix}
\end{equation}
The matrix $\mathbf{A} \in \mathbb{R}^{m \times m}$ is a linear approximation of the nonlinear dynamic system associated with the unsteady flow. Its eigenvalues and eigenvectors provide the dynamic features and spatial structures of the flow, respectively.
If $\mathbf{A}$ is known and its dimension is not too large, the eigenvalues and eigenvectors can be easily determined, such as using the Arnoldi iteration method. When $\mathbf{A}$ is unknown or its dimension is huge, such as in the current case where only a series of flow snapshots are available, DMD can be applied to approximate the dynamic features and spatial structures instead.

Substituting Eq. (\ref{eq:pod_svd}), the POD of the matrix $\mathbf{W}_1^n$, into Eq. (\ref{eq:dmd_linear_relation_matrix}) gives
\begin{equation}
  \mathbf{U}^T\mathbf{A}\mathbf{U} = \mathbf{U}^T \mathbf{W}_2^{n+1} \mathbf{V} \mathbf{\Sigma}^{-1} = \tilde{\mathbf{A}}
  \label{eq:similar_transform}
\end{equation}
In the above equation, by projecting the matrix $\mathbf{A}$ onto $\mathbf{U}$, the POD bases of $\mathbf{W}_1^n$, a reduced-order similar matrix $\tilde{\mathbf{A}} \in \mathbb{R}^{n \times n}$ is obtained. 
The eigenvalues, $\lambda_k \in \mathbb{C}$, and eigenvectors, $\mathbf{y}_k \in \mathbb{C}^n$, of the matrix $\tilde{\mathbf{A}}$, which are easy to determine due to its low dimensionality, form a subset of eigenvalues and eigenvectors of $\mathbf{A}$. 
The $k$th eigenvalue (DMD eigenvalue) of $\mathbf{A}$ is also $\lambda_k$, and the corresponding eigenvector (DMD mode shape) is 
\begin{equation}
  \boldsymbol{\phi}_k = \mathbf{U} \mathbf{y}_k 
\end{equation}
Note that DMD mode shapes, $\boldsymbol{\phi}_k \in \mathbb{C}^n$, are not orthogonal to each other due to the non-orthogonal eigenvectors of $\tilde{\mathbf{A}}$. 
The complex growth rate of the $k$th DMD mode is defined in terms of its eigenvalue $\lambda_k$ as
\begin{equation}
  \mu_k = \frac{\ln\lambda_k}{\Delta t} = \frac{\ln |\lambda_k|}{\Delta t} + \mathrm{i} \frac{\arg (\lambda_k)}{\Delta t} = \sigma_k + \mathrm{i} \omega_k
  \label{eq:dmd_growth_rate}
\end{equation}
where $\sigma_k \in \mathbb{R}$ and $\omega_k \in \mathbb{R}$ represent the temporal growth rate and angular frequency of the $k$th DMD mode, respectively. $\Delta t$ is the time interval between two successive snapshots.

Similar to POD, DMD can also be used as a reduced-order reconstruction model. A general solution to Eq. (\ref{eq:dmd_linear_relation_single}) is
\begin{equation}
  \mathbf{w}_t^\mathrm{DMD} = \sum_{k=1}^{s}{\alpha_k\lambda_k^{t-1} \boldsymbol{\phi}_k}
  \label{eq:dmd_reconstruct_single}
\end{equation}
where $s$ is the number of retained DMD modes, with $s \leq r$. The mode amplitude $\alpha_k \in \mathbb{C}$ represents the contribution of the $k$th mode to the initial snapshot $\mathbf{w}_1$, while its influence on subsequent snapshots is governed by the power of the corresponding DMD eigenvalue $\lambda_k$. If $\alpha_k$ is available, the reduced-order reconstruction, $\mathbf{W}^\mathrm{DMD}$, of the original flow field $\mathbf{W}_1^n$ can be computed by sequentially setting $t = 1, \ 2, \ ..., \ n$ in Eq. (\ref{eq:dmd_reconstruct_single}). Moreover, Eq. \eqref{eq:dmd_reconstruct_single} provides a reduced-order prediction of the system state at future time by allowing $t > n$, a capability that is not available in POD. The mode amplitudes can be determined through an optimization process which minimizes the difference between $\mathbf{W}^\mathrm{DMD}$ and $\mathbf{W}_1^n$. A reduced-order algorithm proposed by Jovanovi{\'c} et al. \cite{2014_Jovanovic_SP-DMD} is applied to solve this optimization problem in the present study. 


\subsection{Mode Selection Criteria in DMD}
It is expected that a flow field can be reconstructed by using the minimum number of modes. Therefore, it is essential to sort these modes based on their significance to the original flow field. In POD, the singular value serves as an effective mode ordering criterion as POD modes are spatially orthogonal to each other. In DMD, however, there is no straightforward and universal criterion to evaluate the contribution of each mode. Based on the practical characteristics of the present problem, the following mode selection criteria for DMD are considered:

\begin{itemize}
    \item \textbf{Amplitude criterion.}
    Modes are sorted in descending order based on their amplitude moduli $|\alpha_k|$.

    \item \textbf{Frequency criterion.}
    Modes are sorted in ascending order by their angular frequencies $|\omega_k|$.

    \item \textbf{Tissot criterion.}
    Tissot et al. \cite{2014_Tissot_mode_selection} proposed using the mode amplitude weighted by the growth rate
    \begin{equation}
      \centering
      E_k = \frac{1}{T}\int_0^T{\Big|\alpha_k\lambda_k^{t/\Delta t} \Big|^2 \, \mathrm{d}t} = \alpha_k^2 \frac{e^{2\sigma_kT}-1}{2\sigma_kT}
      \label{eq:Tissot_criteria}
    \end{equation}
    where $T = n\Delta t$.

    \item \textbf{SP-DMD method.}
    In the SP-DMD method proposed by Jovanovi{\'c} et al. \cite{2014_Jovanovic_SP-DMD}, modes are not sorted. Instead, a penalty function is incorporated into the non-zero mode amplitudes to identify the optimal combination of DMD modes. 
\end{itemize}

\subsection{Code Implementation}
An in-house code is developed to implement the above algorithms of POD and DMD, along with various mode selection criteria for DMD. 
As indicated by Eq. (\ref{eq:pod_svd}), a POD of the matrix $\mathbf{W}_1^n$ is required as a preliminary step before conducting a DMD analysis. Since the row number $m$ of the matrix $\mathbf{W}_1^n$ is usually huge in engineering problems, performing POD directly on a serial machine is often too time-consuming or even impractical. To overcome this, the parallel tall-and-skinny QR (TSQR) decomposition algorithm \cite{2012_Demmel_communication, 2016_Sayadi_parallel}, together with the message passing interface (MPI) technique, is employed to accelerate the computation of POD. $\mathbf{W}_1^n$ is partitioned into $M$ row-wise submatrices, each of which is processed by an individual CPU core under the MPI framework.
In addition, the software package LAPACK \cite{lapack99} is employed for matrix factorizations, matrix productions, and eigenvalue decompositions involved in the POD and DMD procedures.
This code has been applied to extract unsteady acoustic modes in a rocket-engine combustor by using DMD \cite{zhu2025simulation}, yielding primary acoustic modes that are consistent with those obtained by fast Fourier transform (FFT).

\section{Flow Field Generation} \label{sec:flow_generation}
To evaluate the performance of POD and various DMD methods in multi-stage turbines and to compare their similarities and differences, the unsteady flow fields in a 1.5-stage subsonic axial turbine are analyzed in the present study. The turbine, originally designed and tested by RWTH Aachen University \cite{gallus1995endwall, 2000_Stephan_RWTH}, consists of three blade rows: the first stator (Stator 1), the rotor (Rotor), and the second stator (Stator 2). Each blade row features untwisted low-aspect-ratio blades. The Traupel profile is applied to both stators, while a modified VKI (The von Karman Institute) profile is employed for the rotor. 
The primary geometric and aerodynamic parameters of the turbine are listed in Table \ref{table:turbine_info}. 

\begin{table}[htbp]
    \caption{Geometric and aerodynamic parameters of the turbine}
    \centering
    \label{table:turbine_info}
    \begin{tabular}{l c c}
      \hline
      \textit{ }            &Stator      &Rotor       \\
      \hline
      Blade count           &36          &41         \\
      Hub diameter (mm)     &490         &490        \\
      Casing diameter (mm)  &600         &600        \\
      Aspect ratio          &0.887       &0.917      \\
      Relative inlet angle ($^\circ$)  & 0   & 40.7  \\
      Relative outlet angle ($^\circ$) & 70.0 & -61.2\\
      Tip gap (mm)          & /          &0.4        \\
      Rotating speed (rpm)  & /          &3500       \\
      Reynolds number $\mathrm{Re}_c$ & $6.8\times10^5$ & $4.9\times10^5$ \\
      \hline
    \end{tabular}
\end{table}

The unsteady flow fields are generated using a computational fluid dynamics (CFD) approach. An in-house code specifically developed for flow simulation in multi-stage blade rows is used for this purpose. The three-dimensional compressible URANS equations are solved in the rotating frame of reference by using a second-order cell-centered finite-volume method in multi-block structured grids. By expressing the governing equations in terms of absolute flow quantities in the rotating frame of reference, they are valid for both rotor and stator by setting the angular velocity to the rotor value or zero for the stator. The code has been verified and validated with steady and unsteady test cases in a transonic compressor rotor \cite{2018_Zhu_mixing_plane}, a subsonic turbine cascade \cite{zhu2024numerical, walsh2026flamelet}, a transonic turbine stage \cite{zhu2026_turbine_burner}, and the present subsonic turbine \cite{2017_Zhu_clocking, 2018_Zhu_mixing_plane, 2018_Zhu_blade_lean}. 
Details about the governing equations and numerical algorithms used for computing the unsteady flows in the present turbine are given in Refs. \cite{2017_Zhu_clocking, 2018_Zhu_blade_lean, liu2025computational}. 

\begin{figure}[htb!]
  \centering
  \includegraphics[width=0.495\linewidth]{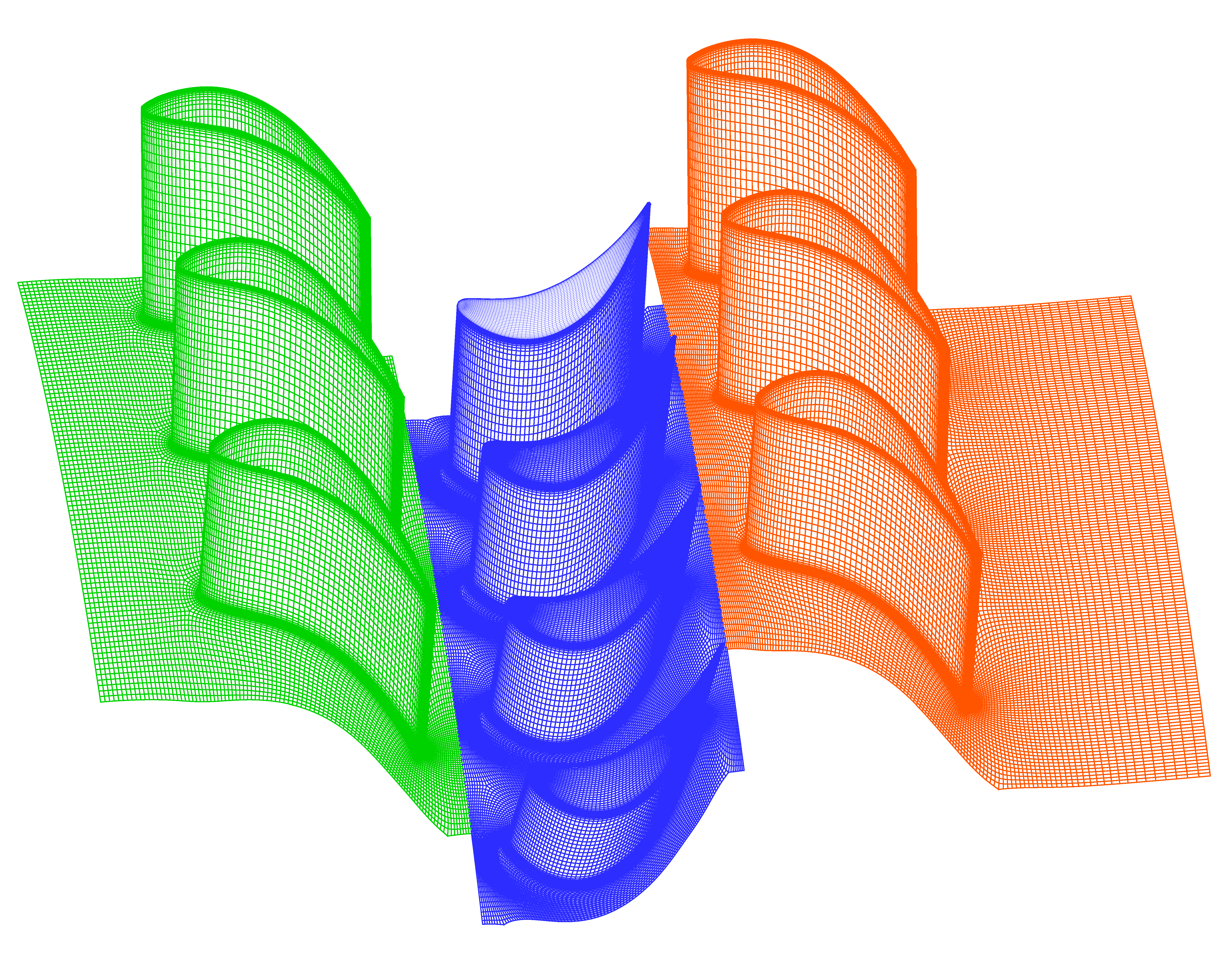}
  \caption{Grids over blade surfaces and hub of the turbine.}
  \label{fig:grid_over_blade_hub}
\end{figure}

For the convenience of unsteady computation, the blade counts of the turbine are scaled from $36\mathbin{:}41\mathbin{:}36$ to $36\mathbin{:}48\mathbin{:}36$ following the blade scaling rule by Rai \cite{1987_Rai_blade_scaling}. Consequently, three passages are configured for each stator, and four passages for the rotor.
Multi-block structured grids are generated for each blade row, as shown in Fig. \ref{fig:grid_over_blade_hub}. The cell vertexes on the interfaces between neighboring blocks are point-to-point matched, except at the interfaces between adjacent blade rows, where a patched grid is employed. The dimensionless wall distance $y^+$ is less than 3 for the first grid point off the wall. The numbers of cells per passage for Stator 1, Rotor, and Stator 2 are 0.35 million, 0.57 million, and 0.41 million, respectively, resulting in a total of over 4.53 million cells.

\begin{figure}[htb!]
  \centering
  \includegraphics[width=0.495\linewidth]{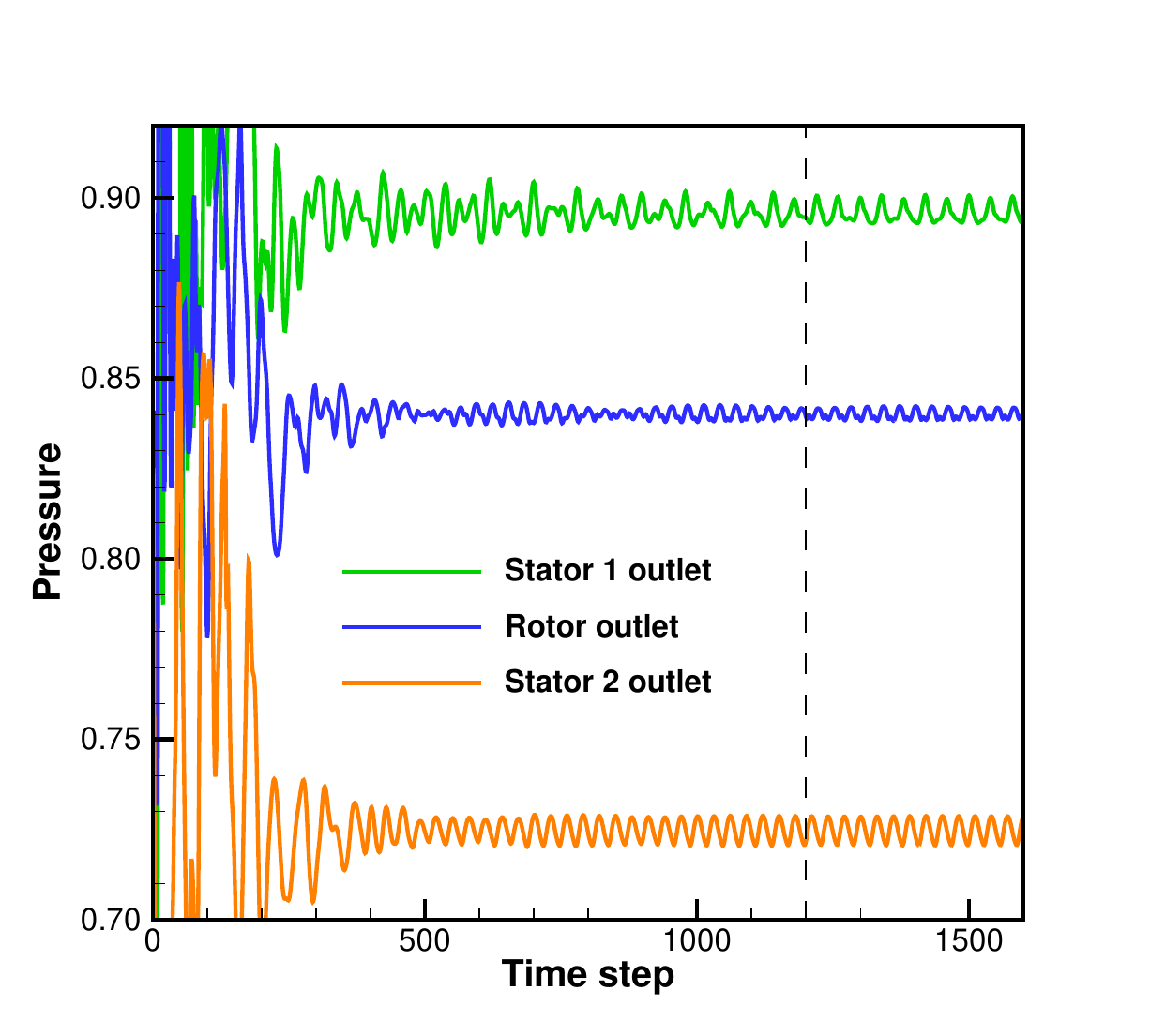}
  \caption{History of static pressure at the outlet of each blade row.}
  \label{fig:p-t_midspan}
\end{figure}

The physical time step for the unsteady computation is $1.2\times10^{-5} \, \mathrm{s}$, resulting in 30 physical time steps per rotor passing period (RPP). The RPP is defined as the time required for a rotor blade to traverse one rotor pitch angle. The unsteady computation is started from a steady solution.
Figure \ref{fig:p-t_midspan} shows the history of static pressure at a selected 50\% span position at the outlet of each blade row. The pressure is normalized by the total pressure at the turbine inlet. At each position, the initial high oscillation gradually diminishes as the flow develops, reaching a periodic state after approximately 1200 time steps. 
The flow fields within the last 120 time steps (1400 to 1520), corresponding to 4 RPPs, are then selected for modal decomposition analysis. That is $n = 120$. Recalling that four rotor passages are included in the computational domain, as shown in Fig. \ref{fig:grid_over_blade_hub}, this implies that within 120 time steps, the turbine undergoes a complete cycle of periodic motion. 

\section{Results and Discussions} \label{sec:results}
In a multi-stage turbomachine, blade-row interactions include potential effects and wake effects \cite{2017_Zhu_clocking}. Potential disturbances propagate both upstream and downstream through pressure fluctuations and primarily affect close neighboring blade rows. In contrast, wake interactions, arising from the vortical and entropic wakes shed from upstream blades, can persist into more downstream blade rows as they are carried by the mean flow convection.
Due to the combined effects of upstream blade wakes and potential disturbances, the flow in Stator 2 exhibits greater unsteadiness and nonlinearity than in Stator 1 and Rotor. Because of this and also the large static pressure drop, the flow in Stator 2 has significant influence on the overall aerodynamic performance of the turbine. Hence, the flow in Stator 2 is analyzed in this study. This results in a total of $1,388,037$ grid points, i.e., $m = 1,388,037$ in both POD and DMD.
To account for the downstream potential influence, such as those from the trailing edge of Stator 2 blade, pressure is chosen for analysis instead of velocity or entropy. The time-averaged value is preserved in the pressure snapshots. 
Since the flow is closely uniform along the span due to the untwisted blades and constant-radius end-walls, the following analysis focuses on the mid-span blade-to-blade (B2B) surface unless stated otherwise.

In DMD, a pair of modes with complex-conjugate eigenvalues are treated as consecutive modes, of which the mode with positive imaginary part is listed first. In this way, the total mode numbers for both POD and DMD are the same and equal to the snapshot number $n$.

\subsection{Reduced-Order Reconstruction} \label{sec:ROR}
\begin{figure}[htb!]
  \centering
  \includegraphics[width=0.495\linewidth]{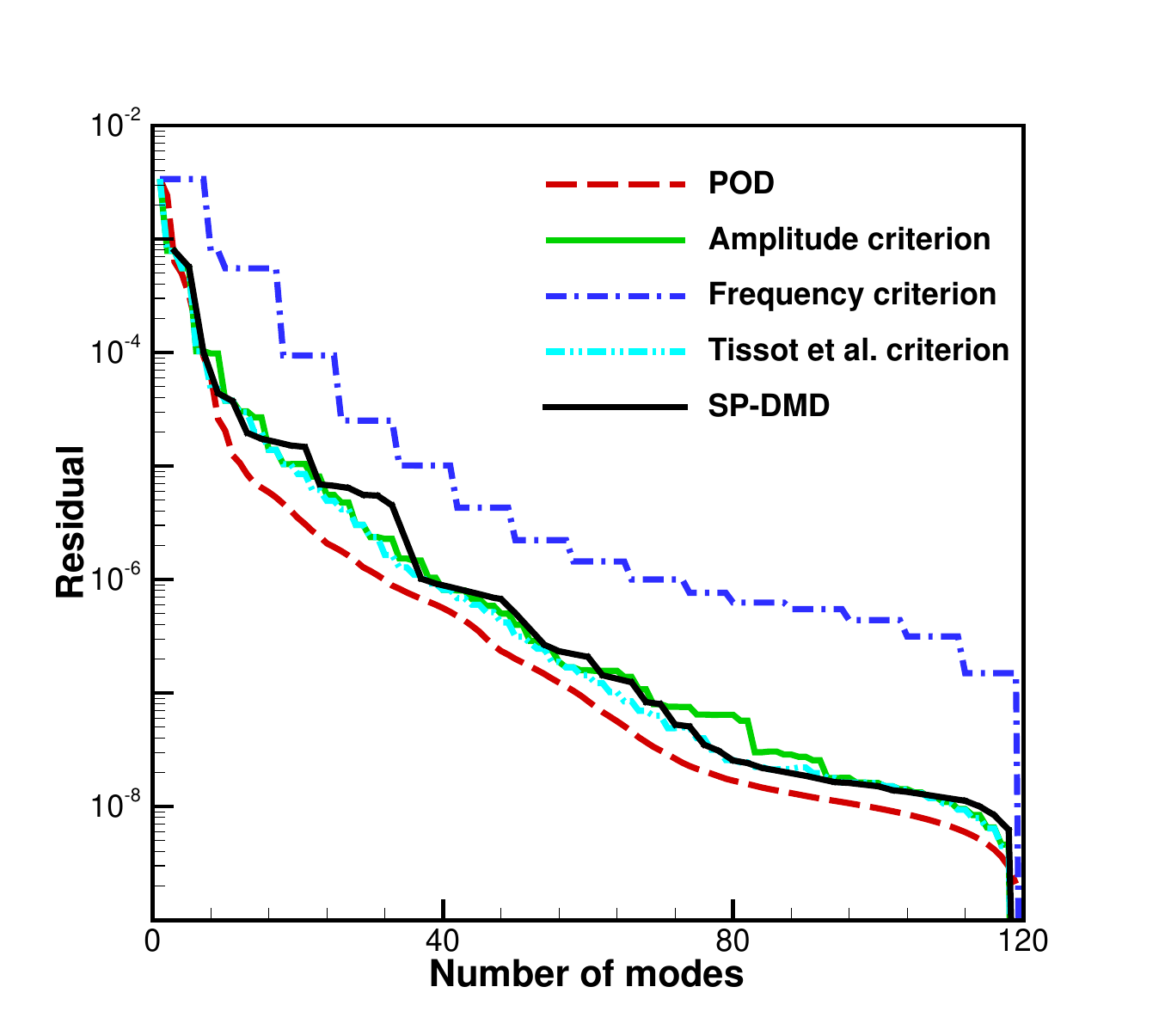}
  \caption{Variation of reconstruction residual with number of retained modes.}
  \label{fig:residual_pod_dmd-mode}
\end{figure}

The accuracy of the modal decomposition methods as reduced-order models for approximating the original flow field is first evaluated. 
Figure \ref{fig:residual_pod_dmd-mode} shows the variation of reconstruction residual with the number of retained modes for each modal decomposition method. The residual is defined as the normalized Frobenius norm of the difference between the reconstructed flow field and the original one, i.e.,
\begin{equation}
  \mathrm{Residual} = \frac{\| \mathbf{W}^\mathrm{Rec} - \mathbf{W}_1^n \|_\mathrm{F}}{\| \mathbf{W}_1^n \|_\mathrm{F}}
  \label{eq:residual_reconstruction}
\end{equation}
where $\mathbf{W}^\mathrm{Rec}$ corresponds to $\mathbf{W}^\mathrm{POD}$ or $\mathbf{W}^\mathrm{DMD}$.

The residuals of all reconstruction methods decrease as the number of retained modes increases, which is consistent with the original intention of both POD and DMD methods. The residuals drop rapidly within the range of small mode numbers. All methods, except the DMD method based on the frequency criterion, achieve a residual level below $10^{-4}$ with seven modes. As the number of modes increases further, the rate of residual reduction decreases. These residual behaviors indicate that there exist several dominant modes within the unsteady flow field. 
For the same number of modes, the residual of POD is lower than those of the DMD methods. This is because DMD relies on a linear-evolution approximation between sequential snapshots, as expressed in Eq. (\ref{eq:dmd_linear_relation_single}), whereas POD, by definition, yields the optimal decomposition of the original flow field in terms of the Frobenius norm, without introducing any additional assumptions. 
The DMD methods based on the amplitude and Tissot criteria as well as the SP-DMD method are consistent with each other. Their residuals are closer to that of POD compared to the DMD method based on the frequency criterion. Quantitatively, for a given mode number, the residual by the frequency criterion is one order of magnitude higher than that of the other methods. This indicates that the first three DMD methods effectively capture the primary features of the unsteady flow field as POD does. The unsteady flow in a multi-stage turbine is dominantly forced by the rotor passing frequency (RPF, defined as the reciprocal of RPP). The DMD modes identified based on frequency order do not effectively capture the dominant features for the present problem.

\begin{figure}[htb!]
  \centering
  \includegraphics[width=0.495\linewidth]{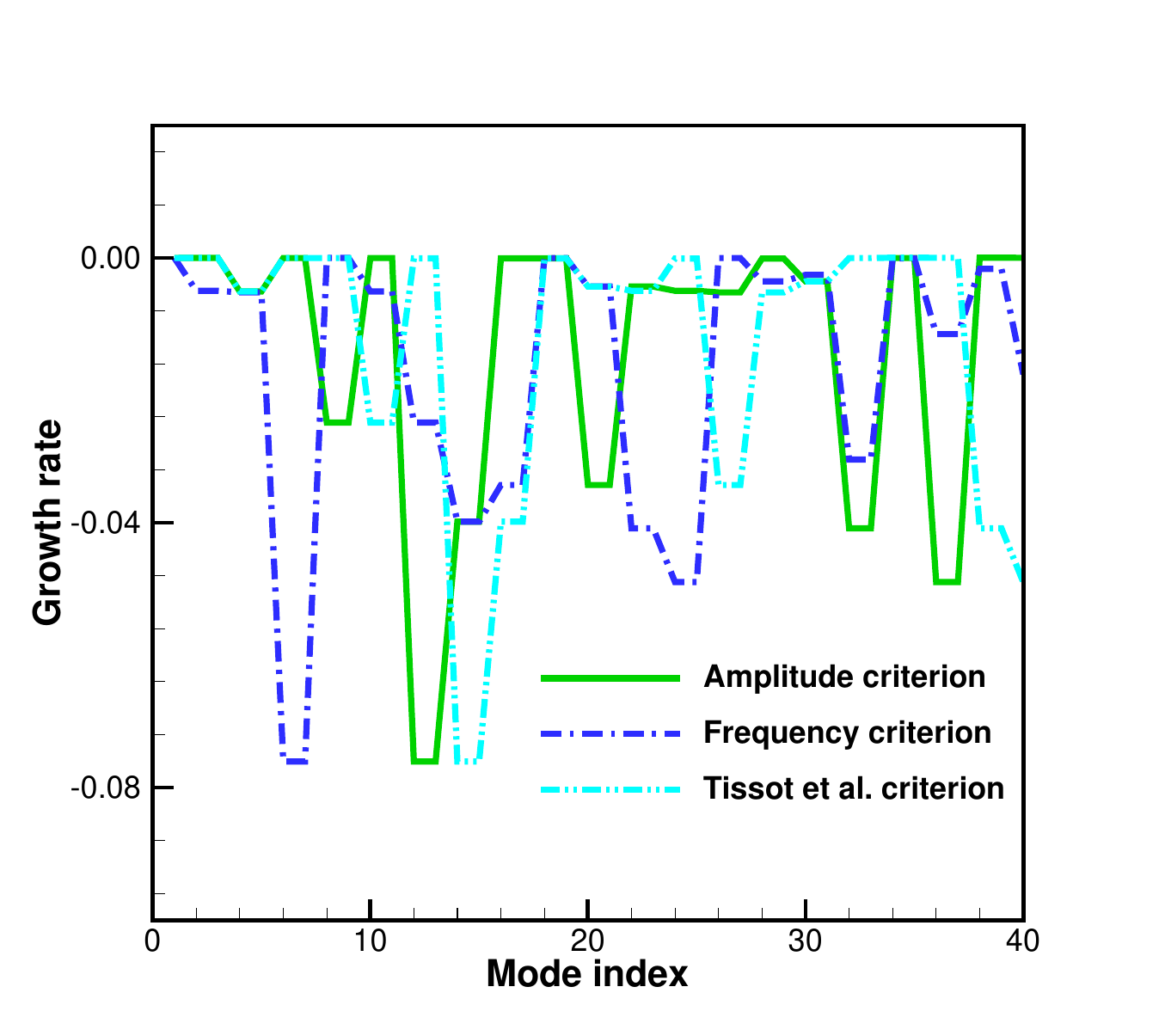}
  \caption{Variation of mode growth rate of DMD with mode index.}
  \label{fig:growth_rate-index_dmd}
\end{figure}

To explore the cause of the significant reconstruction error associated with the frequency criterion, Fig. \ref{fig:growth_rate-index_dmd} compares the variation of the DMD mode growth rate $\sigma_k$ with the mode index for different mode selection criteria. Since the SP-DMD method computes only the amplitudes of non-trivial modes instead of sorting all modes, its growth rate is not shown here. 
As indicated by the consistent growth rates, the mode sequences by the amplitude and Tissot criteria show close agreement, especially for the first seven dominant modes. Specifically, among the seven modes, the fourth and fifth modes exhibit small negative growth rates, while the remaining modes have zero growth rates. In contrast, for the frequency criterion, several modes with high decay rates appear among the leading indices, clearly deviating from the results of the other two methods. 
Strongly decaying modes are unable to persist over time in a globally periodic flow. Hence, they should not be identified as dominant modes in the present unsteady flow field. This explains why the DMD method based on the frequency criterion results in a higher reconstruction error compared to the other methods.

\begin{figure}[htb!]
    \centering
    \begin{subfigure}[b]{0.30\linewidth}
        \centering
        \includegraphics[width=1\linewidth]{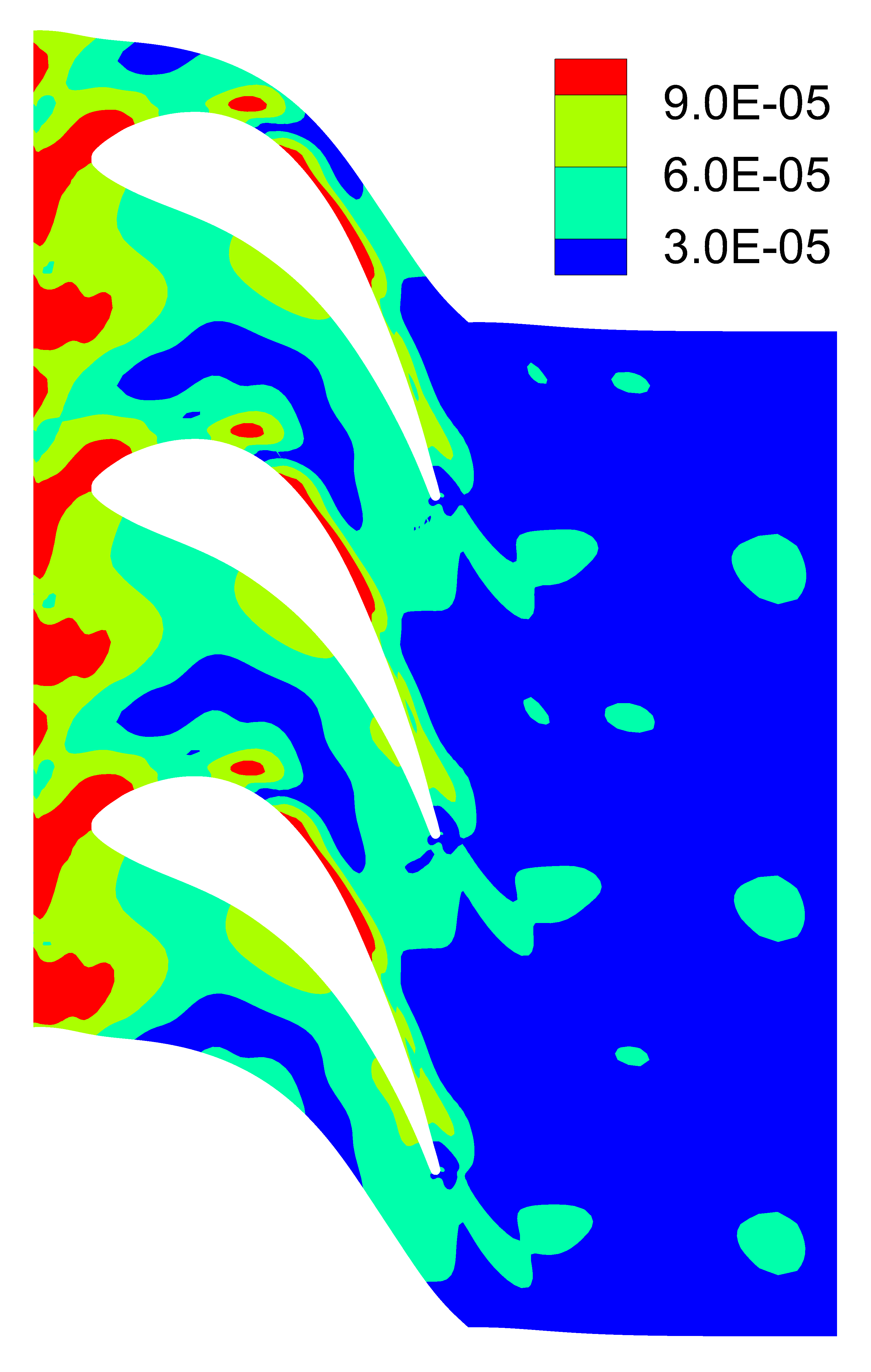}
        \caption{POD}
        \label{fig:contour_error_pod}
    \end{subfigure}
    \begin{subfigure}[b]{0.30\linewidth}
        \centering
        \includegraphics[width=1\linewidth]{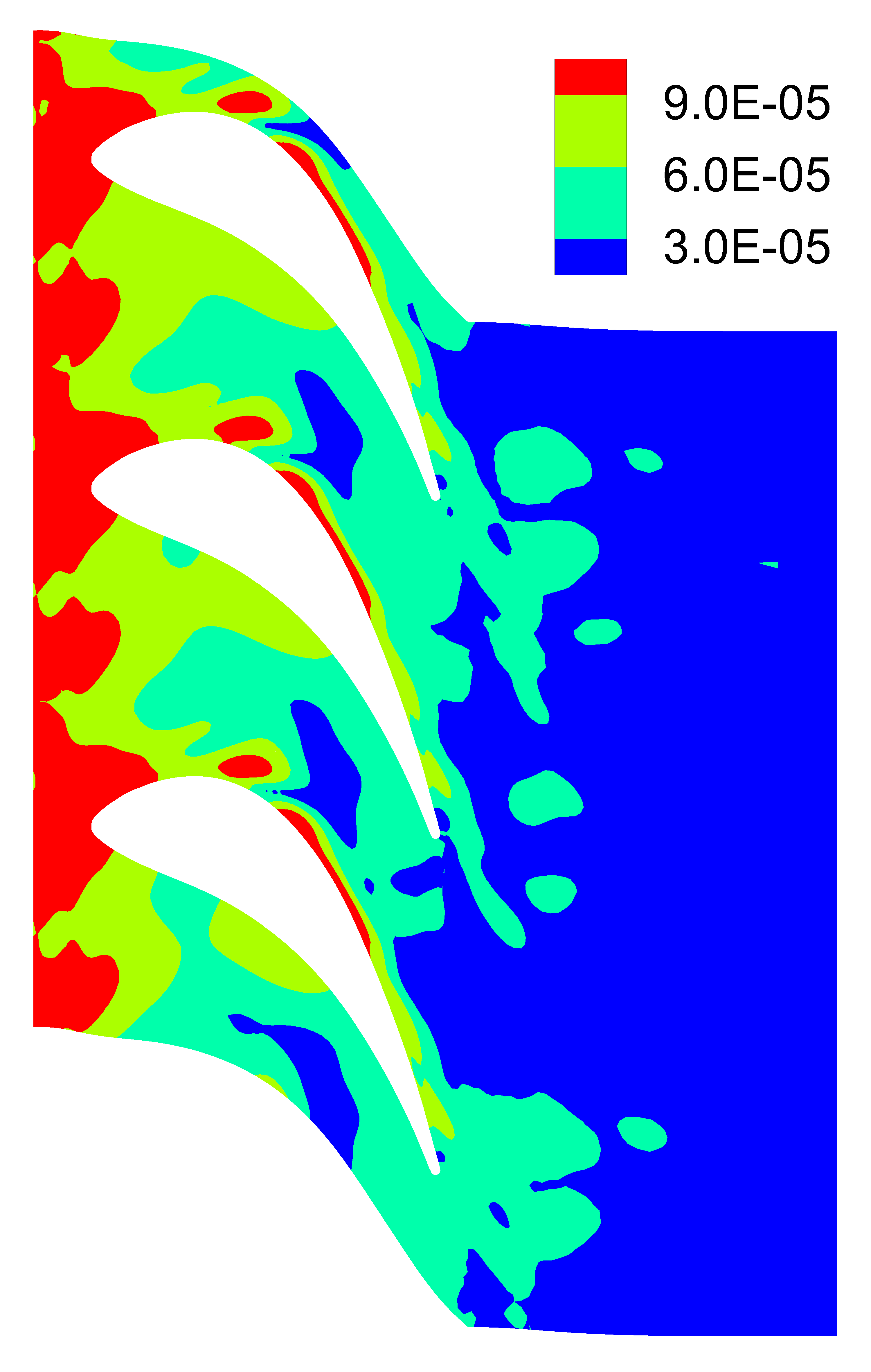}
        \caption{DMD}
        \label{fig:contour_error_dmd}
    \end{subfigure}
    \caption{Contours of relative error of reconstructed pressure by the first seven modes. DMD is based on the Tissot criterion.}
    \label{fig:contour_error_pod_dmd}
\end{figure}

Figure \ref{fig:contour_error_pod_dmd} compares the contours of the relative error between the reconstructed and original pressure fields for both POD and DMD. The first seven modes are used for reconstruction. The Tissot criterion is applied for DMD mode selection. 
The reconstruction errors of both POD and DMD remain below $10^{-4}$ across the entire domain, which is consistent with the global reconstruction residual observed at the mode number of seven in Fig. \ref{fig:residual_pod_dmd-mode}. This further confirms that the first seven modes can be considered as the dominant or leading modes for this problem.
The reconstruction error remains relatively higher near Stator 2 inlet, attributed to the strong unsteady fluctuations caused by upstream wakes from Stator 1 and the Rotor, as well as potential disturbances from Stator 2 blades. The error gradually decreases along the streamwise direction due to the attenuating influence of the upstream fluctuations.
Compared to POD, the reconstruction error of DMD is larger near Stator 2 inlet but quickly decreases to achieve similar levels in the downstream region. This indicates that the linear-evolution approximation in DMD introduces noticeable errors in regions with strong unsteady fluctuations, but achieves acceptable reconstruction accuracy in relatively stable regions when using only a few dominant modes.

\begin{figure}[htb!]
    \centering
    \begin{subfigure}[b]{0.495\linewidth}
        \centering
        \includegraphics[width=1\linewidth]{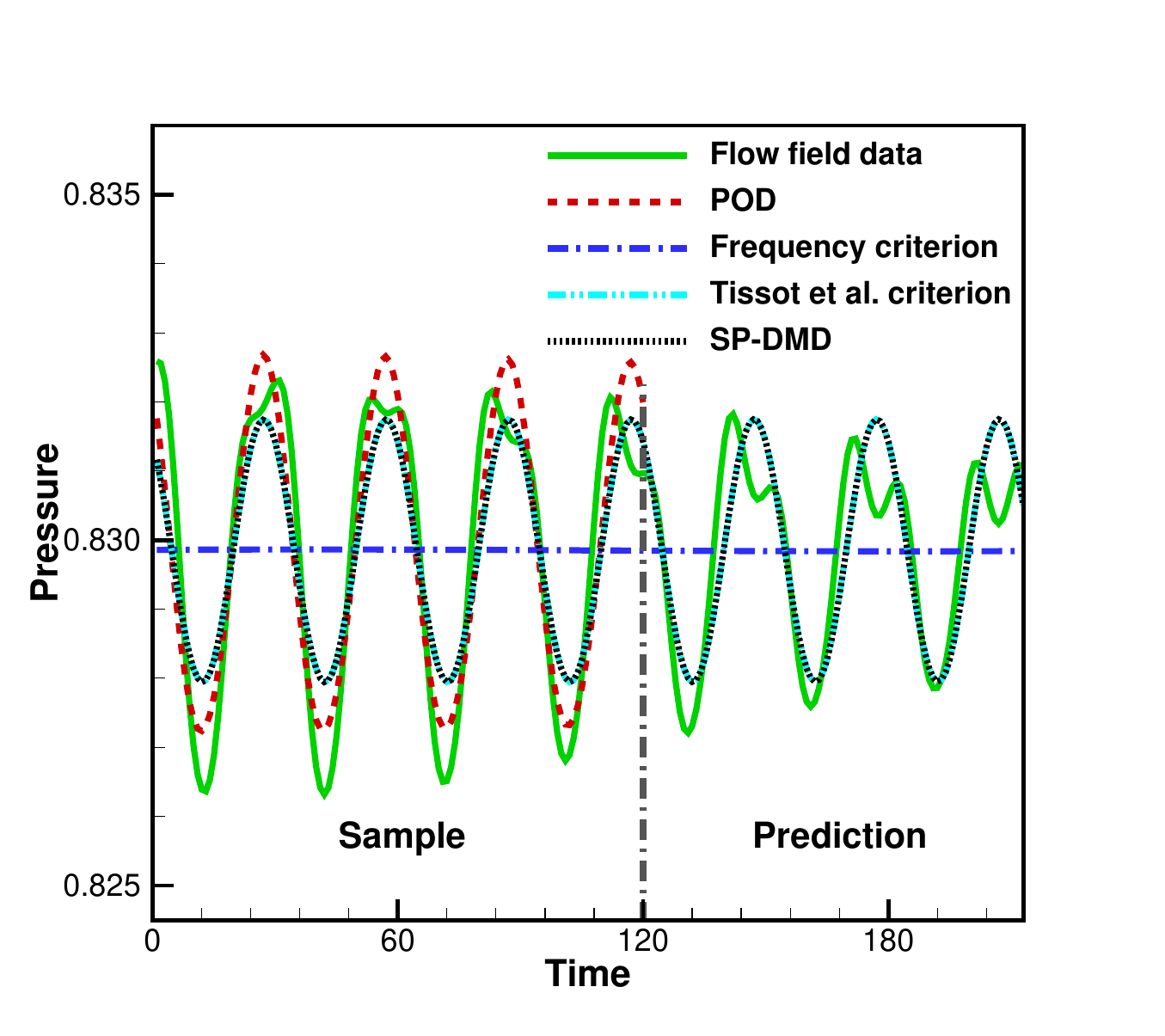}
        \caption{Three modes}
        \label{fig:reconstructed_point_mode=3}
    \end{subfigure}
    \begin{subfigure}[b]{0.495\linewidth}
        \centering
        \includegraphics[width=1\linewidth]{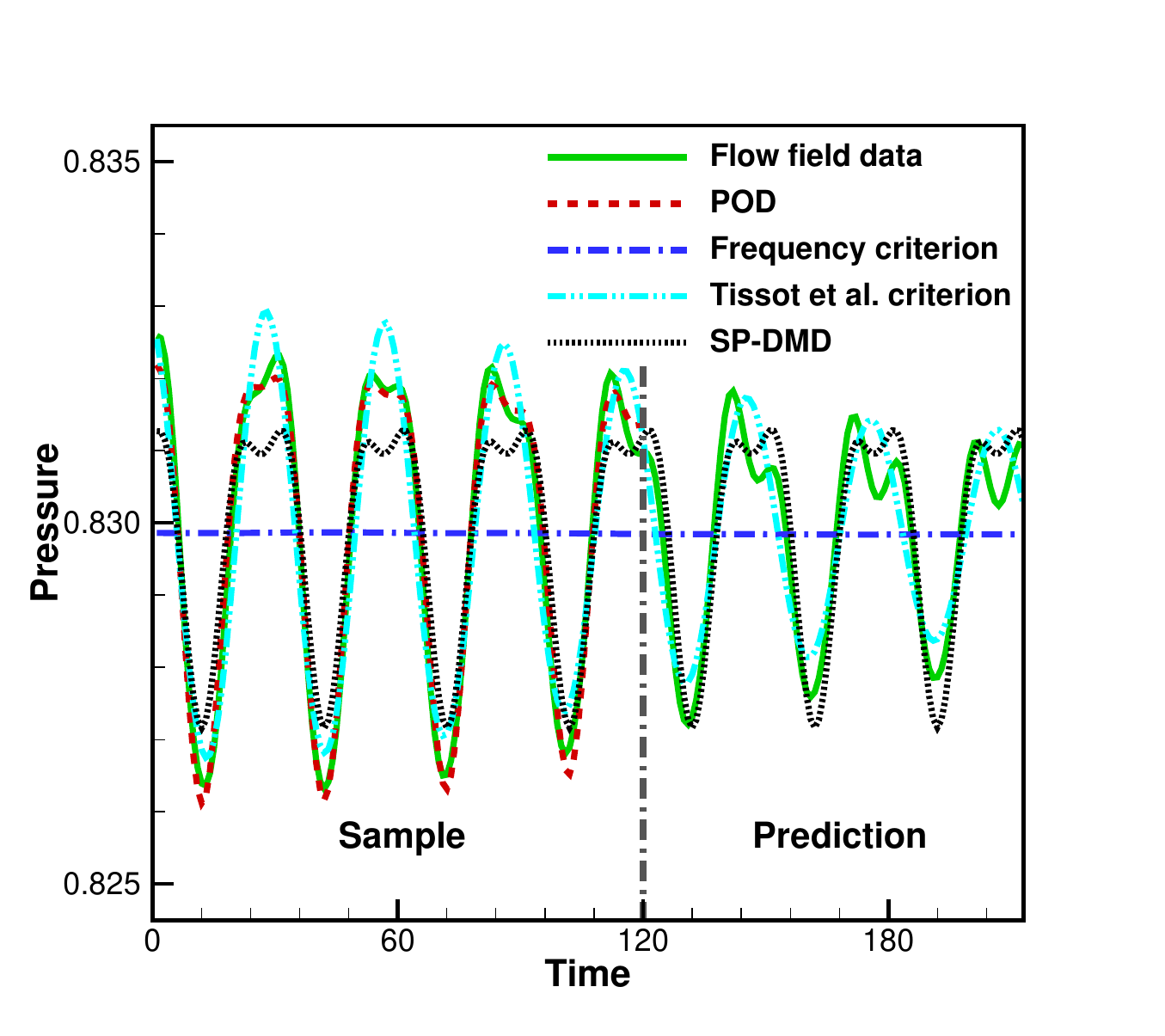}
        \caption{Five modes}
        \label{fig:reconstructed_point_mode=5}
    \end{subfigure}
    \begin{subfigure}[b]{0.495\linewidth}
        \centering
        \includegraphics[width=1\linewidth]{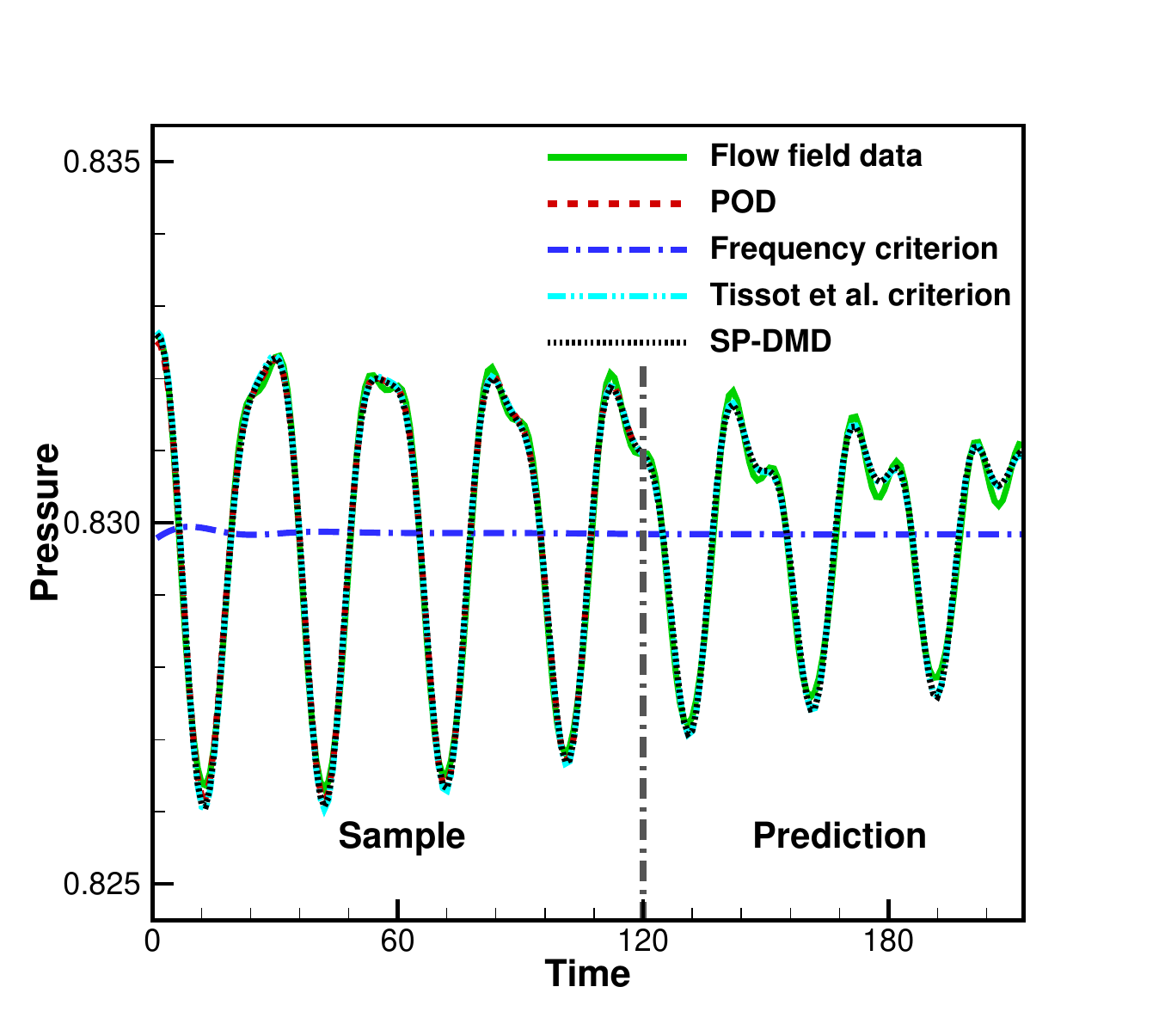}
        \caption{Seven modes}
        \label{fig:reconstructed_point_mode=7}
    \end{subfigure}
    \caption{Variation of reconstructed pressure at a point near the leading edge of Stator 2.}
    \label{fig:reconstructed_point_time_variation}
\end{figure}

The reconstruction error near Stator 2 inlet is further evaluated for different retained mode numbers.
The variation of reconstructed pressure at a grid point near the leading edge of Stator 2 with time step is shown in Fig. \ref{fig:reconstructed_point_time_variation}, along with the original pressure from the CFD results for comparison. The pressure is normalized by the total pressure at the turbine inlet. For DMD, the pressure can not only be reconstructed over the sampling duration but also predicted at future time. In Fig. \ref{fig:reconstructed_point_time_variation}, the first 120 time steps are used for sampling, while the remaining 90 time steps correspond to the prediction intervals. Since the reconstructed fields by using the amplitude and Tissot criteria are almost identical, only the result by the Tissot criterion is presented here for brevity. 
The pressure reconstructed by the frequency criterion rapidly decays to the time-averaged value because the leading modes exhibit high decay rates, as shown in Fig. \ref{fig:growth_rate-index_dmd}. This further demonstrates that the frequency criterion is not suitable for mode selection in the present problem. 

In Fig. \ref{fig:reconstructed_point_time_variation}, all modal decomposition methods, except for the DMD method based on the frequency criterion, effectively reconstruct and predict the evolution of unsteady pressure fluctuations. The weak decaying behavior observed in the original CFD results can be attributed to locally non-perfect periodicity of the flow.
With three retained modes, both POD and DMD methods accurately capture the periodic behavior of pressure fluctuation, and the predicted pressure is in phase with the simulated results. This periodicity represents the most dominant feature induced by the rotating motion of the rotor blades. However, all methods overpredict the pressure value and fail to capture the high-frequency fluctuations and decaying behavior of the original flow field, as only one transient mode pair is retained along with the time-averaged mode. 
With five retained modes, the POD method successfully predicts the high-frequency variation and decaying behavior of pressure. The DMD method based on the Tissot criterion successfully captures the decaying behavior in both sample and prediction regions. However, it still overpredicts the pressure amplitude and lacks high-frequency fluctuation components, similar to the case with only three retained modes. In contrast, the SP-DMD method captures the high-frequency pressure oscillations, but the predicted pressure amplitude is underestimated and does not exhibit the expected decay over time. Both cases indicate that retaining only five modes is insufficient for accurate DMD reconstruction, and additional dominant modes are required to achieve a more precise representation of the unsteady flow.
With seven retained modes, the reconstructed and predicted pressures by both POD and DMD methods, except the DMD method based on the frequency criterion, agree well with the original computational results. For the present problem, seven modes are sufficient to capture the unsteady pressure variations, even in highly unsteady regions.

\subsection{Mode Shapes} \label{sec:ROA}
\begin{figure}[htb!]
    \centering
    \begin{subfigure}[b]{0.24\linewidth}
        \centering
        \includegraphics[width=1\linewidth]{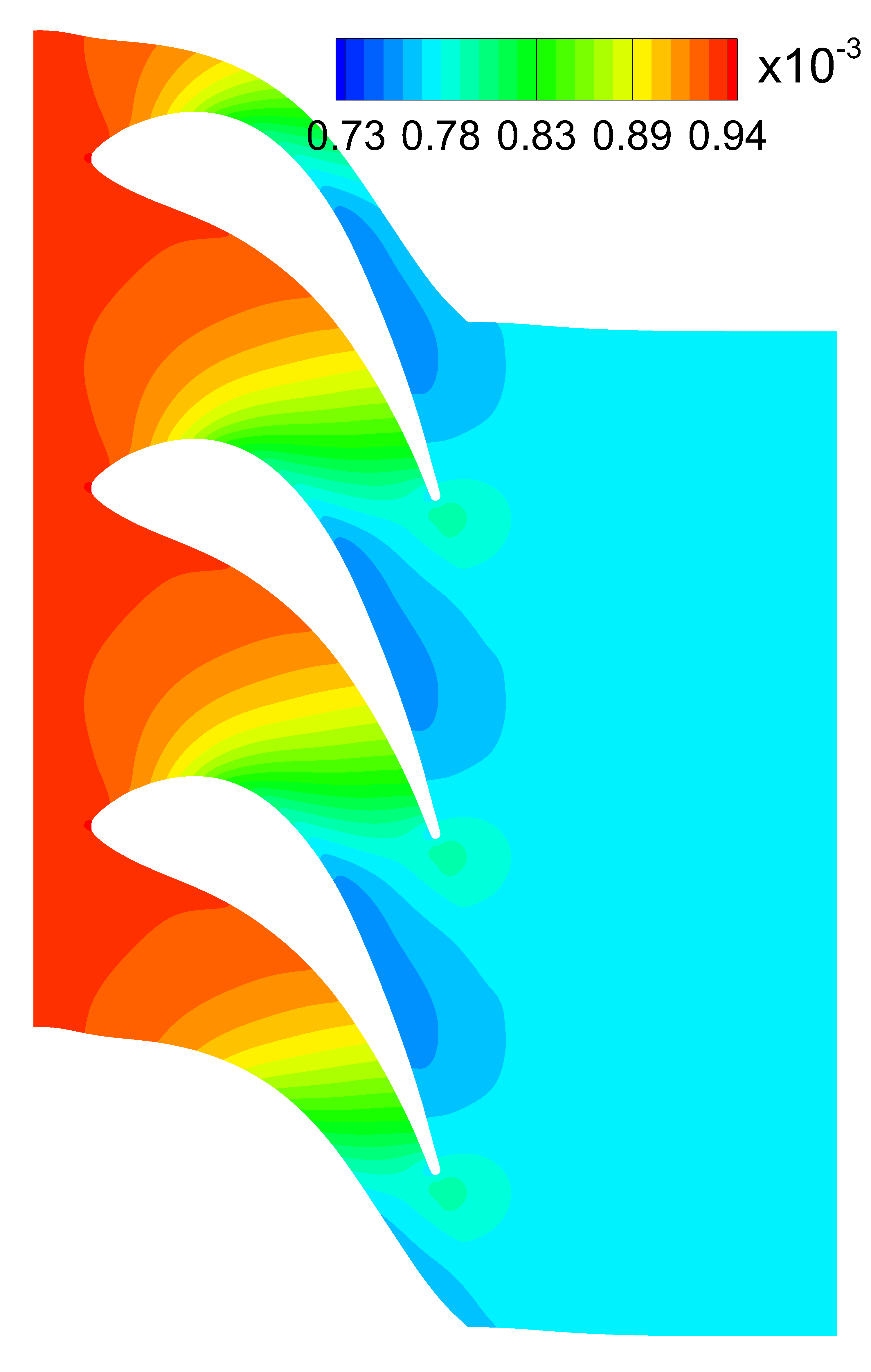}
        \caption{POD}
        \label{fig:contour_mode_pod_mode1}
    \end{subfigure}
    \begin{subfigure}[b]{0.24\linewidth}
        \centering
        \includegraphics[width=1\linewidth]{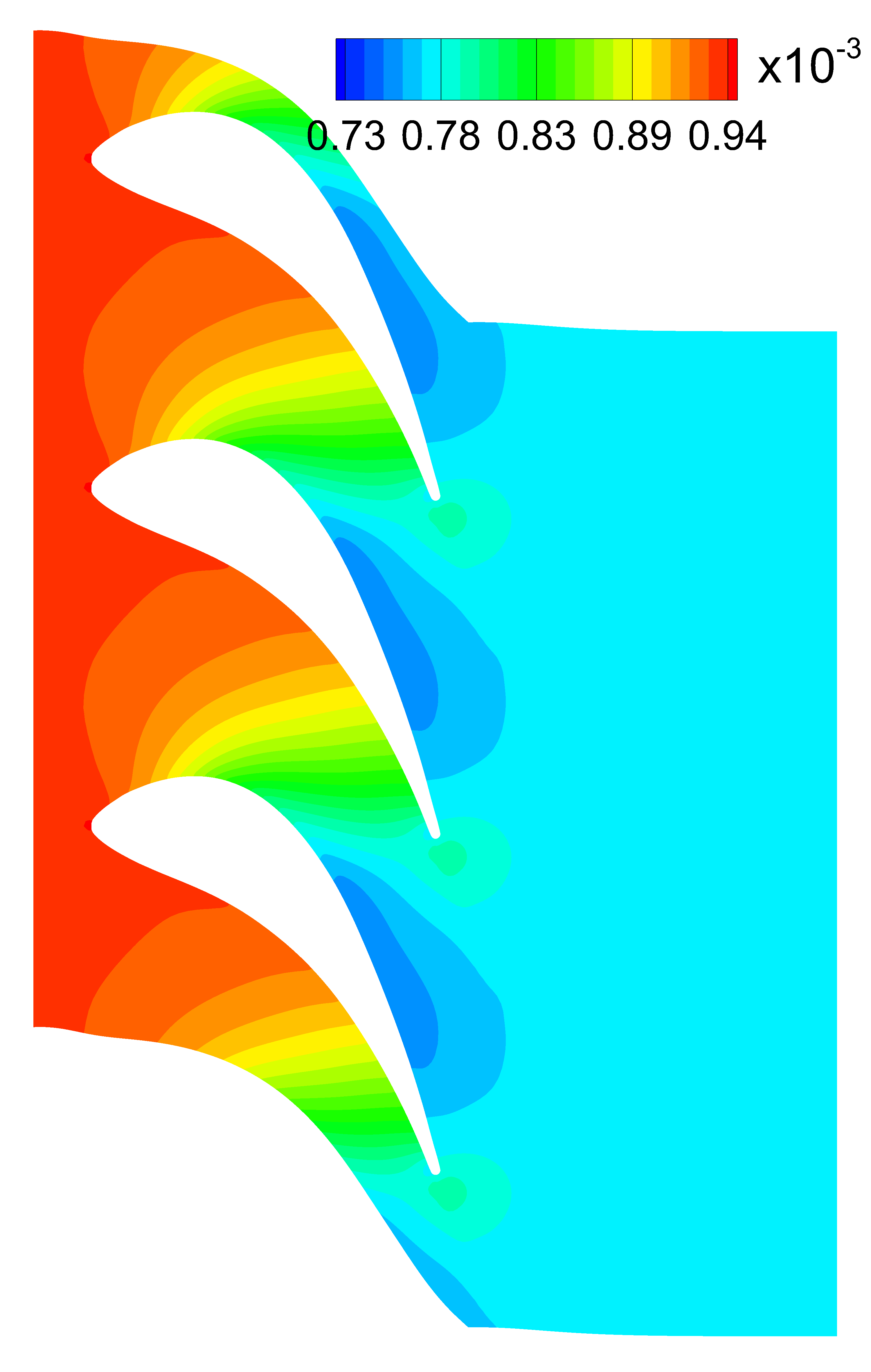}
        \caption{DMD}
        \label{fig:contour_mode_dmd_mode1}
    \end{subfigure}
    \begin{subfigure}[b]{0.24\linewidth}
        \centering
        \includegraphics[width=1\linewidth]{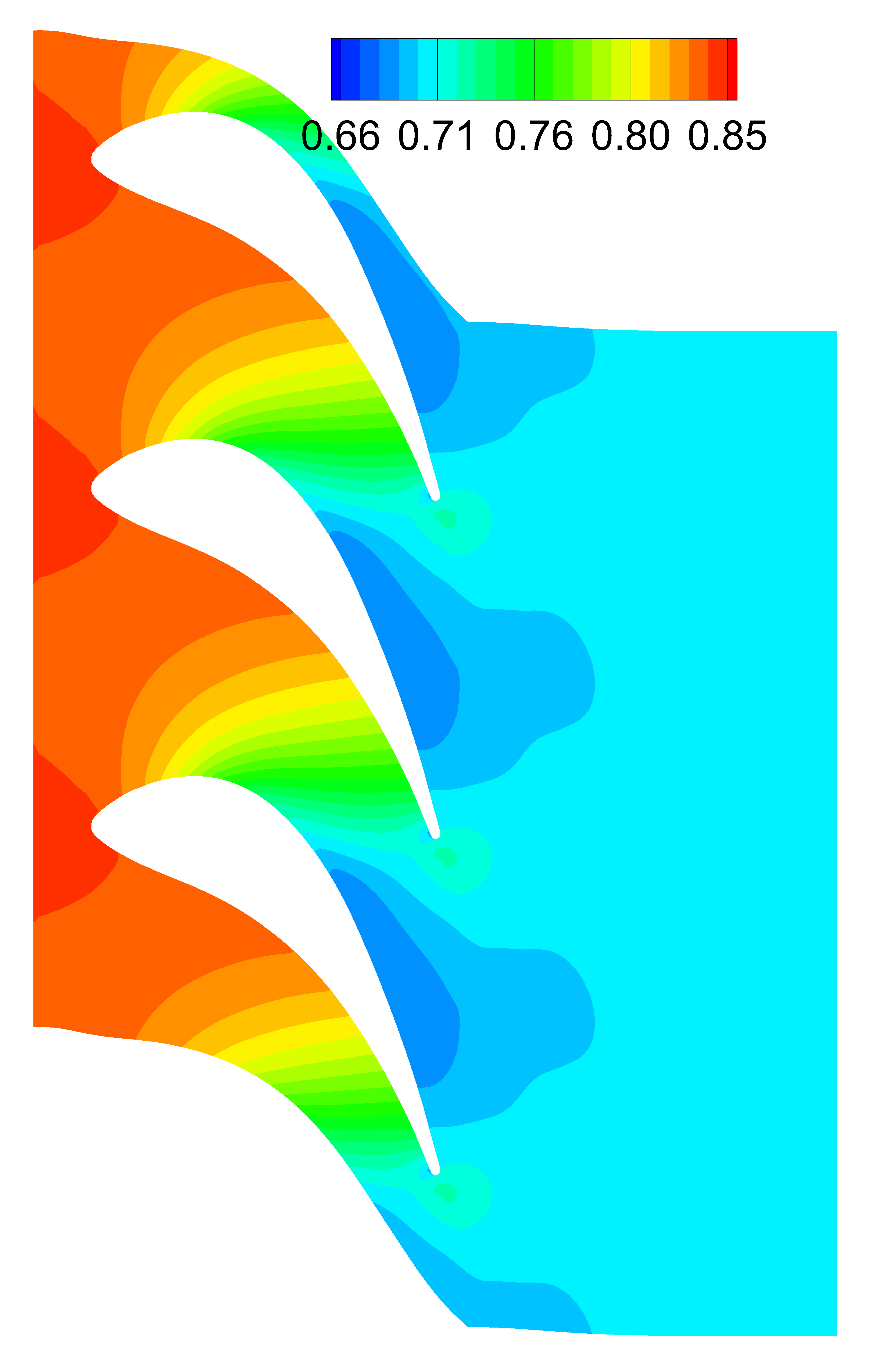}
        \caption{Time-averaged pressure}
        \label{fig:contour_averaged_pressure}
    \end{subfigure}
    \caption{Contours of first mode shapes and time-averaged pressure at 50\% span. DMD is based on the Tissot criterion.}
    \label{fig:contour_mode_pod_dmd_mode1}
\end{figure}

This section focuses on analyzing the spatial structures of the dominant modes and their relationship with the unsteady flow fields. As all DMD methods, except the frequency criterion, yield the same leading modes, only the results based on the Tissot criterion are presented here for brevity.
Figure \ref{fig:contour_mode_pod_dmd_mode1} compares the contours of the first POD and DMD mode shapes at 50\% span with the time-averaged pressure from the CFD solution. Since the time-averaged value is not subtracted from the snapshots before performing decomposition, the first mode should represent the mean flow field. In Fig. \ref{fig:contour_mode_pod_dmd_mode1}, the first mode shapes obtained from POD and DMD are nearly identical, and both closely resemble the time-averaged pressure. This demonstrates that both POD and DMD successfully capture the most important flow structure in the turbine.

\begin{figure}[htb!]
    \centering
    \begin{subfigure}[b]{0.24\linewidth}
        \centering
        \includegraphics[width=1\linewidth]{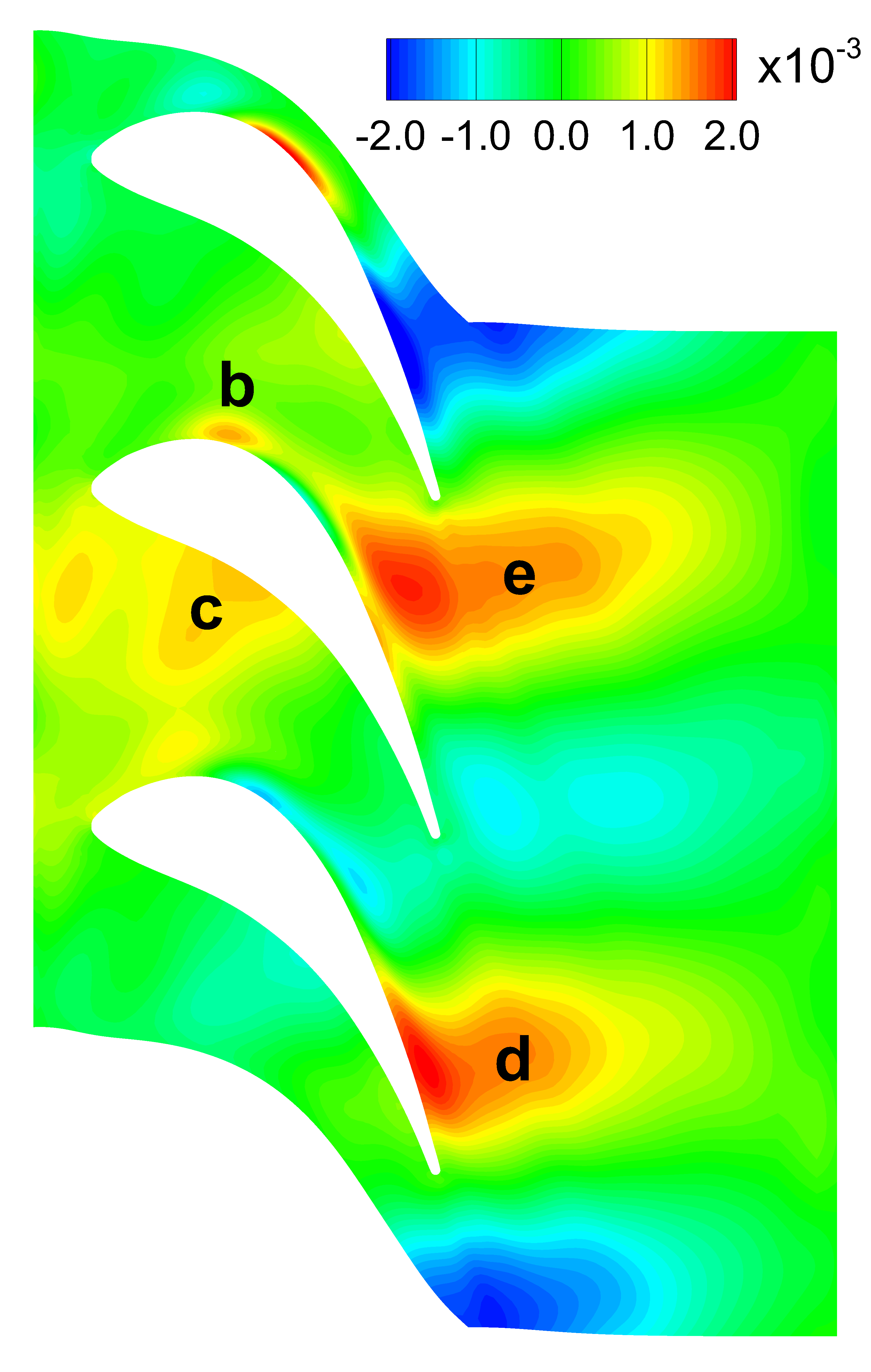}
        \caption{2nd POD mode}
        \label{fig:contour_mode_pod_mode2}
    \end{subfigure}
    \begin{subfigure}[b]{0.24\linewidth}
        \centering
        \includegraphics[width=1\linewidth]{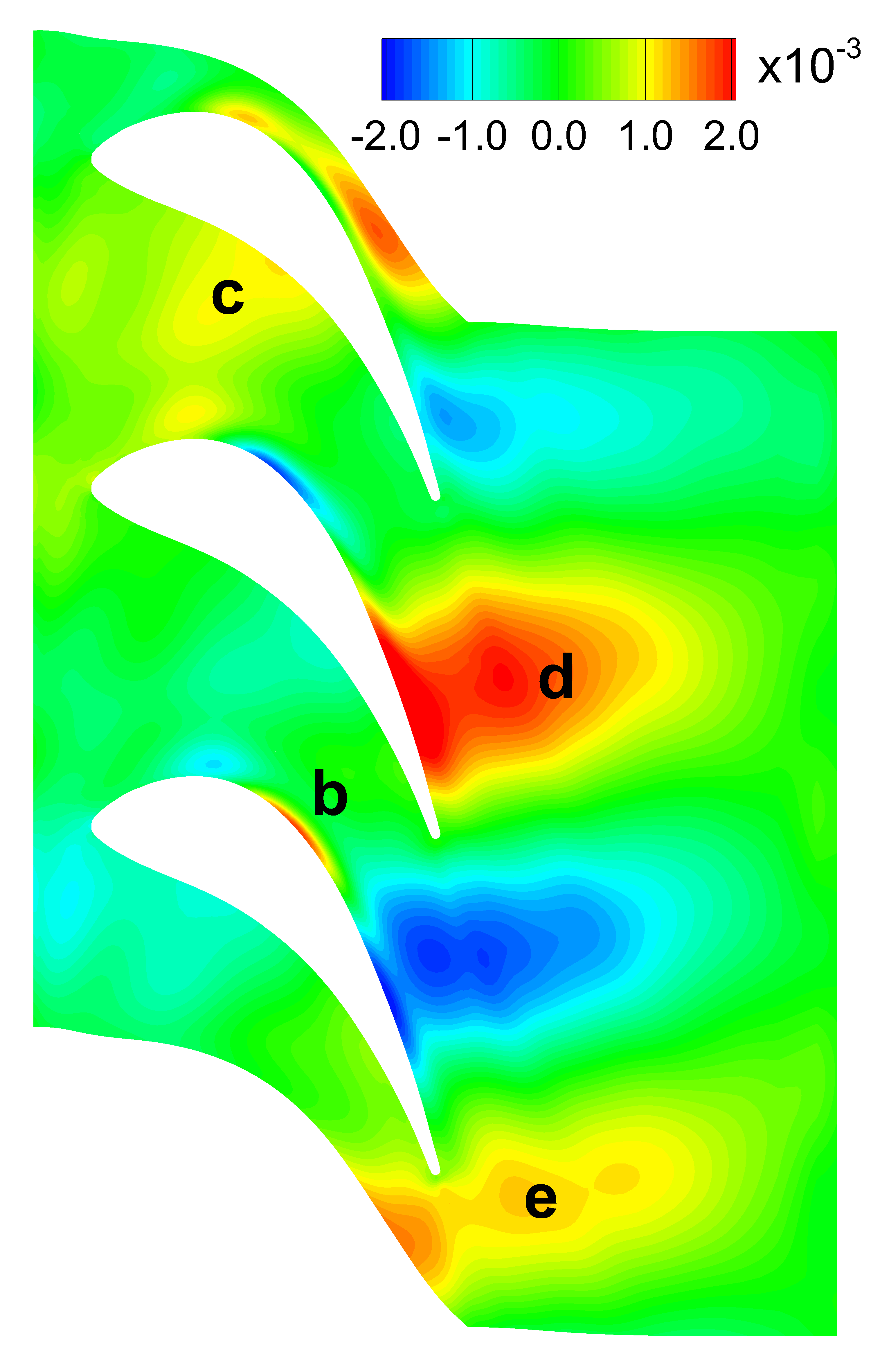}
        \caption{3rd POD mode}
        \label{fig:contour_mode_pod_mode3}
    \end{subfigure}
    \begin{subfigure}[b]{0.24\linewidth}
        \centering
        \includegraphics[width=1\linewidth]{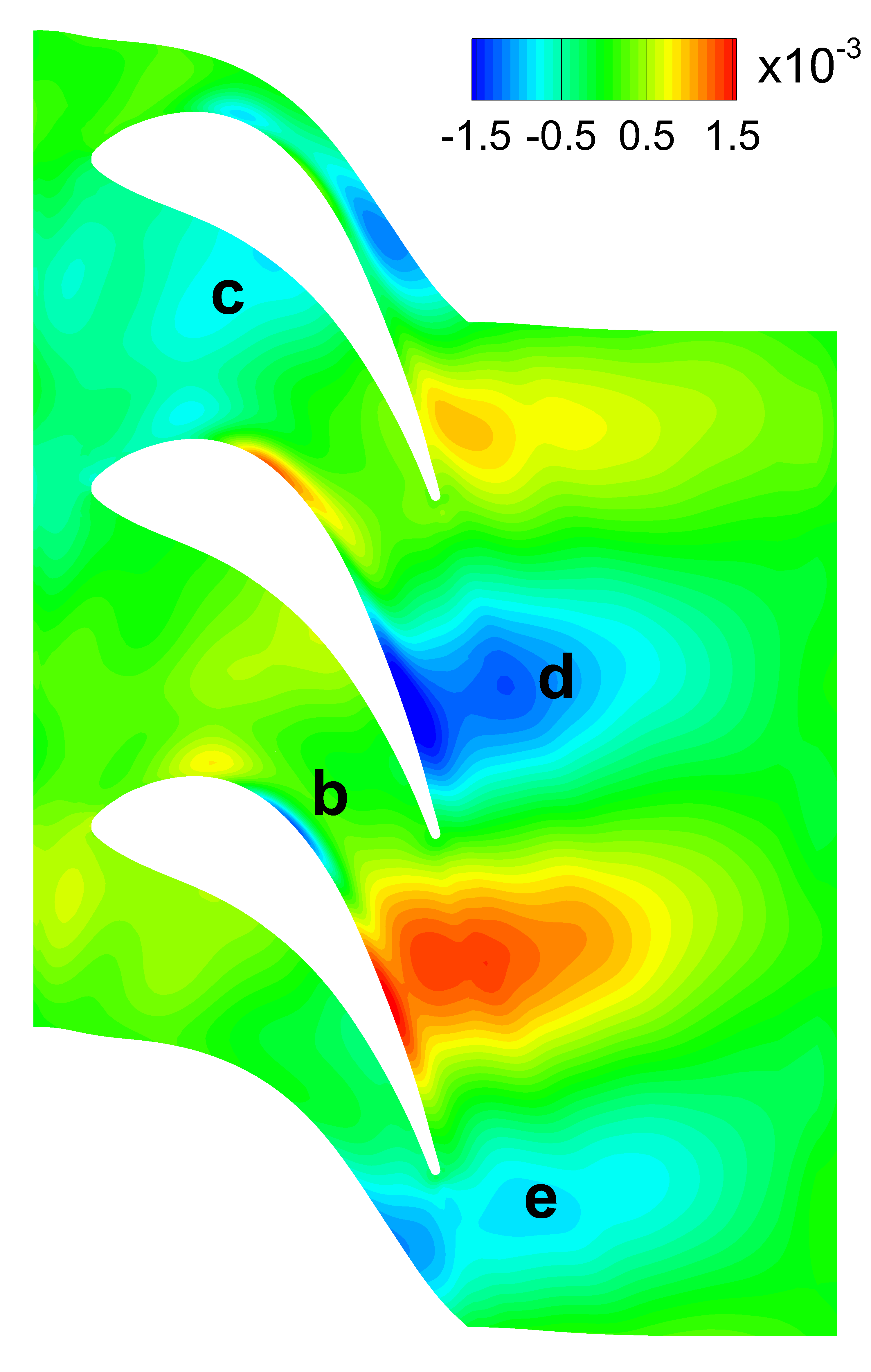}
        \caption{2nd DMD mode}
        \label{fig:contour_mode_dmd_mode2}
    \end{subfigure}
    \begin{subfigure}[b]{0.24\linewidth}
        \centering
        \includegraphics[width=1\linewidth]{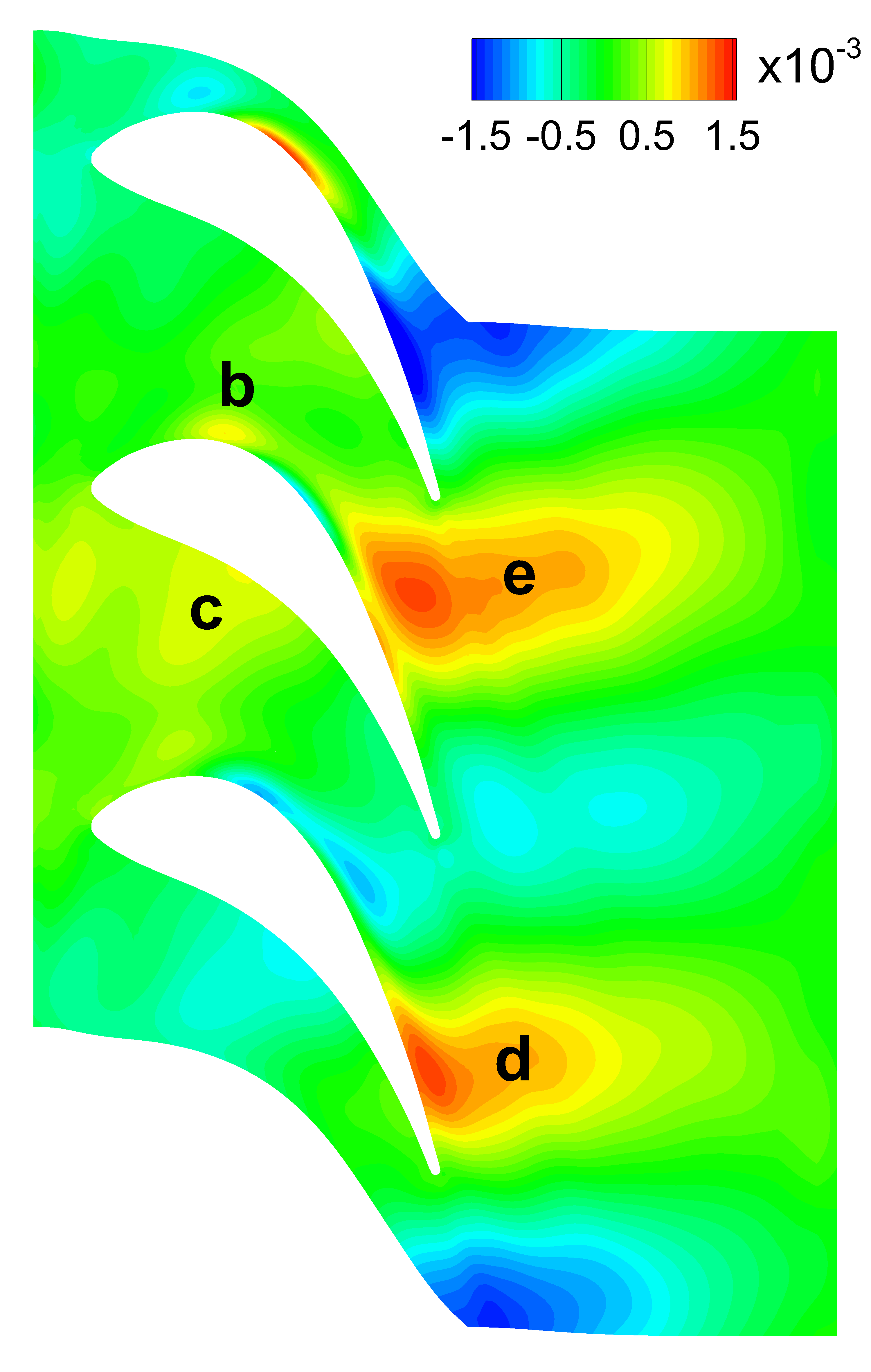}
        \caption{3rd DMD mode}
        \label{fig:contour_mode_dmd_mode3}
    \end{subfigure}
    \caption{Contours of second and third mode shapes at 50\% span. DMD is based on the Tissot criterion. The two DMD modes form a complex-conjugate pair, with the second mode representing the real part.}
    \label{fig:contour_mode_pod_dmd_mode2-3}
\end{figure}

Figure \ref{fig:contour_mode_pod_dmd_mode2-3} shows the contours of the second and third mode shapes at 50\% span. The two DMD modes form a complex-conjugate pair, with the second mode representing the real part. They exhibit similar spatial distributions, differing only slightly in the locations of their peaks and troughs. The second and third POD modes display comparable patterns to those of the DMD modes. Specifically, the second POD mode exhibits strong similarity to the third DMD mode, while the third POD mode is close to the second DMD mode but with an opposite sign. As the real and imaginary parts of a complex-conjugate pair, the ordering of the second and third DMD modes does not imply any precedence between them. This indicates that, for this two leading modes of pressure fluctuations, POD and DMD yield similar spatial structures.

\begin{figure}[htb!]
  \centering
  \includegraphics[width=0.25\linewidth]{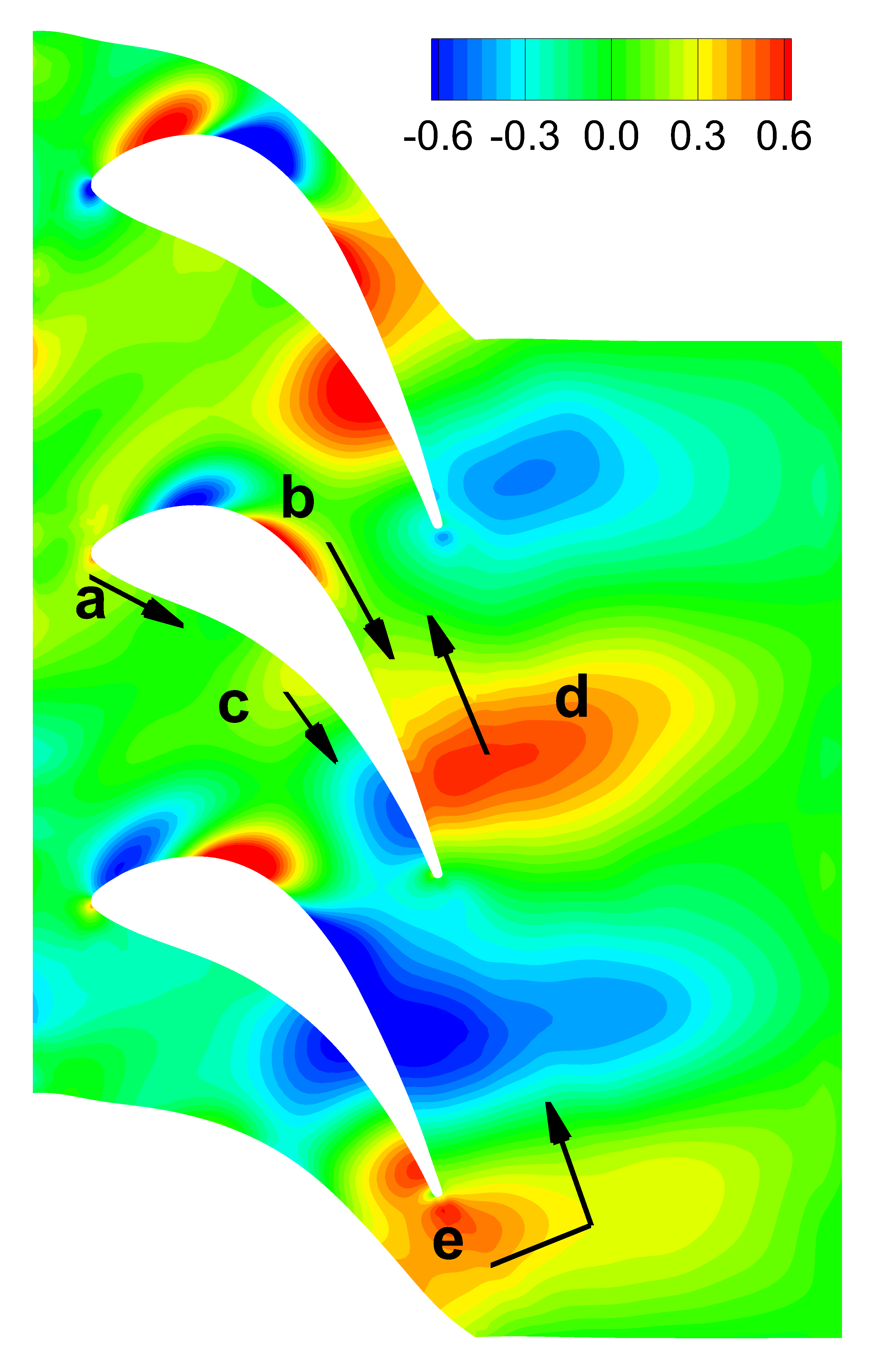}
  \caption{Contours of pressure fluctuation at a time instant at 50\% span in CFD solution.}
  \label{fig:contour_pressure_flucutation}
\end{figure}

The pressure peaks labeled as $b$, $c$, $d$, and $e$ in Fig. \ref{fig:contour_mode_pod_dmd_mode2-3} correspond to the regions of positive pressure fluctuations in the original CFD results shown in Fig. \ref{fig:contour_pressure_flucutation}. Here, pressure fluctuation is defined as the percentage deviation of the instantaneous pressure from its time-averaged value. Arrows adjacent to each label indicate the movement direction of the corresponding pressure fluctuation. The generation and propagation of these pressure structures were thoroughly analyzed in Ref. \cite{2017_Zhu_clocking} and are briefly reviewed here. 
The pressure fluctuation $a$, generated at the leading edge of Stator 2, and fluctuation $c$, originating from the middle of the pressure surface, propagate downstream along the blade passage. On the suction surface, fluctuation $b$ from the leading edge and $d$ from the trailing edge merge at 75\% axial chord. The merged fluctuation leaves the suction surface and interacts with the fluctuation $c$ from the pressure surface of the adjacent blade. The resulting combined fluctuation $e$ then diffracts at the trailing edge of the adjacent blade and propagates upstream along its suction surface.
A comparison between Figs. \ref{fig:contour_mode_pod_dmd_mode2-3} and \ref{fig:contour_pressure_flucutation} confirms that both POD and DMD effectively capture the dominant unsteady pressure features in Stator 2 passage. However, neither method captures pressure fluctuation $a$ shown in Fig. \ref{fig:contour_pressure_flucutation}, as it originates at the blade leading edge and decays rapidly. This transient feature requires more modes for accurate representation.

\begin{figure}[htb!]
    \centering
    \begin{subfigure}[b]{0.24\linewidth}
        \centering
        \includegraphics[width=1\linewidth]{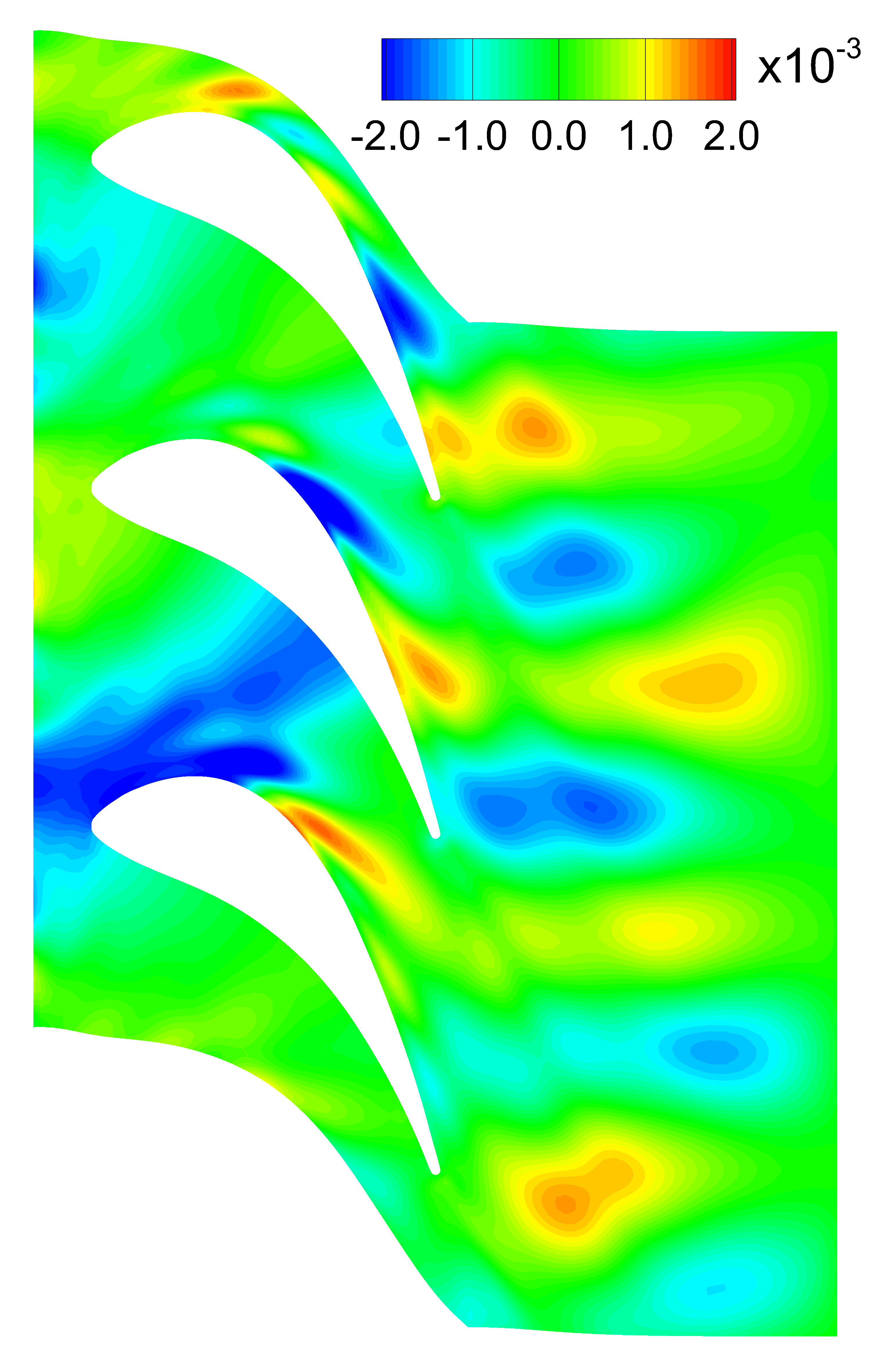}
        \caption{4th POD mode}
        \label{fig:contour_mode_pod_mode4}
    \end{subfigure}
    \begin{subfigure}[b]{0.24\linewidth}
        \centering
        \includegraphics[width=1\linewidth]{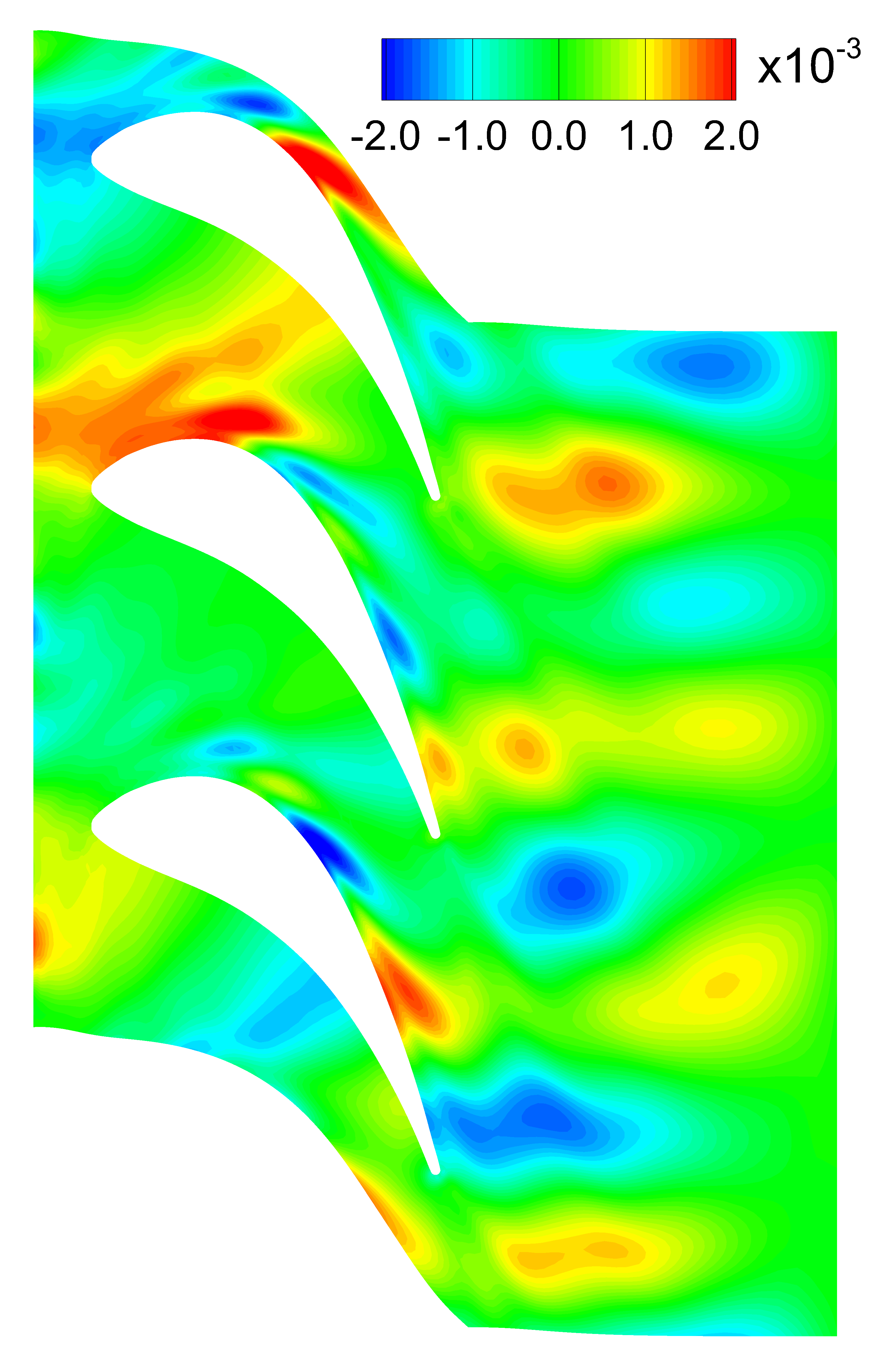}
        \caption{5th POD mode}
        \label{fig:contour_mode_pod_mode5}
    \end{subfigure}
    \begin{subfigure}[b]{0.24\linewidth}
        \centering
        \includegraphics[width=1\linewidth]{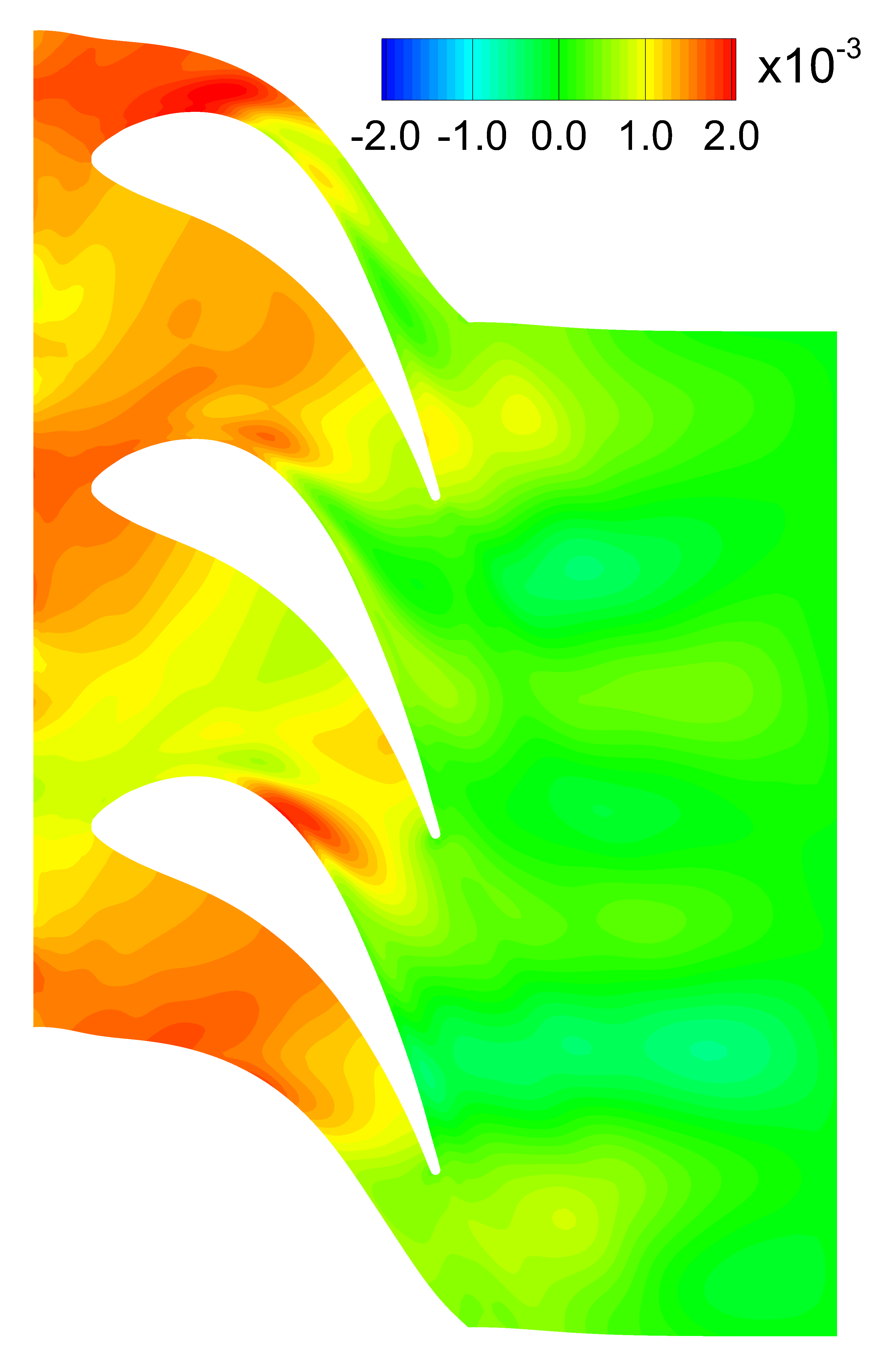}
        \caption{6th POD mode}
        \label{fig:contour_mode_pod_mode6}
    \end{subfigure}
    \begin{subfigure}[b]{0.24\linewidth}
        \centering
        \includegraphics[width=1\linewidth]{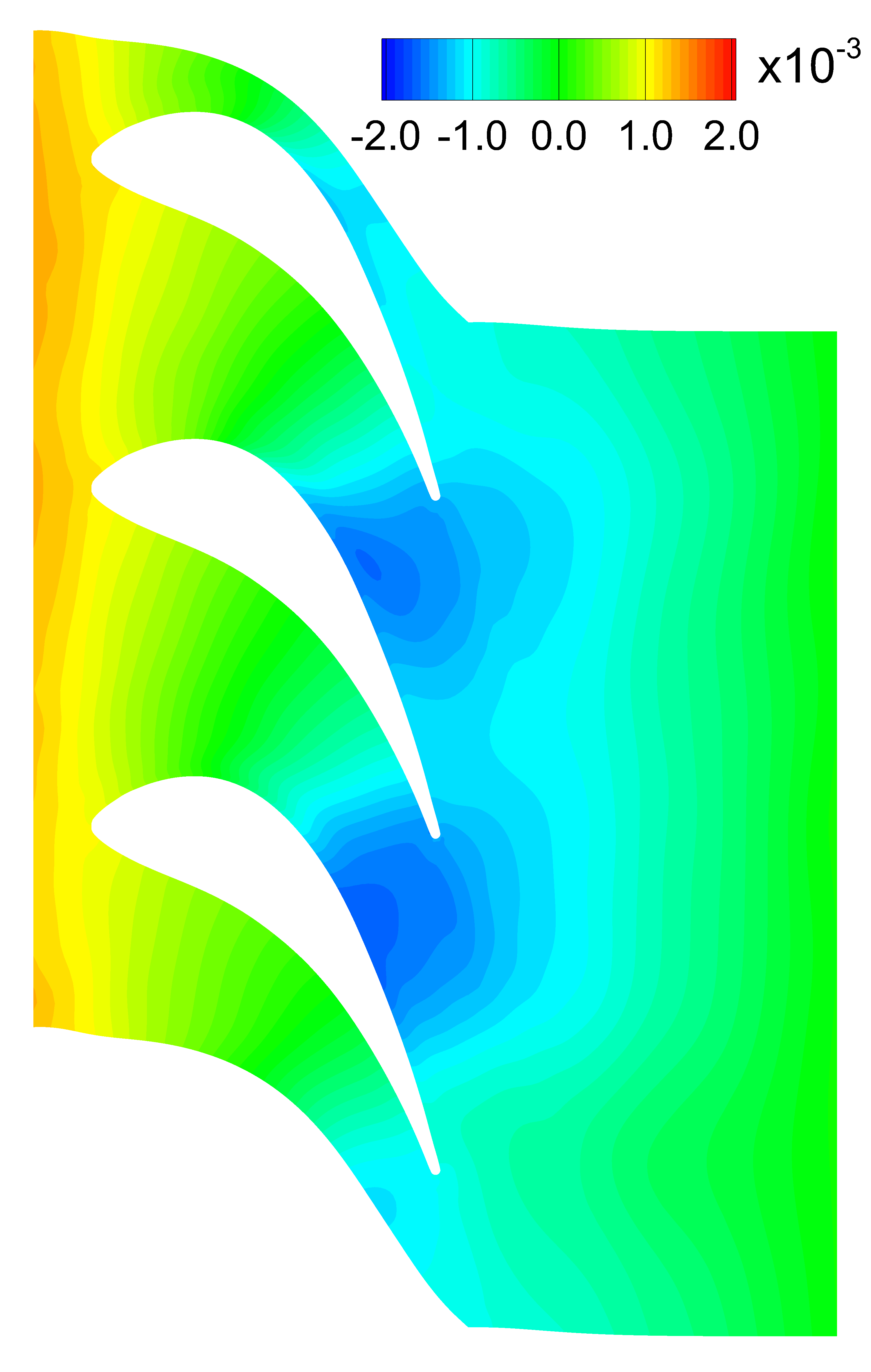}
        \caption{7th POD mode}
        \label{fig:contour_mode_pod_mode7}
    \end{subfigure}
    \begin{subfigure}[b]{0.24\linewidth}
        \centering
        \includegraphics[width=1\linewidth]{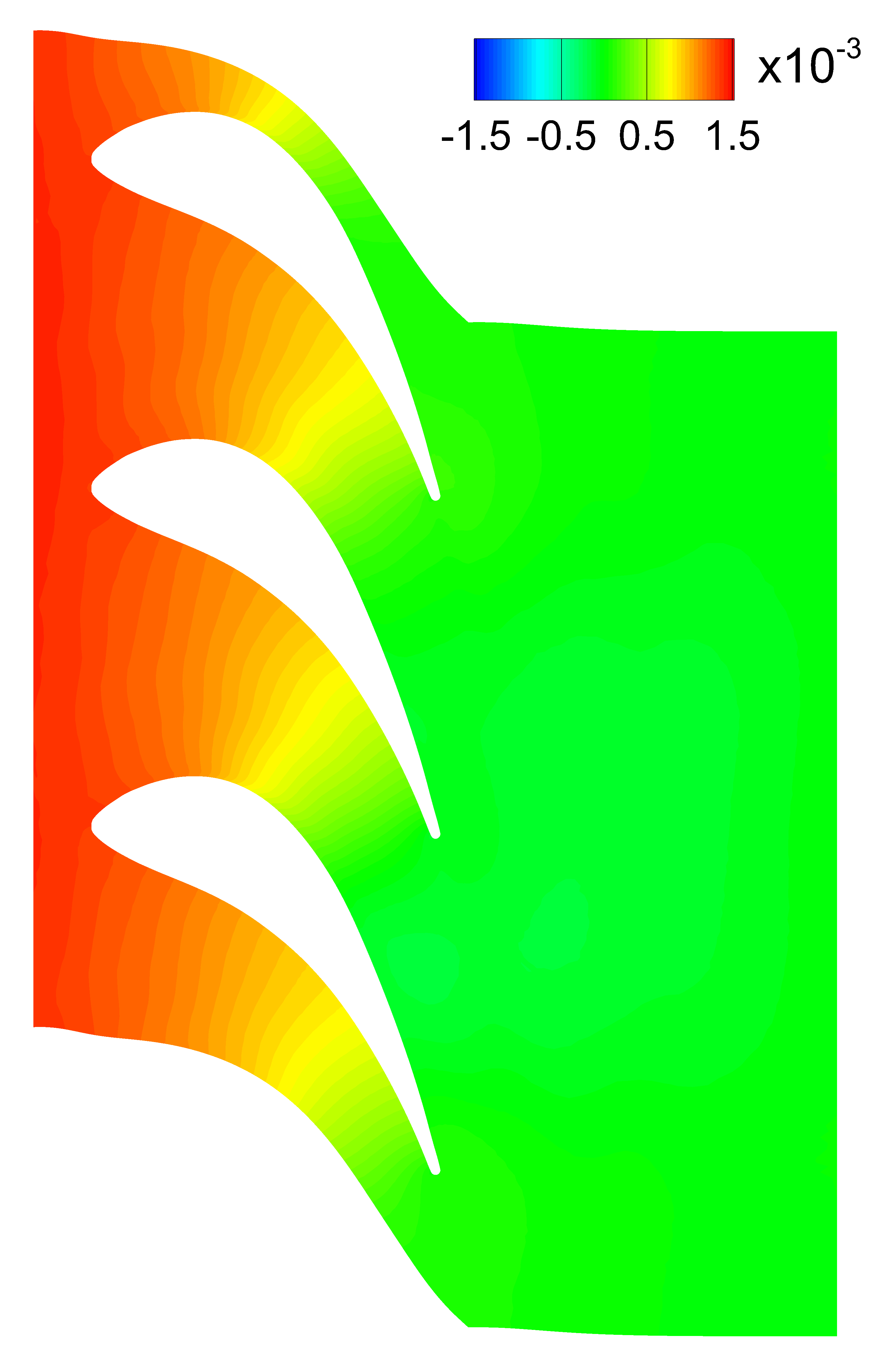}
        \caption{4th DMD mode}
        \label{fig:contour_mode_dmd_mode4}
    \end{subfigure}
    \begin{subfigure}[b]{0.24\linewidth}
        \centering
        \includegraphics[width=1\linewidth]{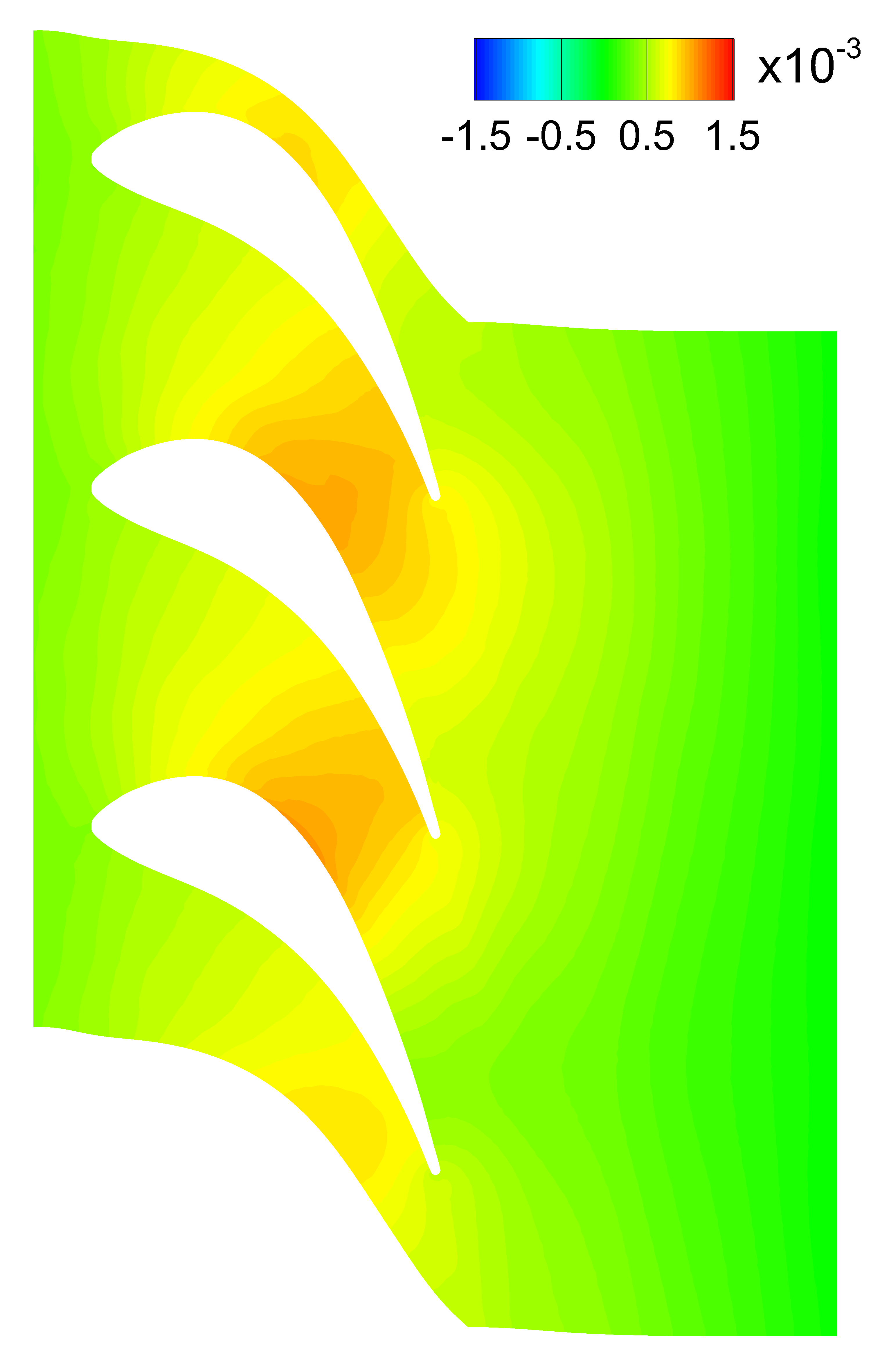}
        \caption{5th DMD mode}
        \label{fig:contour_mode_dmd_mode5}
    \end{subfigure}
    \begin{subfigure}[b]{0.24\linewidth}
        \centering
        \includegraphics[width=1\linewidth]{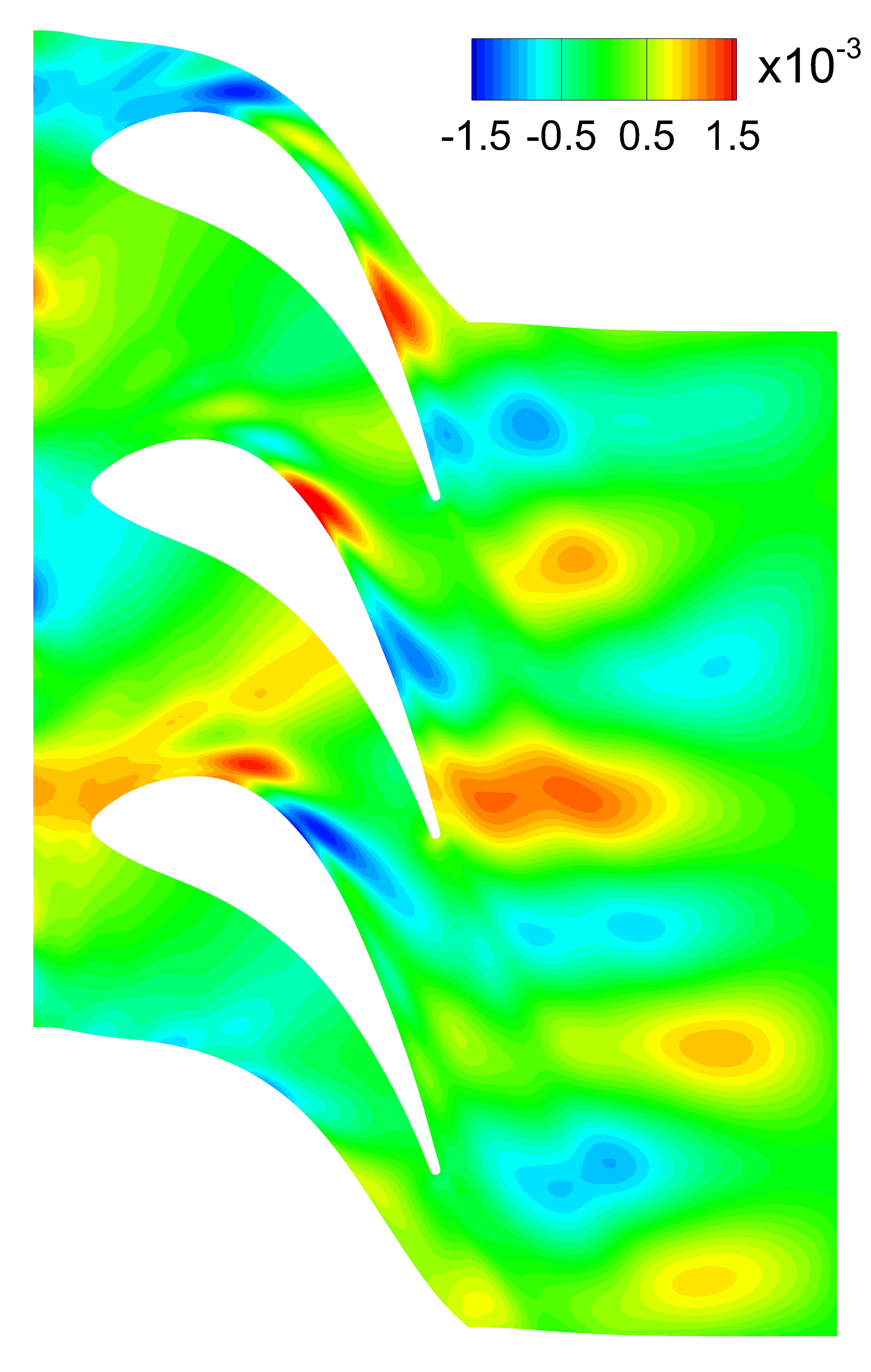}
        \caption{6th DMD mode}
        \label{fig:contour_mode_dmd_mode6}
    \end{subfigure}
    \begin{subfigure}[b]{0.24\linewidth}
        \centering
        \includegraphics[width=1\linewidth]{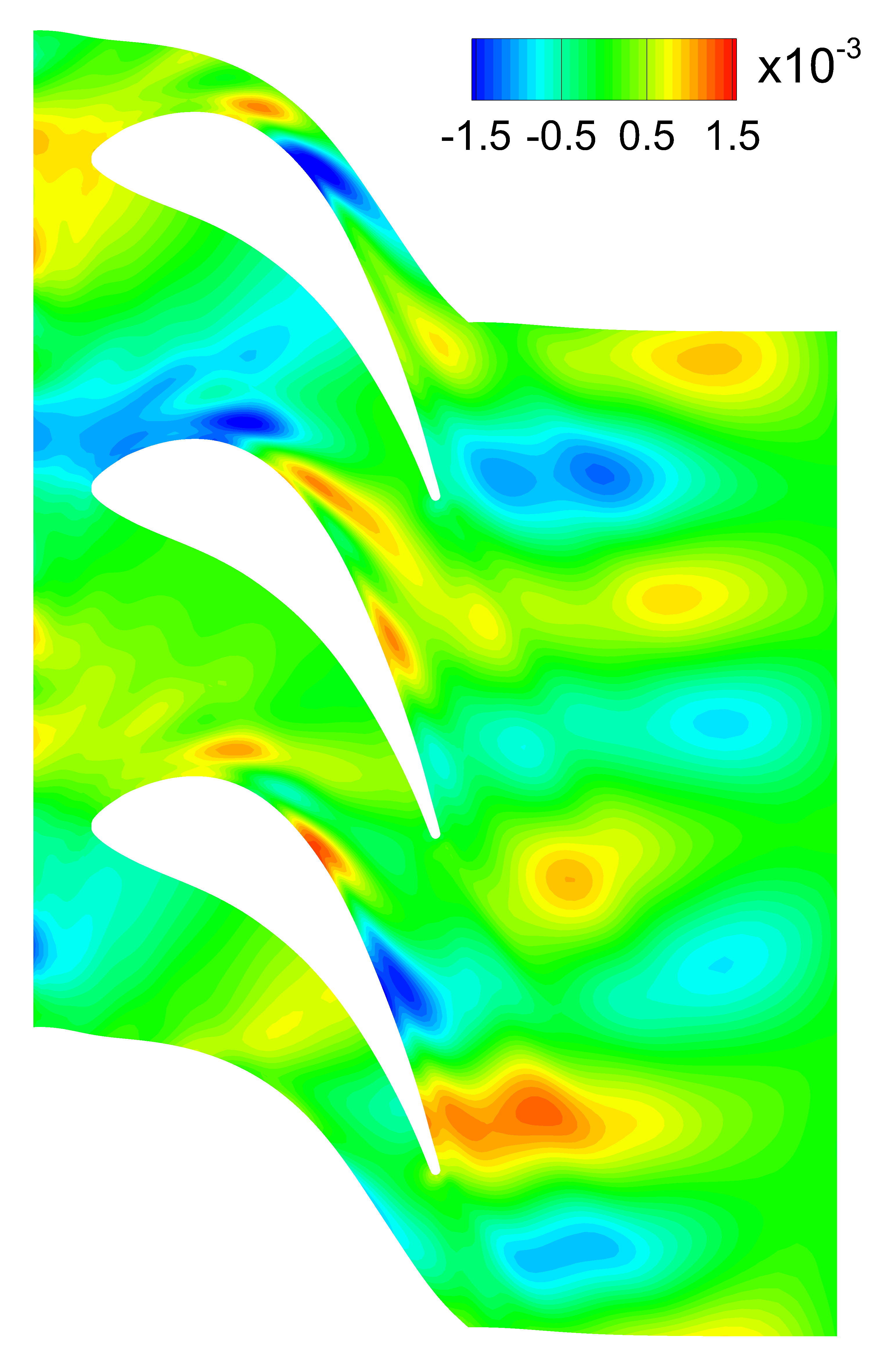}
        \caption{7th DMD mode}
        \label{fig:contour_mode_dmd_mode7}
    \end{subfigure}
    \caption{Contours of fourth to seventh mode shapes at 50\% span. DMD is based on the Tissot criterion. In the DMD results, 4th and 5th modes, and 6th and 7th modes form complex-conjugate pairs, respectively.}
    \label{fig:contour_mode_pod_dmd_mode4-7}
\end{figure}

Figure \ref{fig:contour_mode_pod_dmd_mode4-7} shows the contours of the fourth to seventh mode shapes at 50\% span. 
As in Fig. \ref{fig:contour_mode_pod_dmd_mode2-3}, the real and imaginary parts of each complex-conjugate DMD pair (specifically, the fourth and fifth modes, and the sixth and seventh modes) exhibit similar spatial distributions. 
The fourth and fifth POD mode shapes closely resemble the sixth and seventh DMD mode shapes, respectively, but with opposite signs. These modes exhibit patterns similar to those in Fig. \ref{fig:contour_mode_pod_dmd_mode2-3}, but with higher-order structures of pressure fluctuations, particularly in the region downstream of the trailing edge. The presence of these higher-order components will be further confirmed in Sec. \ref{sec:mode_dynamics}.
The seventh POD mode resembles the spatial structures of the fourth and fifth DMD modes. These modes exhibit nearly uniform distributions in the transverse direction of the blade passage, while showing decaying behavior along the streamwise direction. This is consistent with the negative growth rates of the fourth and fifth DMD modes shown in Fig. \ref{fig:growth_rate-index_dmd}.
The sixth POD mode appears to be a hybrid (not linear combination) of the fifth and seventh POD modes. 

\begin{figure}[htb!]
    \centering
    \begin{subfigure}[b]{0.24\linewidth}
        \centering
        \includegraphics[width=1\linewidth]{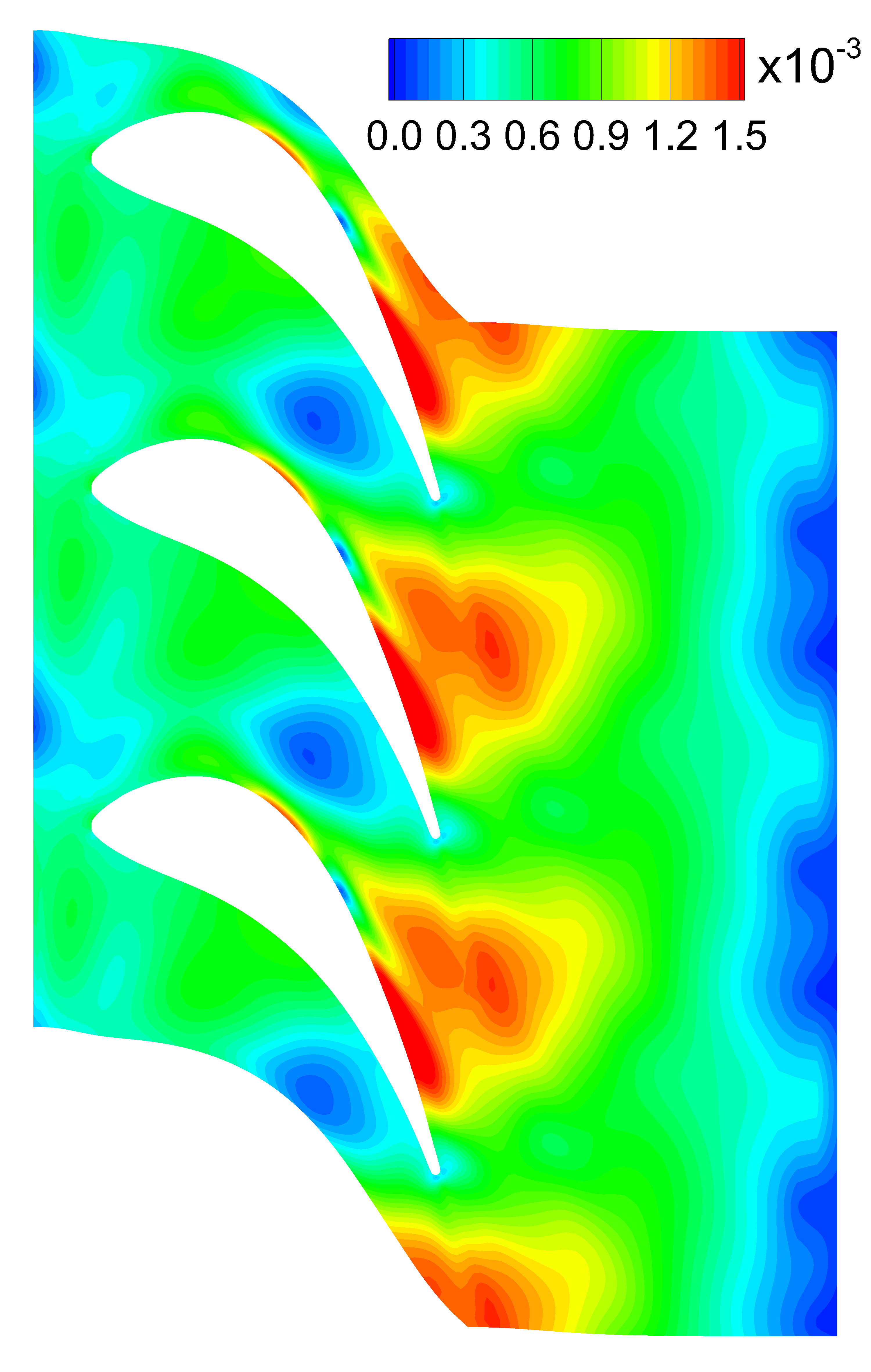}
        \caption{2nd and 3rd mode pair}
        \label{fig:contour_mode_dmd_2_3}
    \end{subfigure}
    \begin{subfigure}[b]{0.24\linewidth}
        \centering
        \includegraphics[width=1\linewidth]{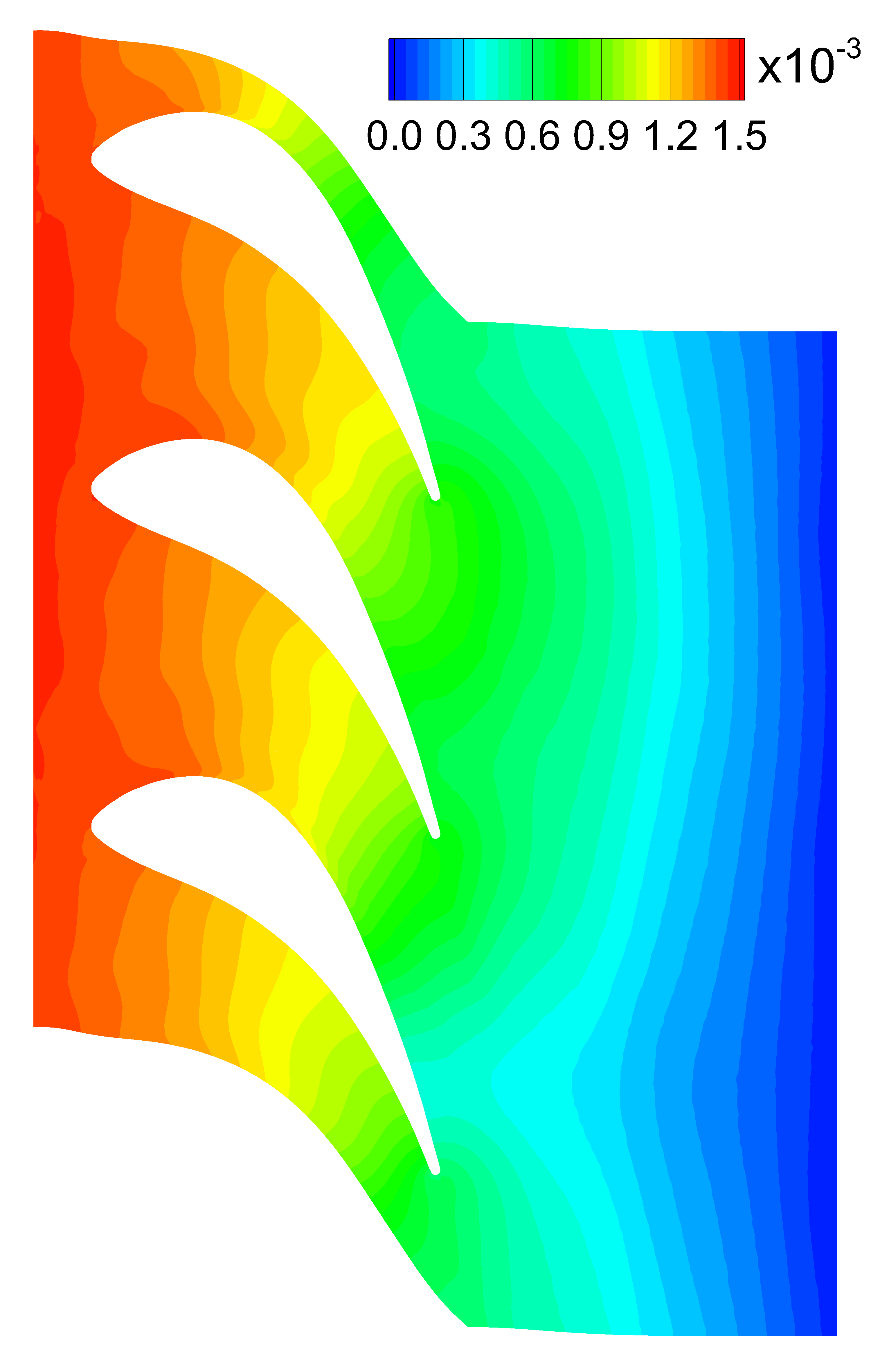}
        \caption{4th and 5th mode pair}
        \label{fig:contour_mode_dmd_4_5}
    \end{subfigure}
    \begin{subfigure}[b]{0.24\linewidth}
        \centering
        \includegraphics[width=1\linewidth]{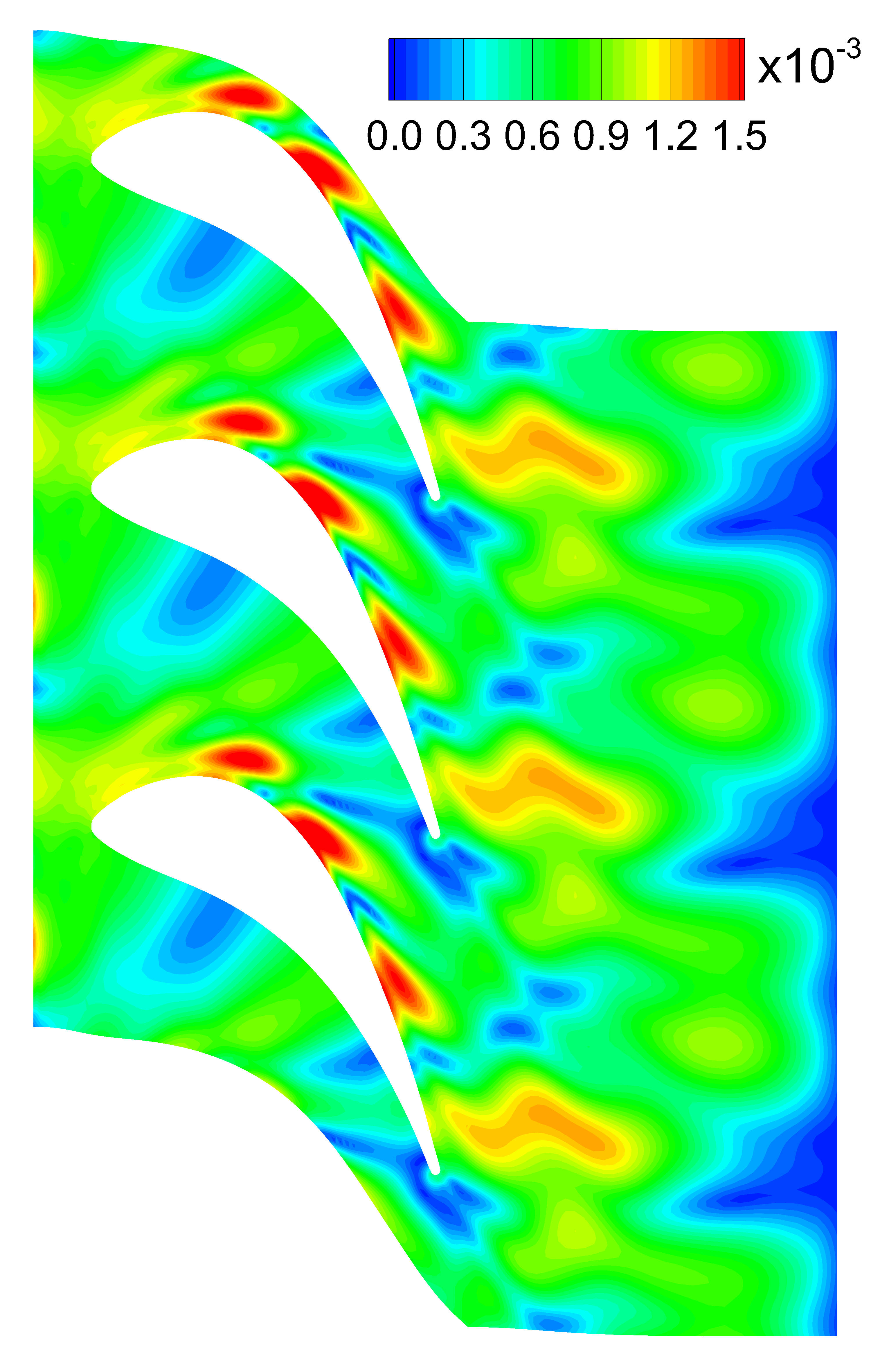}
        \caption{6th and 7th mode pair}
        \label{fig:contour_mode_dmd_6_7}
    \end{subfigure}
    \caption{Contours of modulus of DMD mode pair at 50\% span. DMD is based on the Tissot criterion.}
    \label{fig:contour_mode_dmd_modulus}
\end{figure}

The modulus of each complex-conjugate DMD mode pair discussed in Figs. \ref{fig:contour_mode_pod_dmd_mode2-3} and \ref{fig:contour_mode_pod_dmd_mode4-7} is further examined in Fig. \ref{fig:contour_mode_dmd_modulus}. For each mode pair, the modulus contours are identical across the three blade passages, as each blade experiences the same perturbations from Rotor and Stator 1 during one periodic cycle of the system. The pressure fluctuations near the outlet are negligible because no downstream rotor-stator interaction is considered and a fixed pressure boundary condition is imposed there.
The high-modulus regions in Fig. \ref{fig:contour_mode_dmd_2_3} are consistent with the motion areas of the fluctuations $b$ and $d$ on the suction surface in Fig. \ref{fig:contour_pressure_flucutation}, both of which persist for the longest duration within the stator passage. 
In Fig. \ref{fig:contour_mode_dmd_4_5}, the strong fluctuations originating at Stator 2 inlet decay rapidly along the streamwise direction. This implies that the fourth and fifth modes not only decay temporally, as indicated by their negative growth rates in Fig. \ref{fig:growth_rate-index_dmd}, but also exhibit spatially decaying behavior.
The behavior in Fig. \ref{fig:contour_mode_dmd_6_7} is similar to that in Fig. \ref{fig:contour_mode_dmd_2_3}, but features four high-modulus regions in the stator passage compared with two in Fig. \ref{fig:contour_mode_dmd_2_3}.
This higher-order behavior will be revealed by examining the mode dynamics in Sec. \ref{sec:mode_dynamics}.

\subsection{Mode Dynamics} \label{sec:mode_dynamics}
\begin{figure}[htb!]
    \centering
    \begin{subfigure}[b]{0.495\linewidth}
        \centering
        \includegraphics[width=1\linewidth]{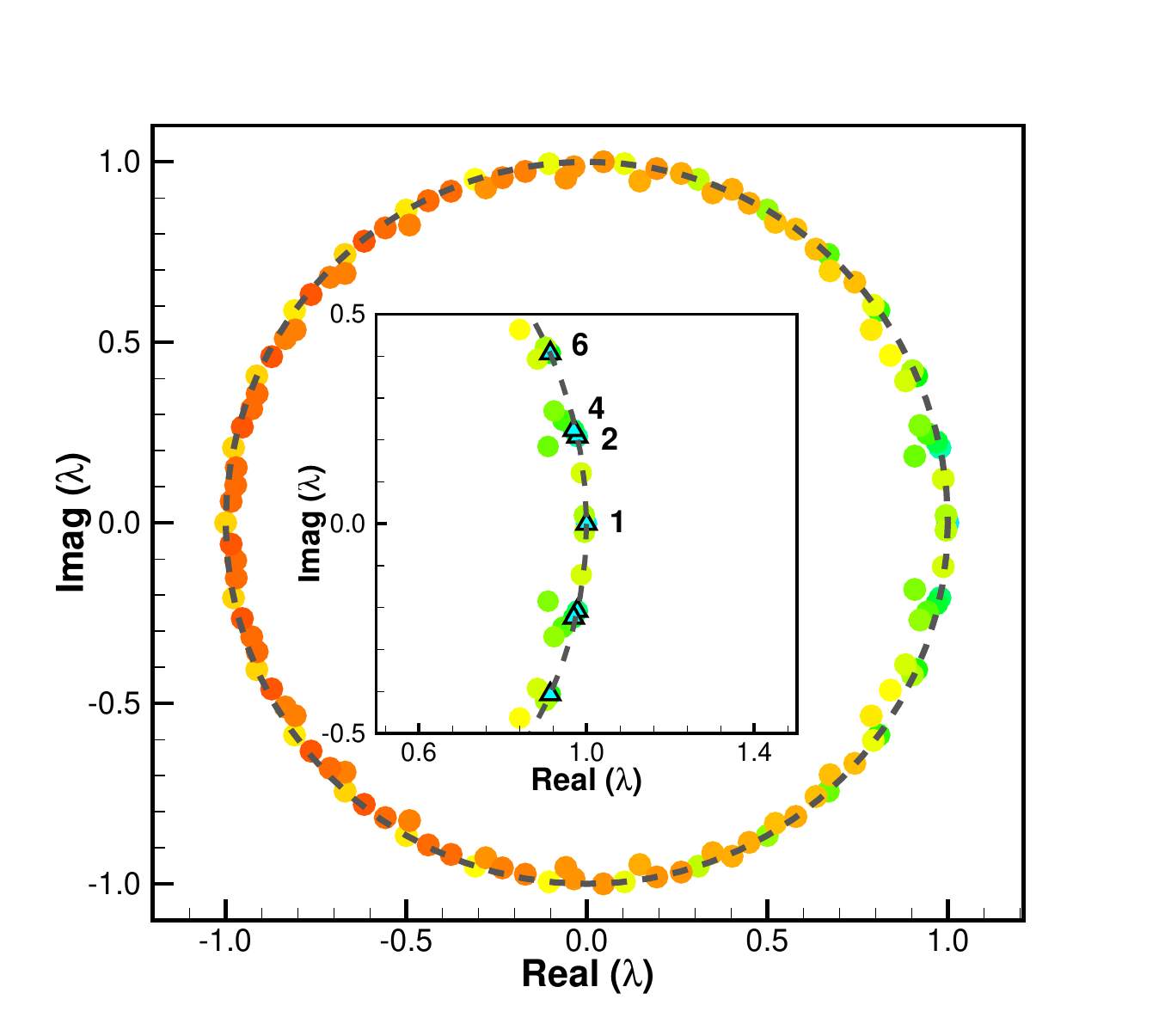}
        \caption{Real and imaginary parts}
        \label{fig:scatter_imag-real}
    \end{subfigure}
    \begin{subfigure}[b]{0.495\linewidth}
        \centering
        \includegraphics[width=1\linewidth]{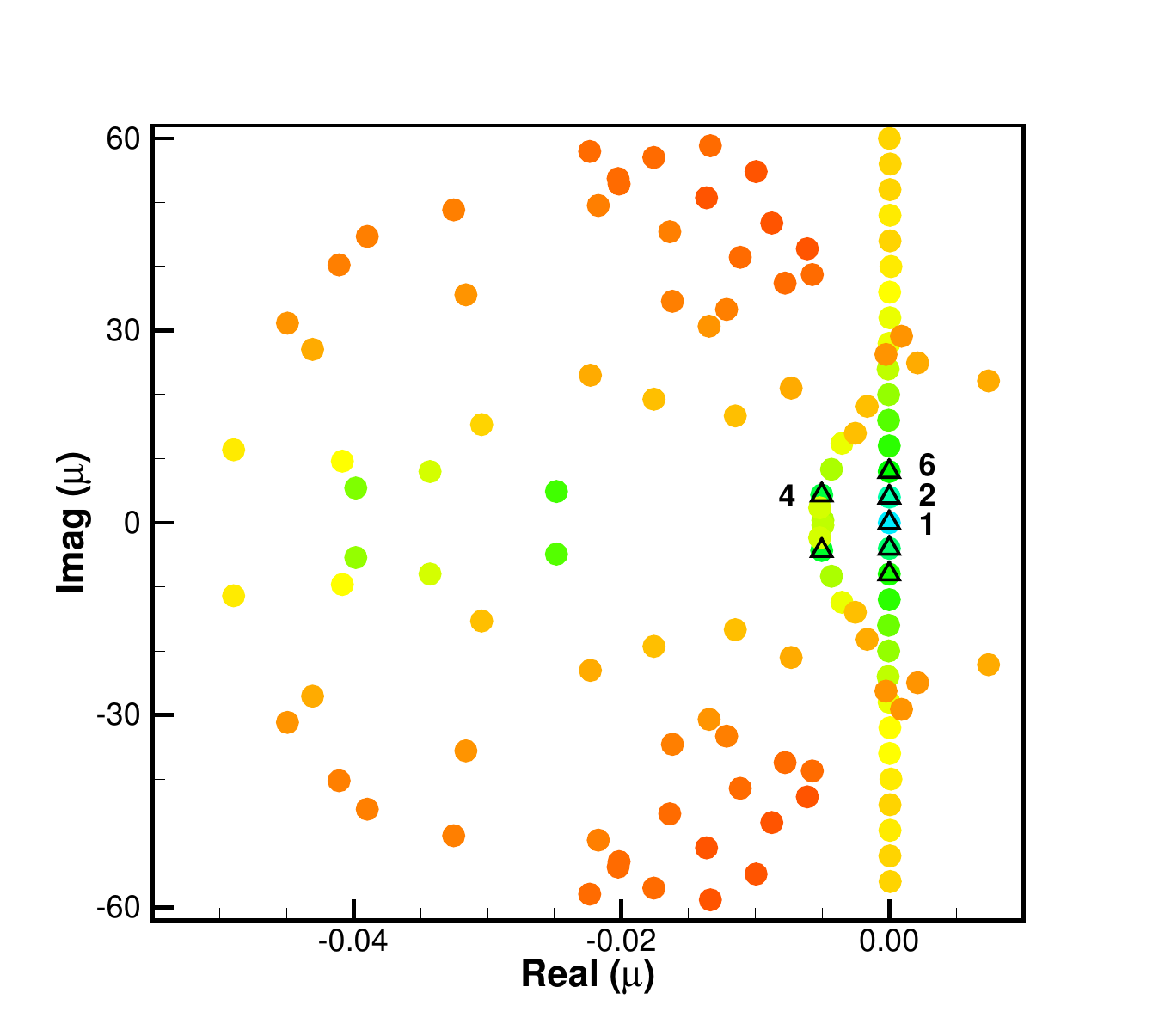}
        \caption{Growth rate and angular frequency}
        \label{fig:scatter_sigma-omega}
    \end{subfigure}
    \caption{Scatters of DMD eigenvalues obtained by the Tissot criterion.}
    \label{fig:scatter_dmd_eigenvalues}
\end{figure}

Besides the spatial mode shapes, modal decomposition methods, particularly DMD, can provide the dynamic evolution of each mode, which is essential for understanding of unsteady flow behaviors.
The scatters of DMD eigenvalues, $\lambda_k$, obtained by the Tissot criterion are shown in Fig. \ref{fig:scatter_imag-real}, where the abscissa and ordinate are the real and imaginary parts of these eigenvalues, respectively. A unit circle centered at the origin, as shown by the dashed curve, is also included for reference. The corresponding complex growth rates, $\mu_k$, of Eq. \eqref{eq:dmd_growth_rate} are shown in Fig. \ref{fig:scatter_sigma-omega}. The mode index is represented by the scatter color, transitioning from light to dark shades.


Each mode pair is symmetrically distributed about the real axis, reflecting the complex-conjugate nature of the eigenvalues in DMD. The exception is the first mode, which lies independently at the real axis. All modes are clustered near the unit circle in Fig. \ref{fig:scatter_imag-real} and correspondingly along the imaginary axis in Fig. \ref{fig:scatter_sigma-omega}, indicating that they exhibit either sustained or slightly decaying/growing dynamics over time. Modes located exactly on the unit circle in Fig. \ref{fig:scatter_imag-real} and on the imaginary axis in Fig. \ref{fig:scatter_sigma-omega} represent neutral dynamics with zero growth rates and correspond to the potential effects primarily associated with the rotor blade passing frequency. Modes inside the unit circle in Fig. \ref{fig:scatter_imag-real} and on the left side of the imaginary axis in Fig. \ref{fig:scatter_sigma-omega} are decaying modes, indicating attenuation of pressure fluctuations due to viscous dissipation. A few growing modes lie slightly outside the circle in Fig. \ref{fig:scatter_imag-real} and on the right side of the imaginary axis in Fig. \ref{fig:scatter_sigma-omega}, despite the overall flow being globally periodic and stable. These may result from transient features not fully damped within the sampling window.
The argument of each complex eigenvalue in Fig. \ref{fig:scatter_imag-real} and the imaginary value in Fig. \ref{fig:scatter_sigma-omega} imply the frequency of the corresponding mode, as indicated by Eq. (\ref{eq:dmd_growth_rate}). In general, the frequency tends to increase with the mode index, as indicated by the mode color. However, several higher modes exhibit relatively low frequencies. This further suggests that although low-frequency modes dominate the unsteady flow, frequency alone is not a sufficient criterion for mode selection.

The eigenvalues with real parts close to one are magnified in the center of Fig. \ref{fig:scatter_imag-real}. The first seven modes are marked with triangles and labeled by their respective indices in Figs. \ref{fig:scatter_imag-real} and \ref{fig:scatter_sigma-omega}.
All seven modes lie nearly on the unit circle and close to the real axis in Fig. \ref{fig:scatter_imag-real}, while they are located near the origin in Fig. \ref{fig:scatter_sigma-omega}. This indicates that the unsteady flow in the downstream stator is primarily governed by low-frequency, neutrally stable fluctuations. Actually, these fluctuations are driven by the motion of the upstream rotor blades, as will be shown later. The fourth and fifth modes lie slightly on the left side of the imaginary axis in Fig. \ref{fig:scatter_sigma-omega}. This explains the damping behavior of pressure in Fig. \ref{fig:reconstructed_point_time_variation}.

\begin{figure}[htb!]
  \centering
  \includegraphics[width=0.495\linewidth]{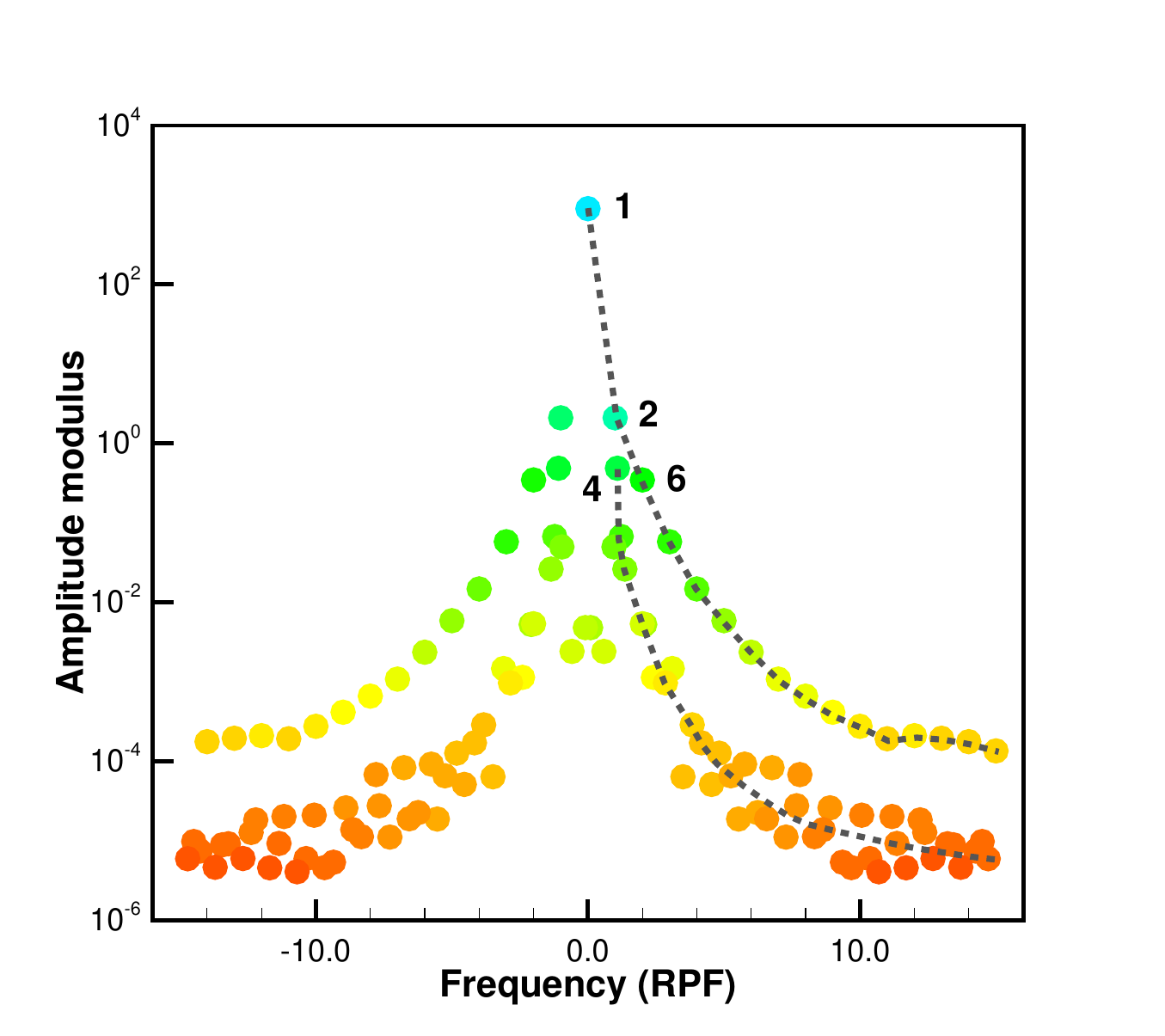}
  \caption{Amplitude-frequency scatter of DMD modes obtained by the Tissot criterion.}
  \label{fig:scatter_amplitude-frequency}
\end{figure}

According to Eq. (\ref{eq:dmd_reconstruct_single}), the mode amplitude $\alpha_k$ represents the contribution of the $k$th mode to the reconstruction. The DMD eigenvalues are thus analyzed in the amplitude-frequency space. 
Figure \ref{fig:scatter_amplitude-frequency} shows the amplitude-frequency scatter plot of all DMD modes obtained by the Tissot criterion, where frequency is normalized by the RPF. As in Fig. \ref{fig:scatter_dmd_eigenvalues}, the color of each scatter point represents the mode index, and the first seven modes are additionally labeled by their corresponding indices. 
All modes are symmetrically distributed on the two sides of the vertical axis, except for the first mode, which lies exactly on the axis. The first mode, with a zero frequency, represents the time-averaged pressure and exhibits a significantly larger amplitude than the other modes. This implies the particularly dominant role of the time-averaged pressure among all modes.
The second to fifth modes have frequencies equal to one RPF, while the sixth and seventh modes correspond to two RPFs. This indicates that the pressure fluctuations in Stator 2 are primarily driven by the motion of the upstream rotor blades. Furthermore, the presence of the second RPF harmonics in the leading modes explains the high-frequency pressure oscillations observed in Fig. \ref{fig:reconstructed_point_time_variation}.

In Fig. \ref{fig:scatter_amplitude-frequency}, the descending order of mode amplitudes closely aligns with the ascending order of mode indices, implying a strong consistency between the Tissot criterion and the amplitude criterion for the present flow problem. This consistency explains the nearly identical reconstruction errors and reconstructed pressure fields obtained by the two criteria, as discussed in Sec. \ref{sec:ROR}. Therefore, it can be concluded that for a globally periodic flow dominated by specific forcing frequencies, the amplitude criterion can serve as an effective alternative to the Tissot criterion and the SP-DMD method for mode selection.
As the mode index increases, the DMD modes separate into two distinct branches in Fig. \ref{fig:scatter_amplitude-frequency}, characterized by high and low amplitudes, respectively, as indicated by the dashed curves. Both branches trend toward higher frequencies with increasing mode index. At any given frequency, the high-amplitude modes appear earlier in the mode sequence than the low-amplitude ones.
A detailed analysis of the growth rates shows that the high-amplitude branch consists of neutral modes, while the low-amplitude branch includes both decaying and growing modes. This suggests that, for a given frequency, neutral modes are more dominant than those with nonzero growth rates. This observation is consistent with the fact that the flow exhibits a globally periodic limit-cycle oscillation.

\begin{figure}[htb!]
    \centering
    \begin{subfigure}[b]{0.495\linewidth}
        \centering
        \includegraphics[width=1\linewidth]{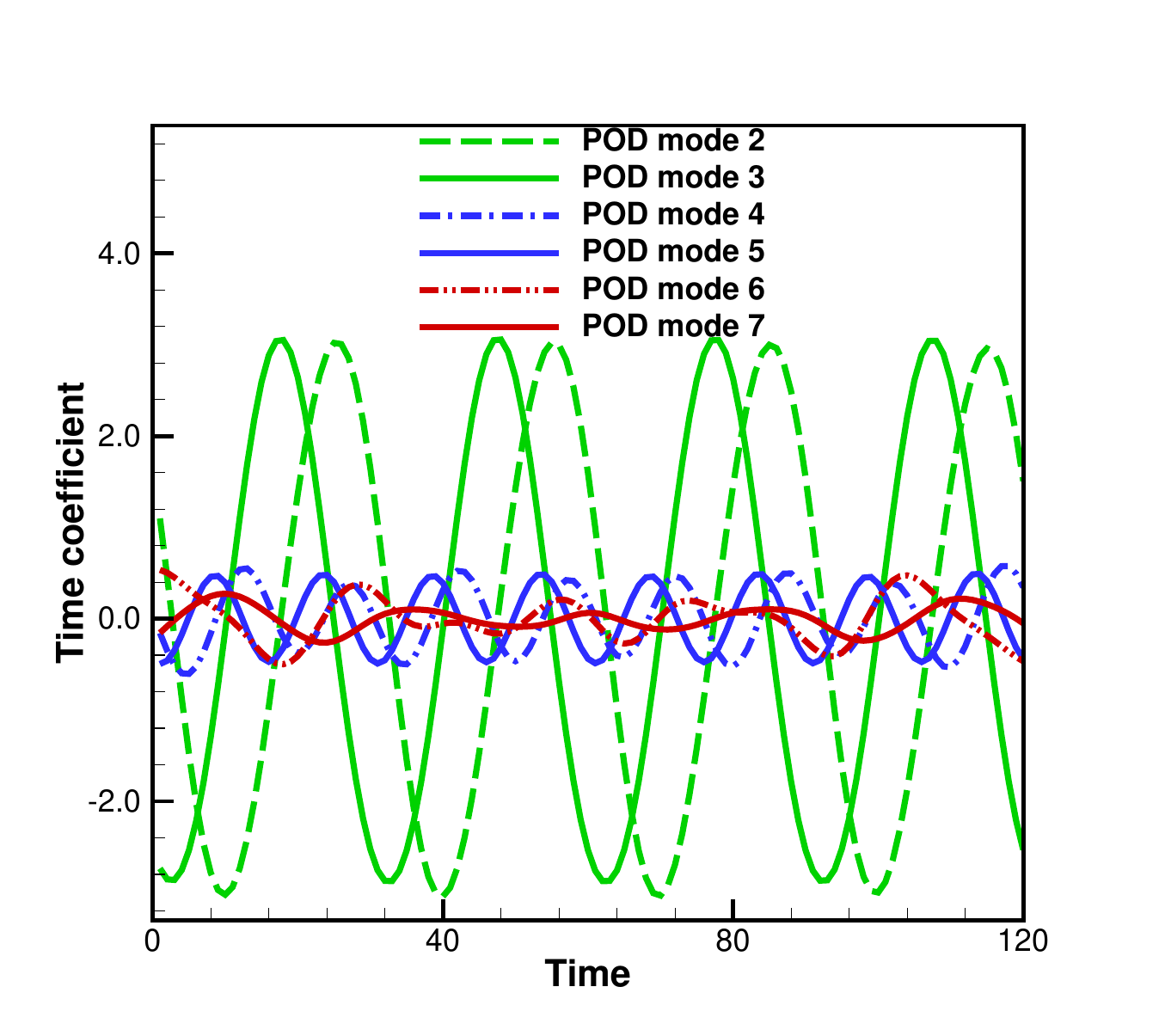}
        \caption{POD}
        \label{fig:time_coefficient_pod}
    \end{subfigure}
    \begin{subfigure}[b]{0.495\linewidth}
        \centering
        \includegraphics[width=1\linewidth]{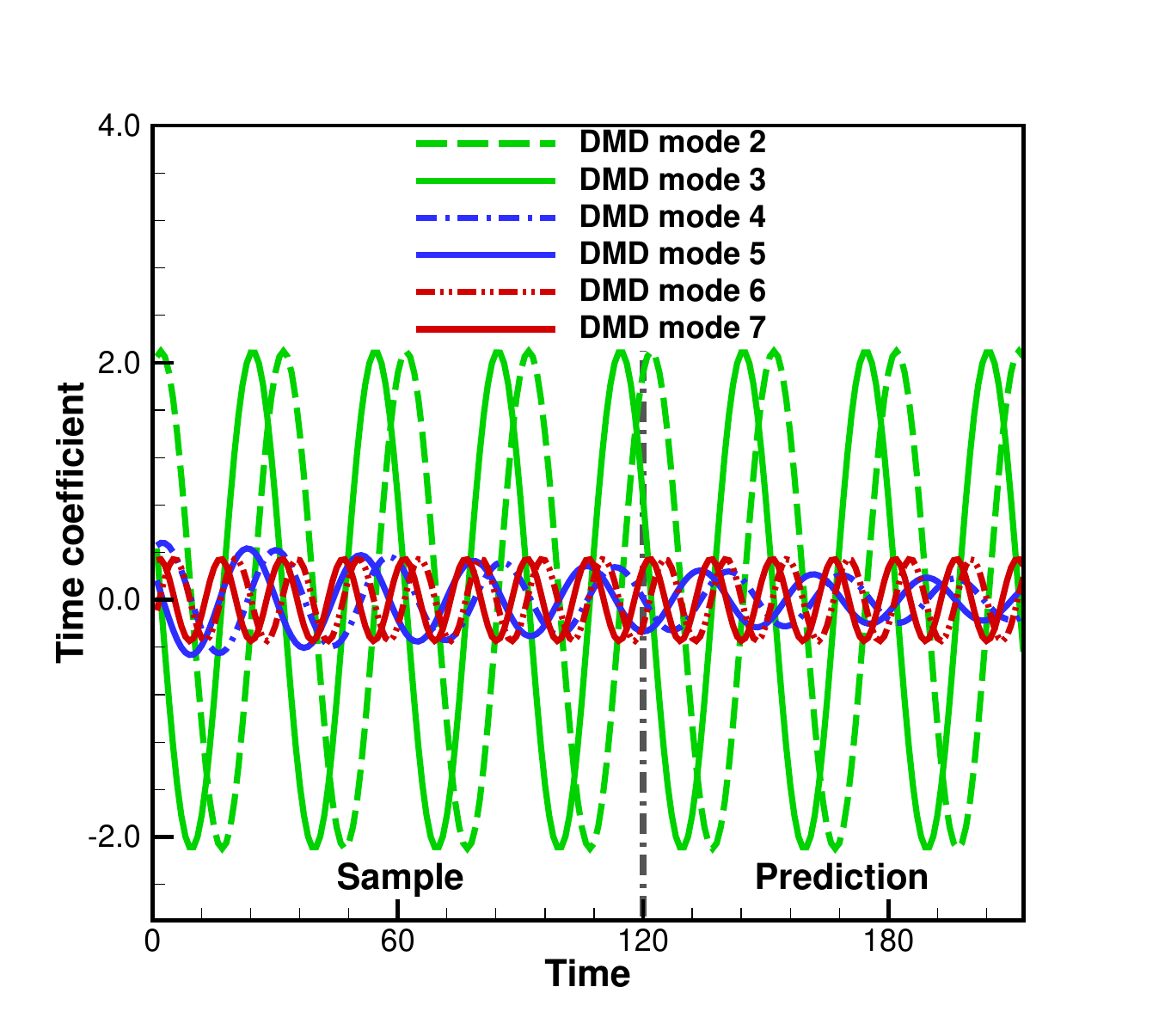}
        \caption{DMD}
        \label{fig:time_coefficient_dmd}
    \end{subfigure}
    \caption{Variations of time coefficients of dominant modes. DMD is based on the Tissot criterion. In the DMD results, 2nd and 3rd modes, 4th and 5th modes, and 6th and 7th modes form complex-conjugate pairs, respectively.}
    \label{fig:time_coefficient_pod_dmd}
\end{figure}

The mode shape describes the spatial structure of a mode, while the time coefficient reflects its temporal evolution. From Eqs. (\ref{eq:pod_svd}) and (\ref{eq:dmd_reconstruct_single}), the time coefficients at the $n$ snapshot time instants of the $k$th POD and DMD modes are, respectively, given by
\begin{equation}
    \mathbf{c}_k^\mathrm{POD} = \sigma_k \mathbf{v}_k^T, \
    \mathbf{c}_k^\mathrm{DMD} = \alpha_k \left[ \lambda_k^0, \lambda_k^1, ..., \lambda_k^{n-1} \right]   
\end{equation}
The variations of the time coefficients of the dominant modes from POD and the DMD method based on the Tissot criterion are shown in Fig. \ref{fig:time_coefficient_pod_dmd}. In the DMD results, the second, fourth, and sixth modes are the complex conjugates of the third, fifth, and seventh modes, respectively. 
For each conjugate pair, the time coefficients of the real and imaginary components exhibit similar trends, with the real part lagging the imaginary part by one quarter of the oscillation period. This can be explained by a theoretical examination of the time coefficient. By Eq. (\ref{eq:dmd_growth_rate}), the $k$th DMD time coefficient at time $t$ can be expressed as
\begin{equation} \label{eq:time_coefficient_dmd}
    c_k^\mathrm{DMD}(t) = \alpha_k \lambda_k^{t/\Delta t} = \left( |\alpha_k| \mathrm{e}^{\psi_\alpha} \mathrm{e}^{\sigma_k t} \right) \mathrm{e}^{\mathrm{i}\omega_k t}  
\end{equation}
where $\psi_\alpha$, representing the argument of $\alpha_k \in \mathbb{C}$, serves as the initial phase of $c_k^\mathrm{DMD}(t)$. Equation \eqref{eq:time_coefficient_dmd} indicates that the real and imaginary parts of $c_k^\mathrm{DMD}(t)$ are phase-shifted by $\pi/2$, corresponding to one quarter of the period.
In the POD results, similar phase-lag behavior is observed between the second and third modes, as well as between the fourth and fifth modes, even though all time coefficients are real-valued. However, the sixth and seventh POD modes do not exhibit a fixed phase difference. These behaviors are expected since the spatial structures of the second and third modes are highly similar---analogous to a complex-conjugate pair---as shown in Figs. \ref{fig:contour_mode_pod_mode2} and \ref{fig:contour_mode_pod_mode3}. A similar relationship is observed for the fourth and fifth modes in Figs. \ref{fig:contour_mode_pod_mode4} and \ref{fig:contour_mode_pod_mode5}. In contrast, the sixth and seventh mode shapes differ significantly, as shown in Figs. \ref{fig:contour_mode_pod_mode6} and \ref{fig:contour_mode_pod_mode7}.

The oscillation amplitudes of the time coefficients for both POD and DMD decrease as the mode index increases. This is because the time coefficients of POD and DMD are scaled by singular values and mode amplitudes, respectively, both of which decrease with the mode index. 
Consistent with the dynamic features of DMD eigenvalues, the time coefficients of the fourth and fifth DMD modes exhibit a decaying trend, whereas the remaining DMD modes show periodic behavior. In contrast, all POD modes display purely periodic variations, indicating that POD does not give the dynamic features of an unsteady flow field.

\begin{figure}[htb!]
    \centering
    \begin{subfigure}[b]{0.495\linewidth}
        \centering
        \includegraphics[width=1\linewidth]{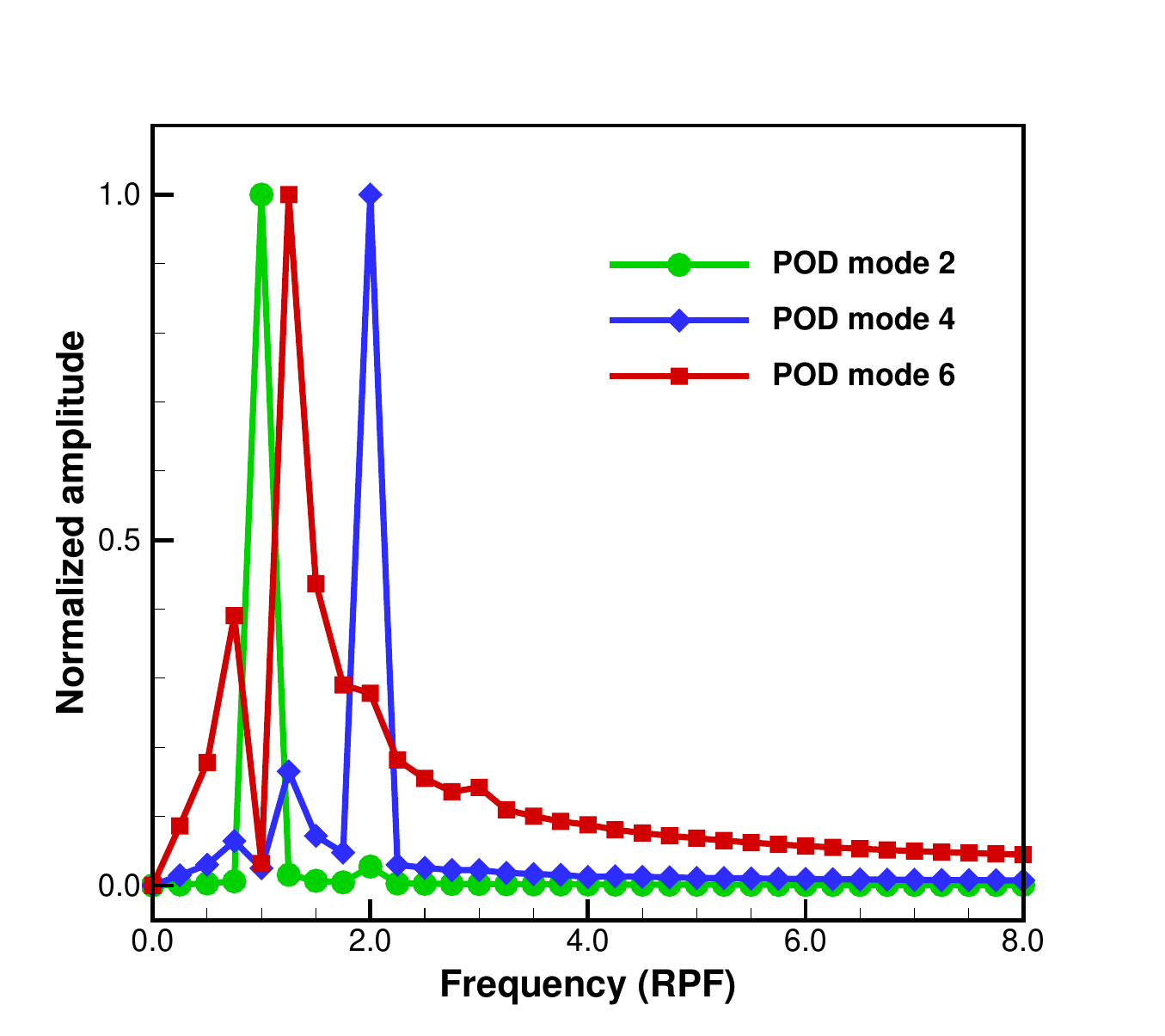}
        \caption{POD}
        \label{fig:energy_spect_coeff_pod}
    \end{subfigure}
    \begin{subfigure}[b]{0.495\linewidth}
        \centering
        \includegraphics[width=1\linewidth]{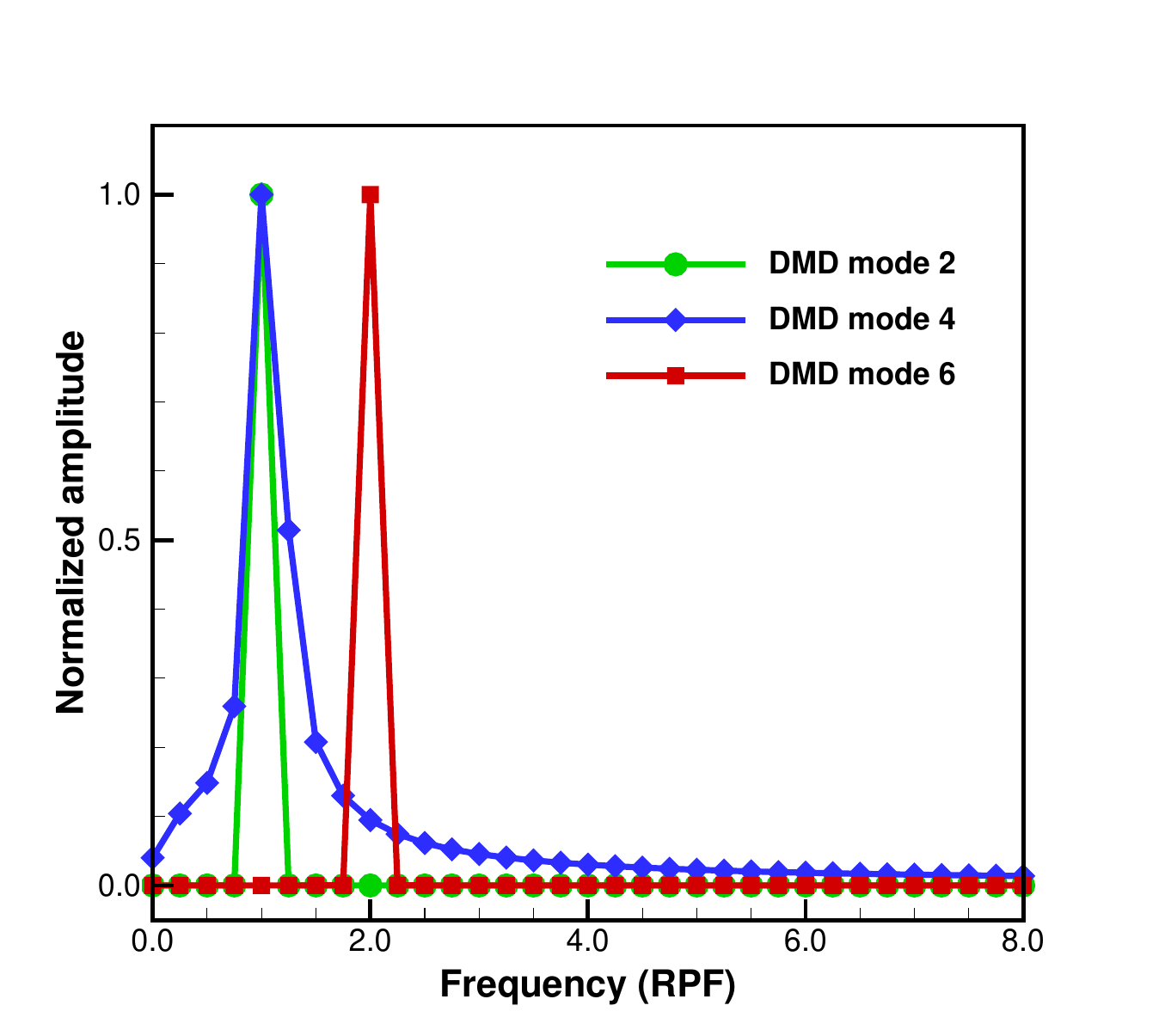}
        \caption{DMD}
        \label{fig:energy_spect_coeff_dmd}
    \end{subfigure}
    \caption{Fourier amplitude spectra of time coefficients. DMD is based on the Tissot criterion.}
    \label{fig:energy_spect_coeff_pod_dmd}
\end{figure}

To further compare the differences between POD and DMD, Fig. \ref{fig:energy_spect_coeff_pod_dmd} shows the Fourier amplitude spectra of the second, fourth, and sixth time coefficients. The results of the third, fifth, and seventh time coefficients, not shown here for brevity, exhibit similar trends. The frequency and Fourier amplitude are normalized by the RPF and the maximum spectral component, respectively. 
For DMD, each time coefficient is dominated by a single frequency, corresponding to a harmonic of the RPF. In contrast, the POD time coefficients contain not only the leading frequency but also multiple higher and lower frequency components. Moreover, for the sixth POD mode, the leading frequency of 1.25 RPFs itself arises from the superimposition of several RPF harmonics, further emphasizing the mixed-frequency content of POD modes.
The dominant frequency of the fourth POD mode and the sixth DMD mode is 2 RPFs. This is consistent with the observation that these mode shapes capture high-order pressure fluctuations within the passage, as shown in Figs. \ref{fig:contour_mode_pod_dmd_mode4-7} and \ref{fig:contour_mode_dmd_modulus}.

\subsection{Influence of Stator Clocking} \label{sec:clocking_config}
\begin{figure}[htb!]
  \centering
  \includegraphics[width=0.495\linewidth]{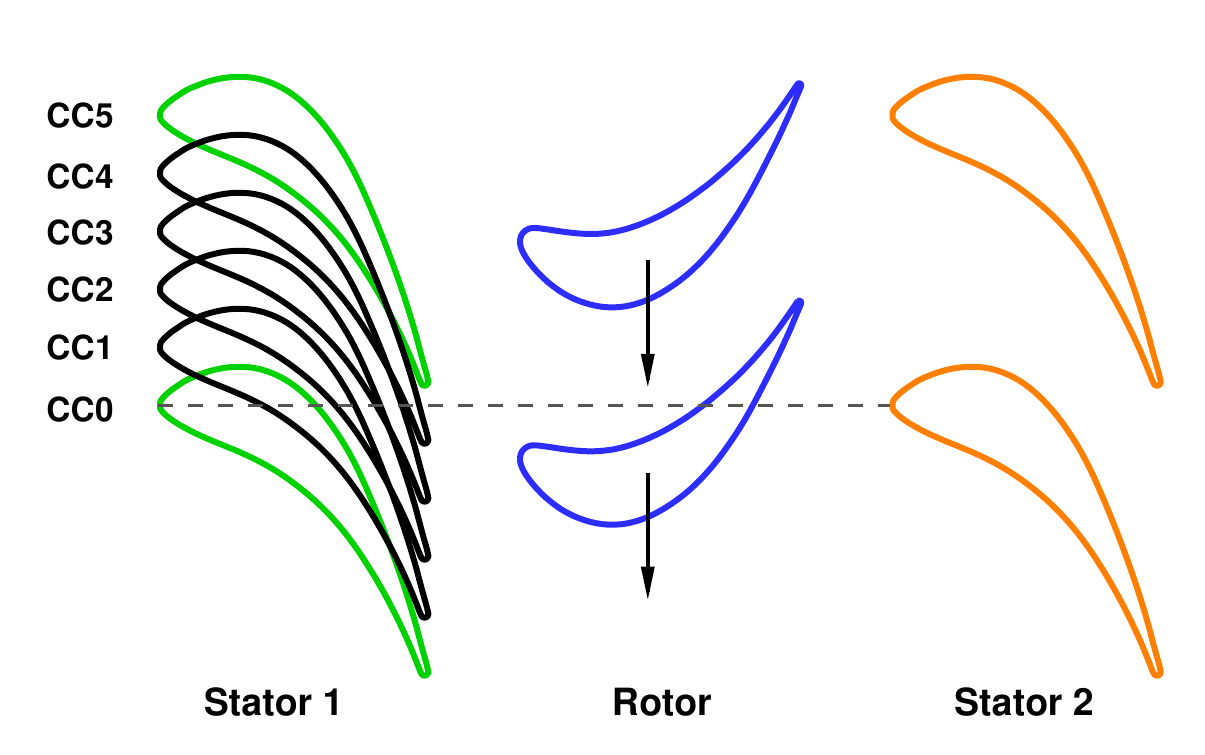}
  \caption{Sketch of stator clocking configurations.}
  \label{fig:sketch_clocking_configuration}
\end{figure}

In a multi-stage turbomachine, the relative pitchwise positioning of adjacent-stage stator/rotor blades affect the aerodynamic performance, a phenomenon known as clocking effects. A previous study on the stator clocking effects in the same turbine \cite{2017_Zhu_clocking} demonstrated that the clocking positions of Stator 1 and Stator 2 blades have evident influence on both the pressure fluctuations in the downstream stator passage and the overall adiabatic efficiency of the turbine.  It is therefore anticipated that the stator clocking also affects the pressure modes within Stator 2 passage. 
In this study, the same stator clocking configurations used in Ref. \cite{2017_Zhu_clocking} are adopted, as illustrated in Fig. \ref{fig:sketch_clocking_configuration}. Five configurations, labeled $CC0$ through $CC4$, are defined by rotating Stator 1 blades in $2^\circ$ increments within one stator pitch. The baseline case, $CC0$, corresponds to the aligned configuration where the leading edges of Stator 1 and Stator 2 are in line. Configuration $CC5$, which is identical to $CC0$, is also shown for completeness.

\begin{figure}[htb!]
    \centering
    \begin{subfigure}[b]{0.24\linewidth}
        \centering
        \includegraphics[width=1\linewidth]{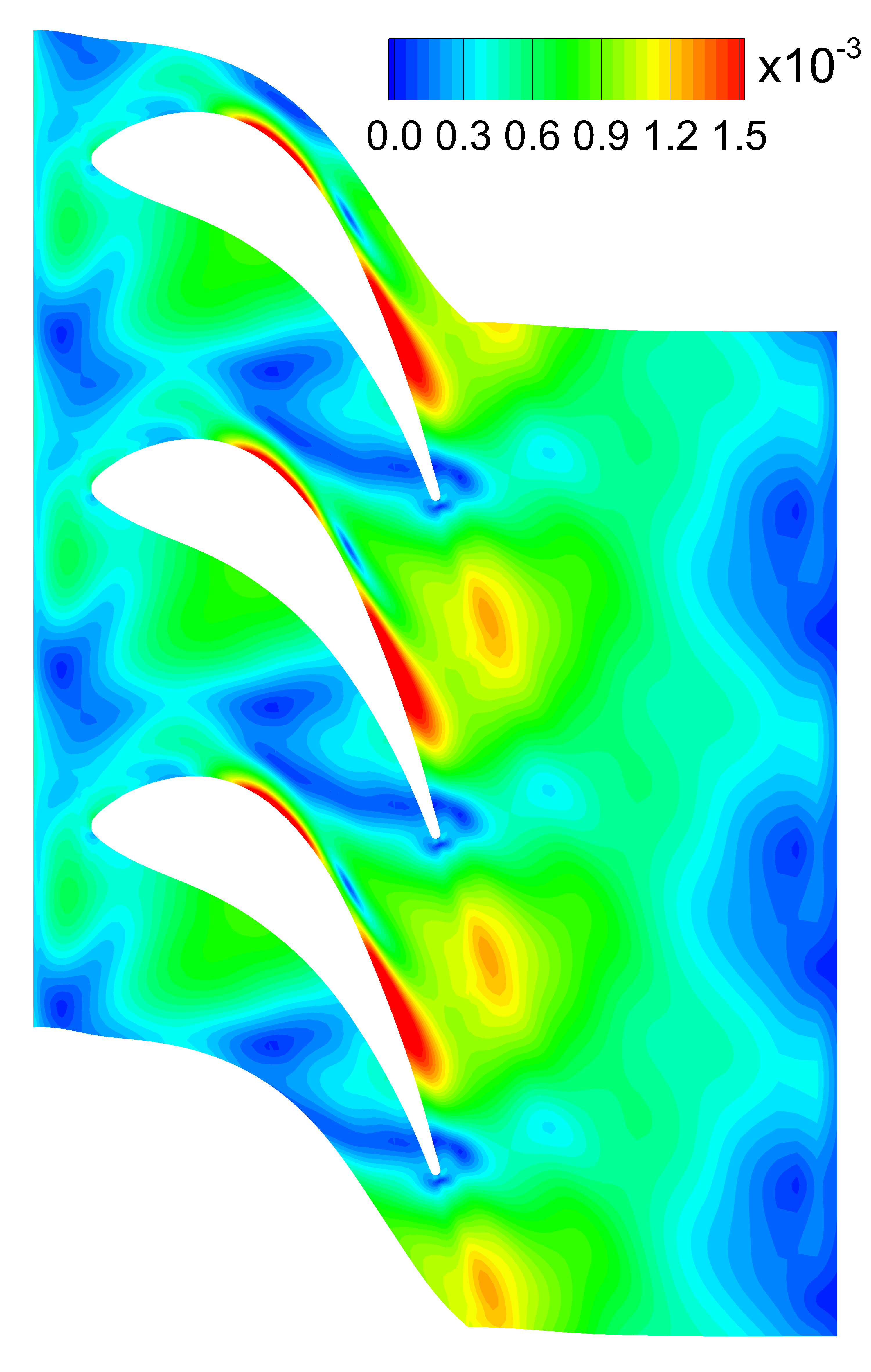}
        \caption{$CC1$}
        \label{fig:contour_mode_dmd_modulus_2_3_CC1}
    \end{subfigure}
    \begin{subfigure}[b]{0.24\linewidth}
        \centering
        \includegraphics[width=1\linewidth]{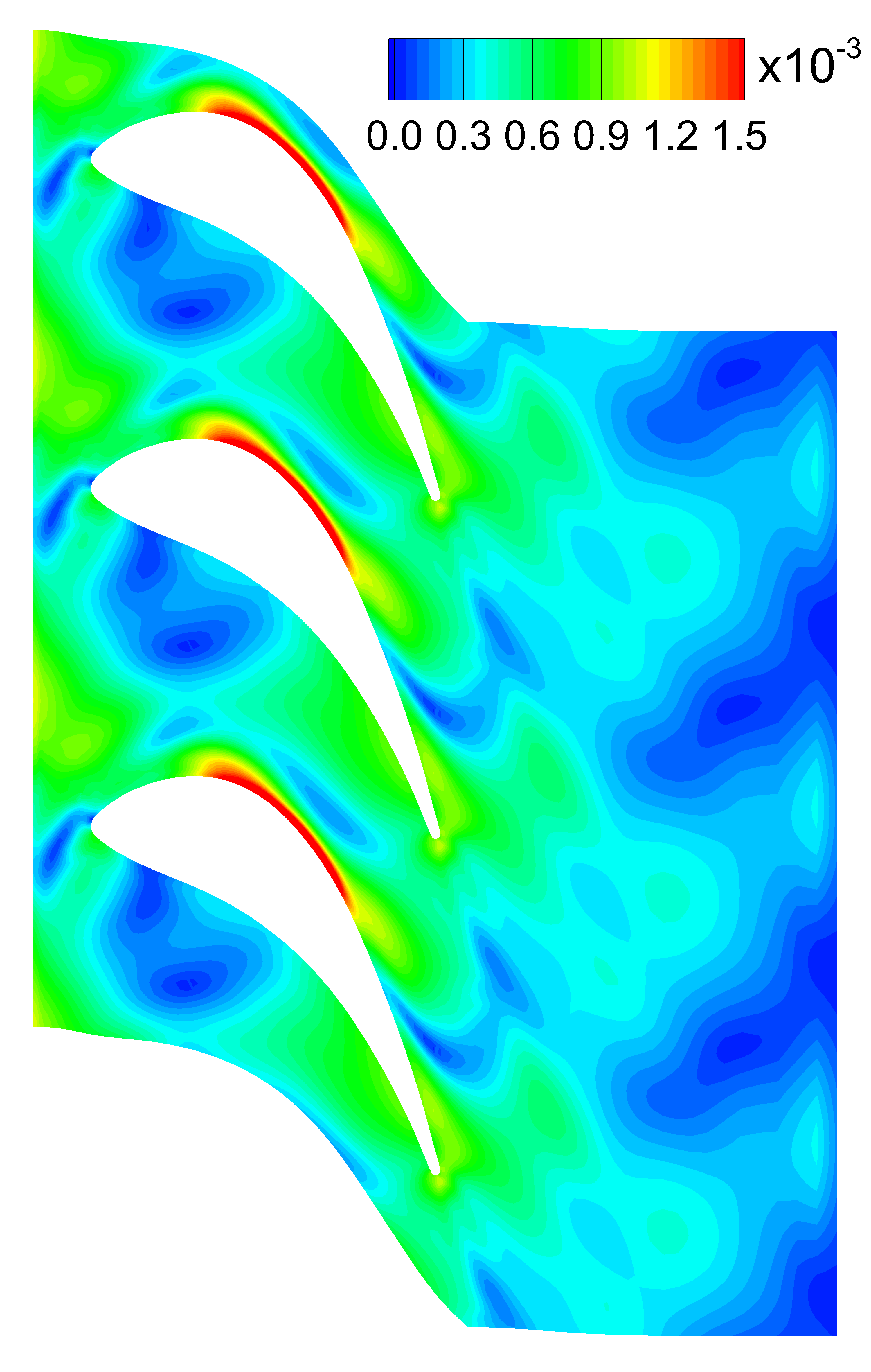}
        \caption{$CC2$}
        \label{fig:contour_mode_dmd_modulus_2_3_CC2}
    \end{subfigure}
    \begin{subfigure}[b]{0.24\linewidth}
        \centering
        \includegraphics[width=1\linewidth]{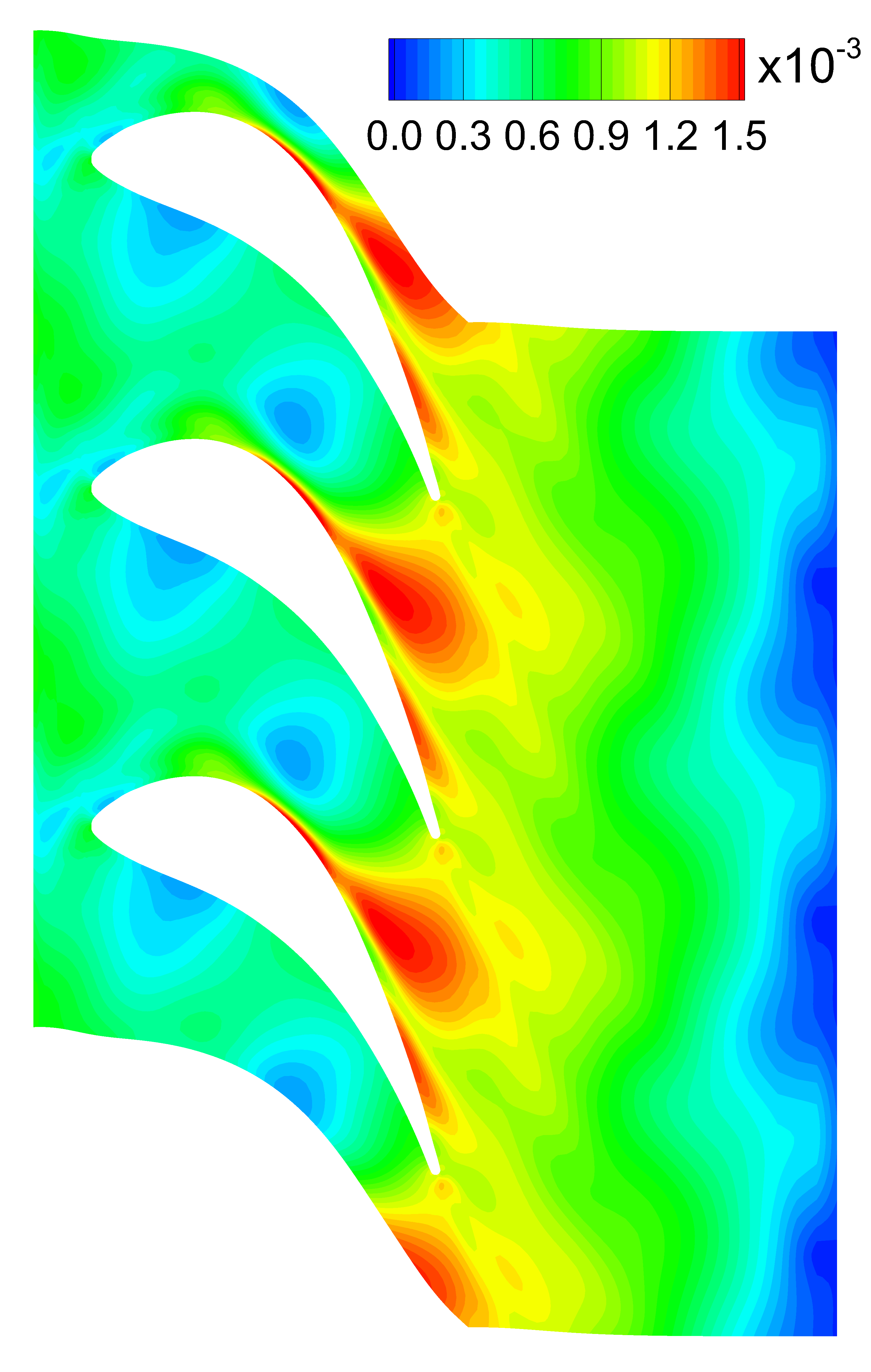}
        \caption{$CC3$}
        \label{fig:contour_mode_dmd_modulus_2_3_CC3}
    \end{subfigure}
    \begin{subfigure}[b]{0.24\linewidth}
        \centering
        \includegraphics[width=1\linewidth]{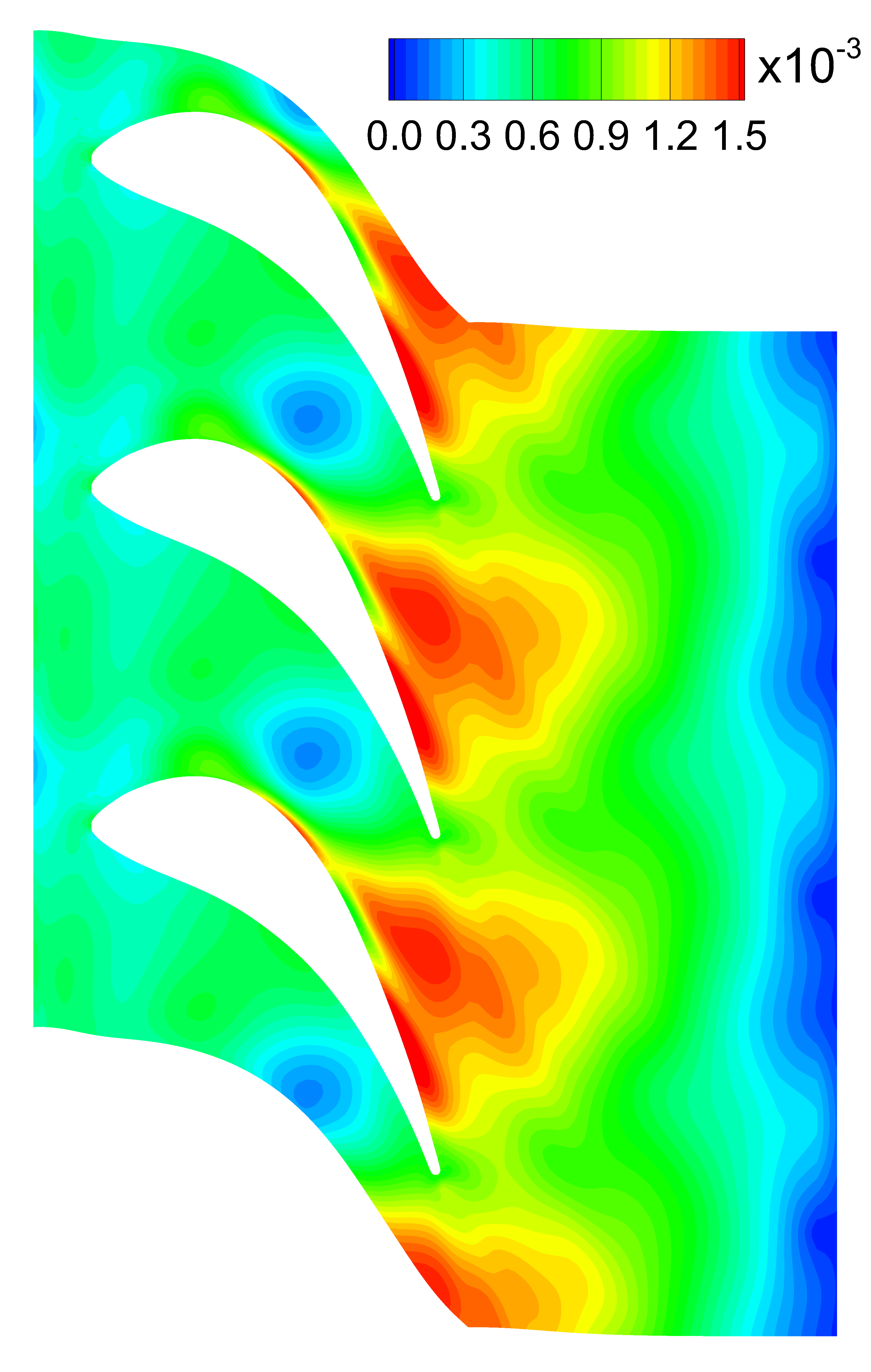}
        \caption{$CC4$}
        \label{fig:contour_mode_dmd_modulus_2_3_CC4}
    \end{subfigure}
    \caption{Contours of the modulus of the second and third DMD mode pair for different clocking configurations at 50\% span. DMD is based on the Tissot criterion.}
    \label{fig:contour_mode_dmd_modulus_2_3_clocking}
\end{figure}

\begin{figure}[htb!]
  \centering
  \includegraphics[width=0.495\linewidth]{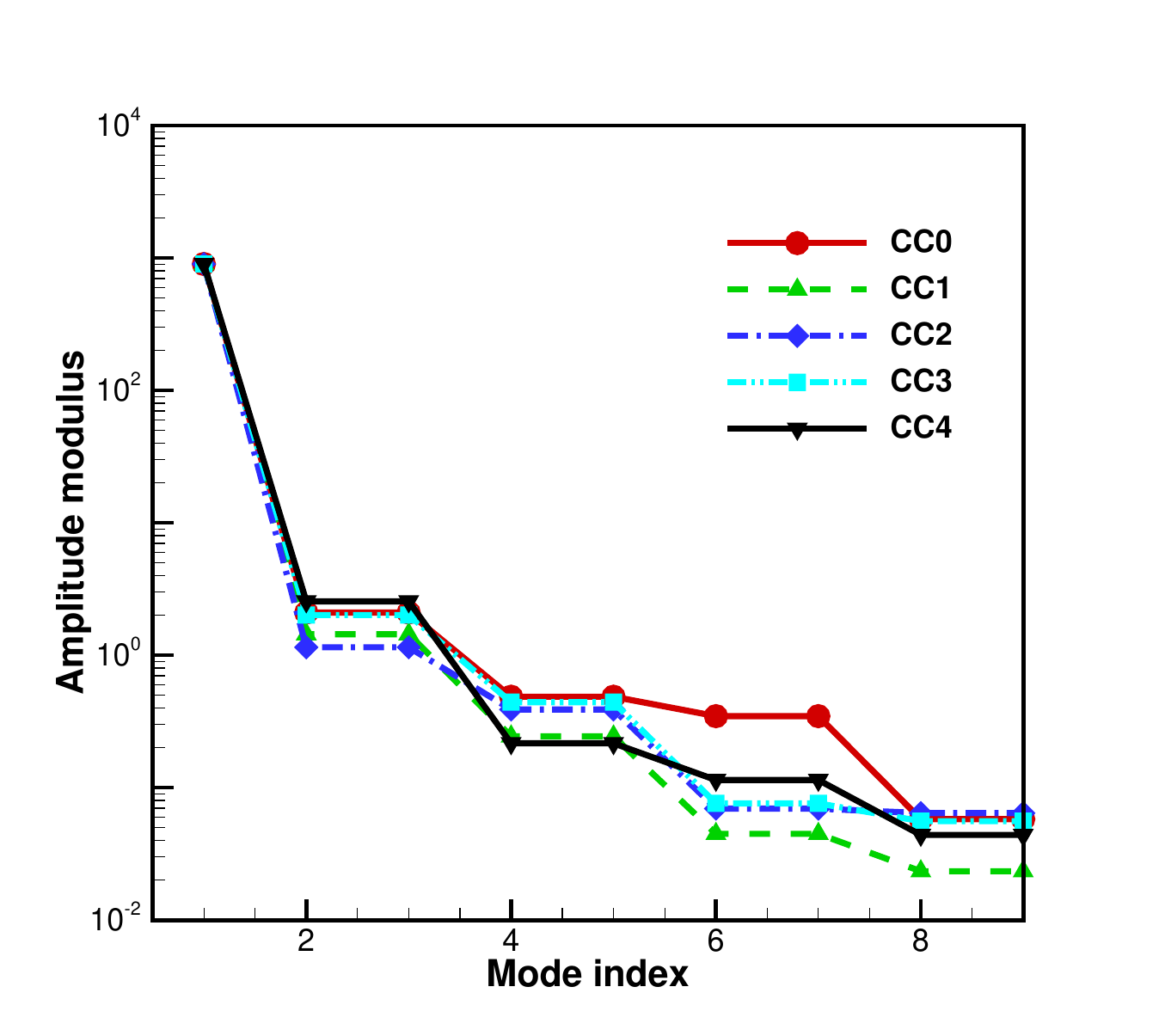}
  \caption{Variation of DMD amplitude modulus with mode index for different clocking configurations. DMD is based on the Tissot criterion.}
  \label{fig:amplitude_modulus-clocking}
\end{figure}

Figure \ref{fig:contour_mode_dmd_modulus_2_3_clocking} shows the contours of the modulus of the second and third DMD mode pair at 50\% span for different clocking configurations. The results of $CC0$, the baseline case, can be found in Fig. \ref{fig:contour_mode_dmd_2_3}. 
Across all clocking configurations, the mode shapes show similar spatial distributions. They capture the dominant pressure fluctuation structures within Stator 2 passage, consistent with the baseline configuration shown in Fig. \ref{fig:contour_mode_dmd_mode2}. However, the mode strengths vary across the clocking configurations. 
Based on the extent of high-modulus regions (Red regions in Fig. \ref{fig:contour_mode_dmd_modulus_2_3_clocking}), the mode strength is ranked as $CC4 > CC0 > CC3 > CC1 > CC2$. This trend is consistent with the ordering of the amplitude moduli of the second and third DMD mode pair shown in Fig. \ref{fig:amplitude_modulus-clocking}. Both mode strength and amplitude modulus rankings coincide with the ordering of adiabatic efficiency at 50\% span reported in Ref. \cite{2017_Zhu_clocking}.
This correlation suggests that clocking configurations with higher adiabatic efficiency exhibit stronger second and third DMD modes. This is attributed to the larger pressure fluctuation amplitudes associated with higher-efficiency configurations, as previously concluded in Ref. \cite{2017_Zhu_clocking}. 
For higher DMD modes, whose amplitude moduli are orders of magnitude lower than those of the second and third modes, no clear correlation with the aerodynamic parameters of the turbine is observed.

\begin{figure}[htb!]
  \centering
  \includegraphics[width=0.495\linewidth]{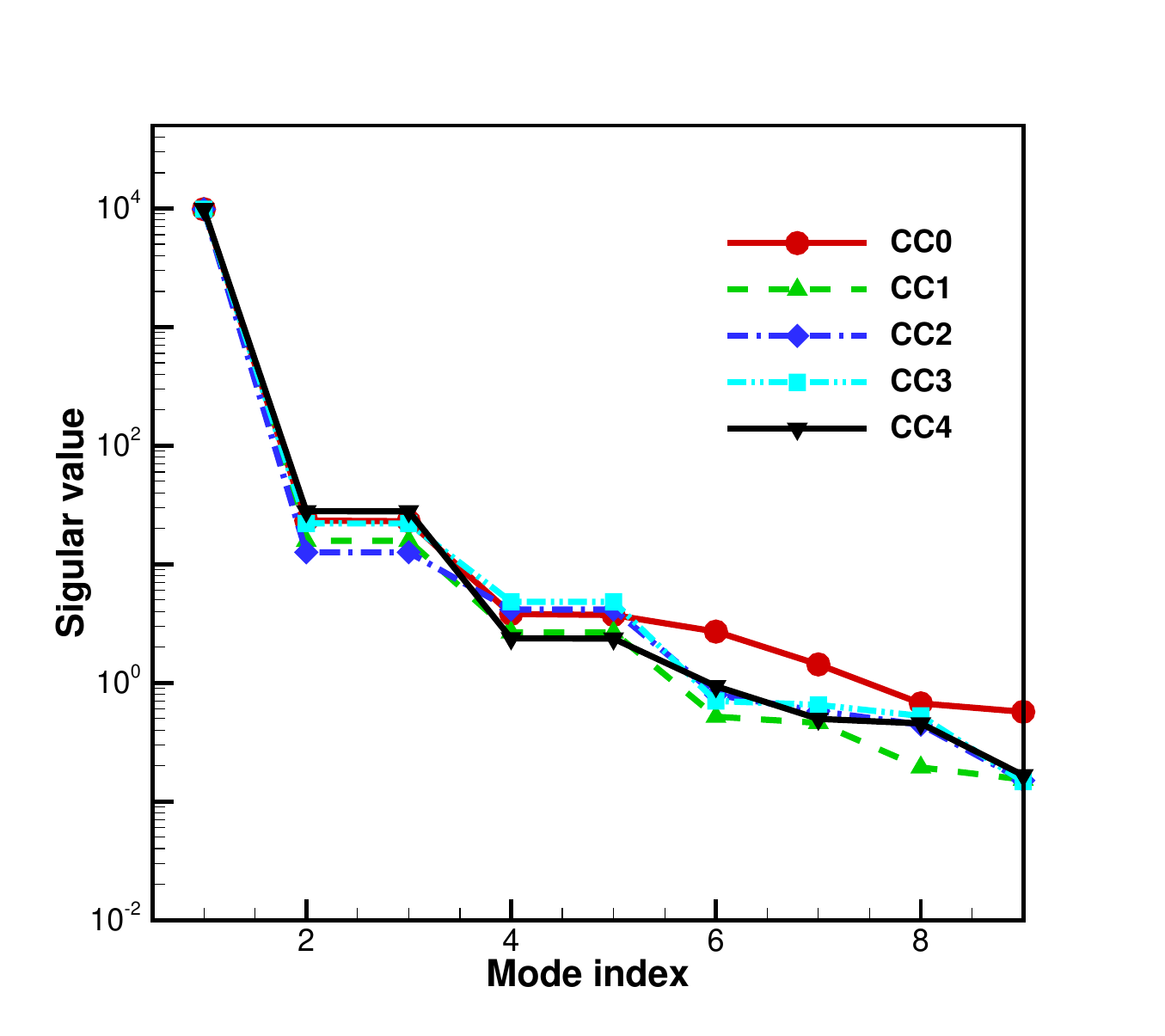}
  \caption{Variation of POD singular value with mode index for different clocking configurations.}
  \label{fig:singular_value-clocking}
\end{figure}

\begin{figure}[htb!]
    \centering
    \begin{subfigure}[b]{0.24\linewidth}
        \centering
        \includegraphics[width=1\linewidth]{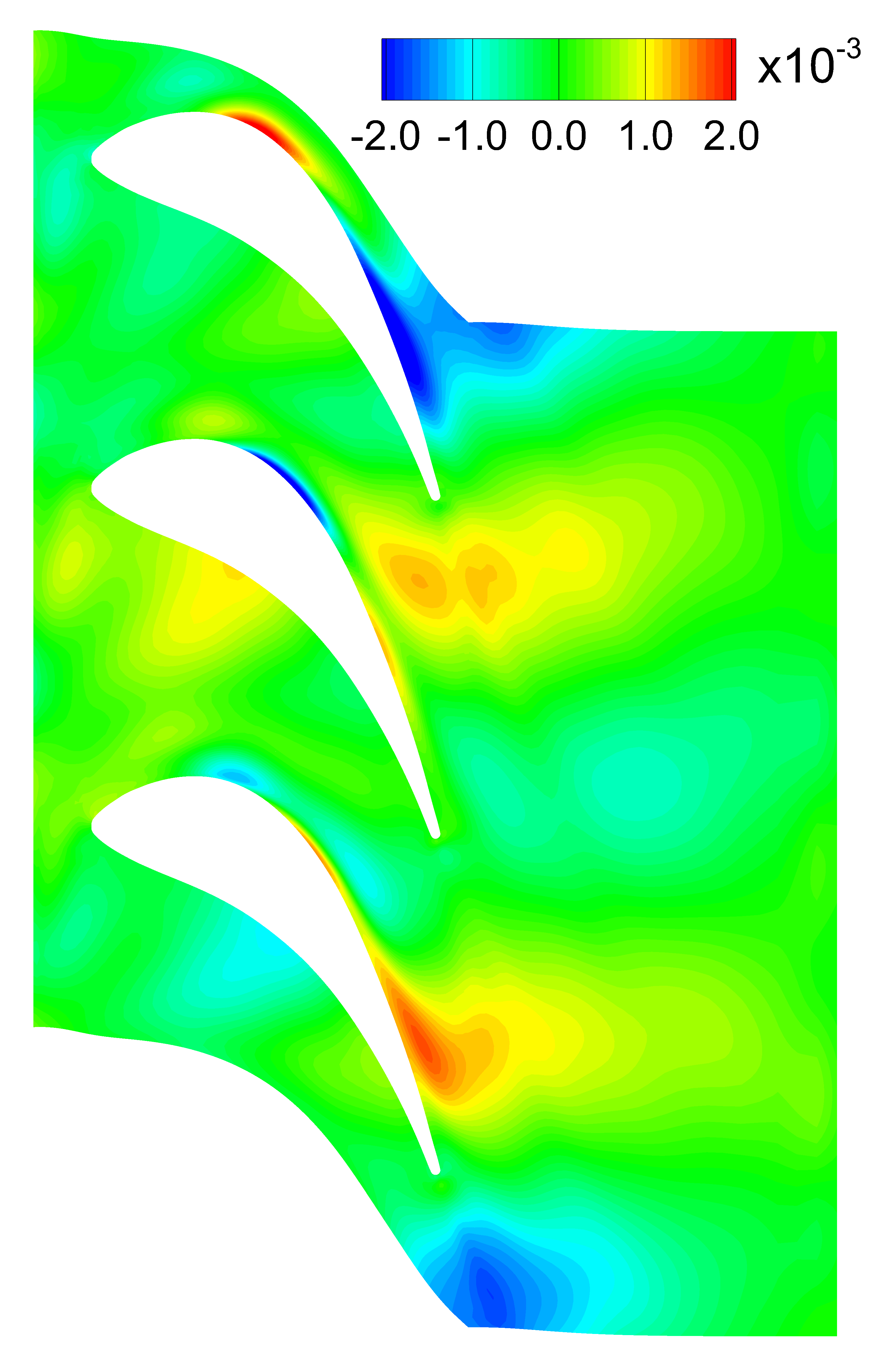}
        \caption{$CC1$}
        \label{fig:contour_mode_pod_mode2_CC1}
    \end{subfigure}
    \begin{subfigure}[b]{0.24\linewidth}
        \centering
        \includegraphics[width=1\linewidth]{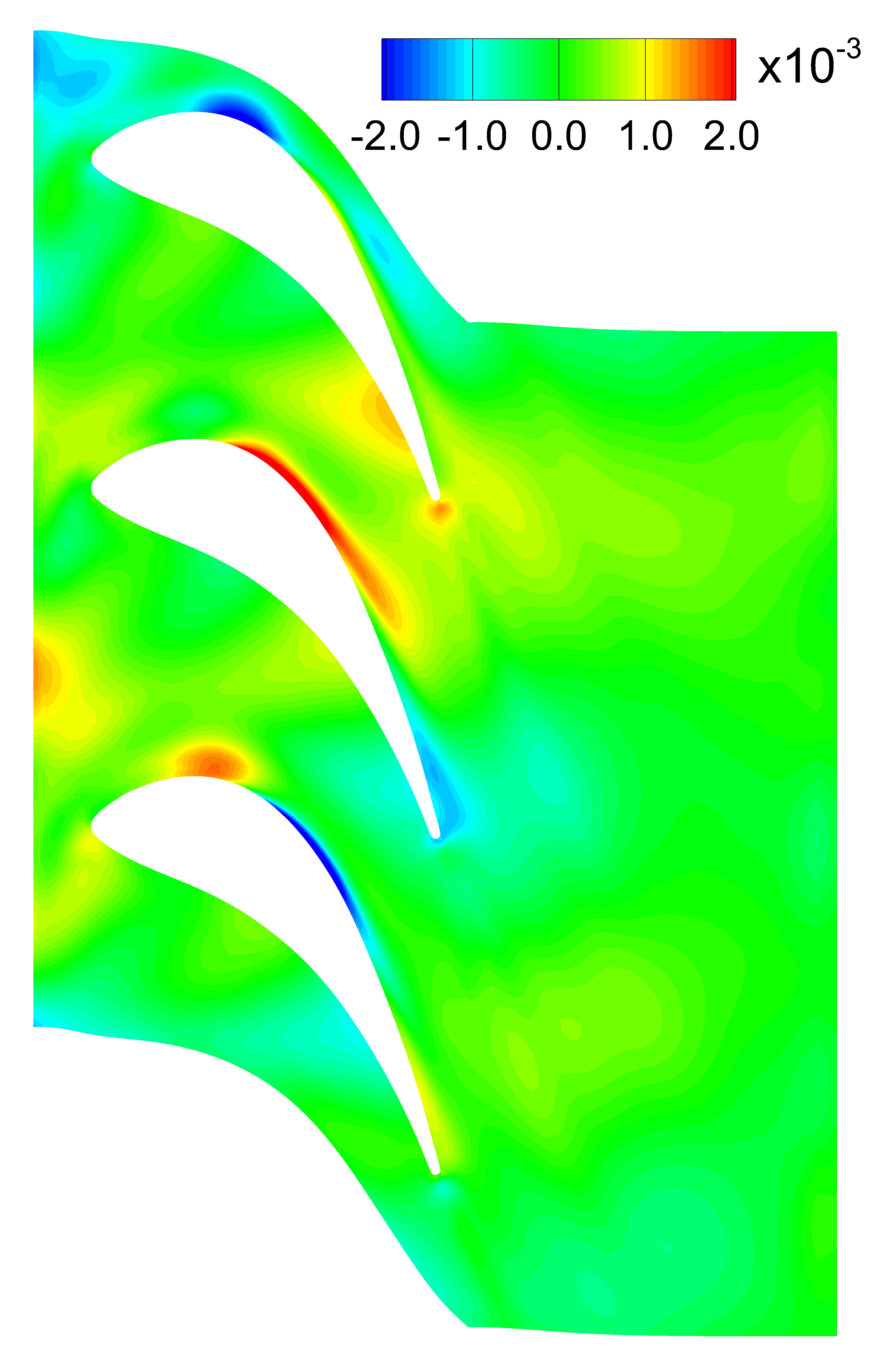}
        \caption{$CC2$}
        \label{fig:contour_mode_pod_mode2_CC2}
    \end{subfigure}
    \begin{subfigure}[b]{0.24\linewidth}
        \centering
        \includegraphics[width=1\linewidth]{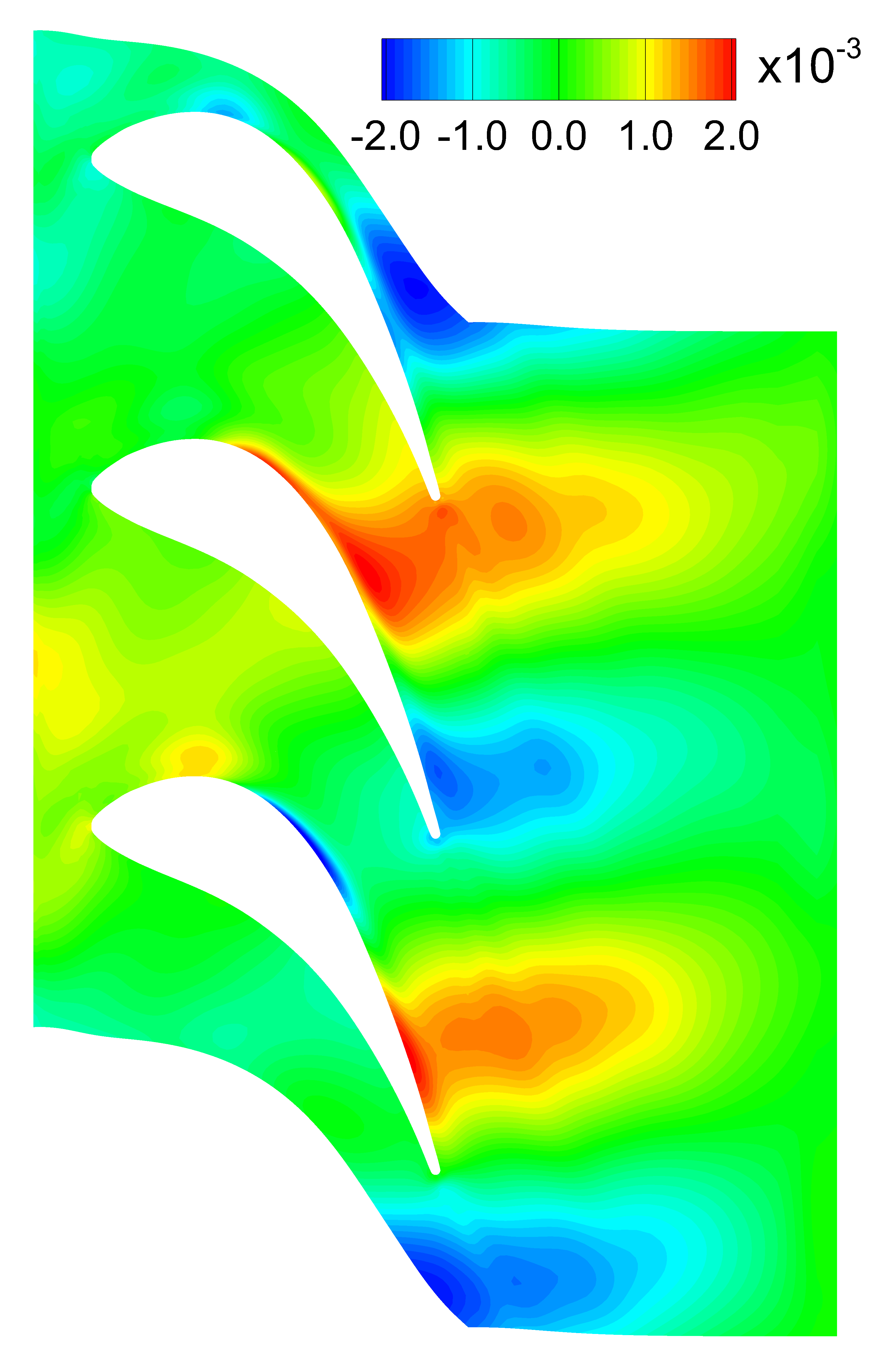}
        \caption{$CC3$}
        \label{fig:contour_mode_pod_mode2_CC3}
    \end{subfigure}
    \begin{subfigure}[b]{0.24\linewidth}
        \centering
        \includegraphics[width=1\linewidth]{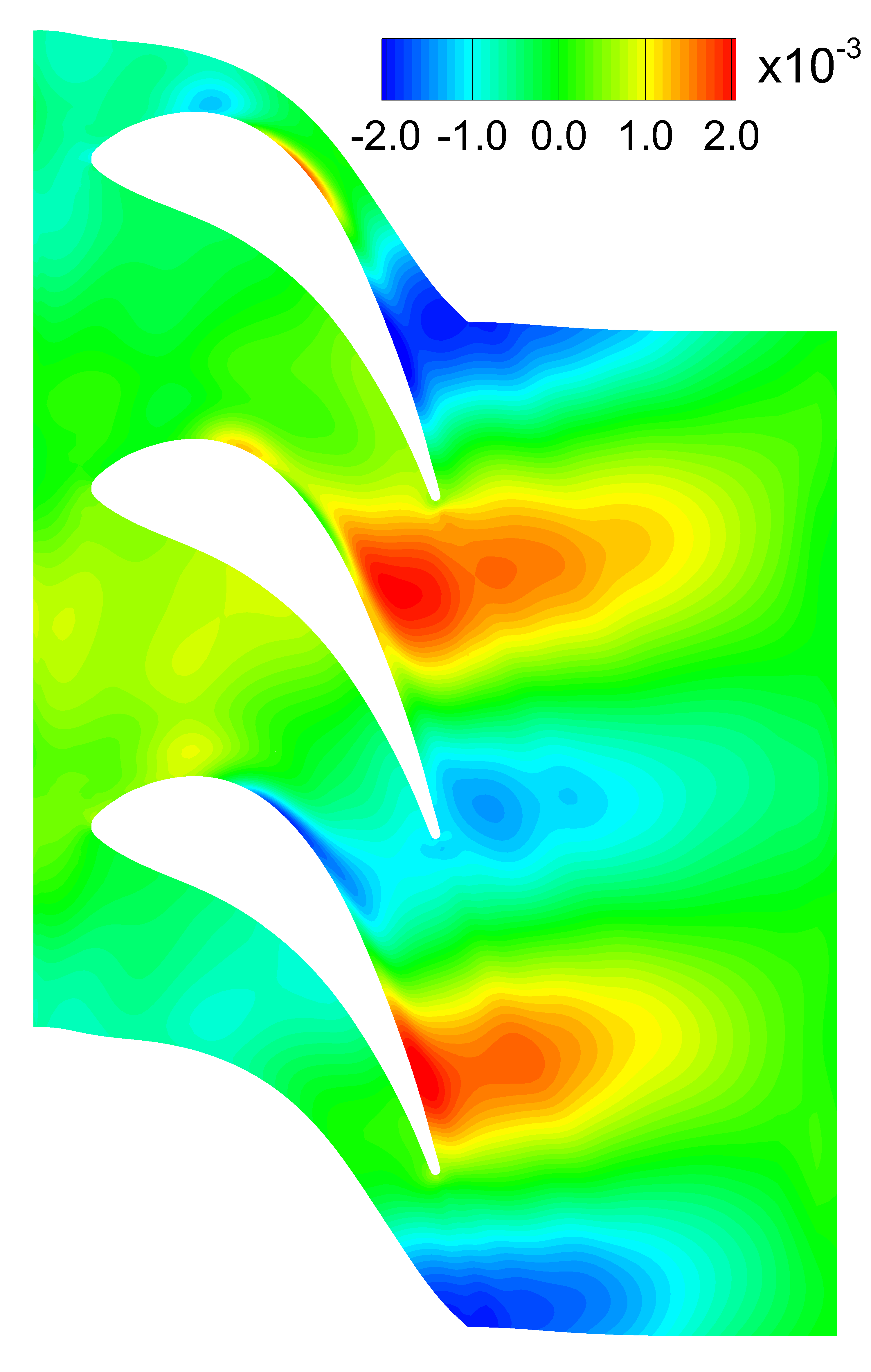}
        \caption{$CC4$}
        \label{fig:contour_mode_pod_mode2_CC4}
    \end{subfigure}
    \caption{Contours of second POD mode shape for different clocking configurations at 50\% span.}
    \label{fig:contour_mode_pod_mode2_clocking}
\end{figure}

Figure \ref{fig:singular_value-clocking} shows the variation of POD singular value with mode index for different clocking configurations. As with the DMD amplitude modulus in Fig. \ref{fig:amplitude_modulus-clocking}, the POD singular values show similar trends across the clocking configurations. For each clocking configuration, the second and third POD modes, representing the dominant pressure fluctuations in the unsteady flow, have nearly identical singular values. Across the different clocking configurations, the singular values follow the ordering $CC4 > CC0 > CC3 > CC1 > CC2$. This trend is also reflected by the strengths of the second POD modes, as shown by the red and blue regions in Fig. \ref{fig:contour_mode_pod_mode2_clocking}. The rankings of both POD mode strength and singular value are the same as those of the DMD mode strength in Fig. \ref{fig:contour_mode_dmd_modulus_2_3_clocking} and amplitude modulus in Fig. \ref{fig:amplitude_modulus-clocking}, as well as the adiabatic efficiency at 50\% span in Ref. \cite{2017_Zhu_clocking}.
Similar to the DMD modes, no correlation is found between the higher POD modes and the aerodynamic parameters of the turbine.

\section{Conclusions} \label{sec:conclusions}
In this study, POD and four DMD variants are applied to analyze the unsteady pressure fields in the downstream stator of a 1.5-stage subsonic axial turbine. The flow fields are obtained by solving the URANS equations.

The performance of POD and DMD in reduced-order pressure reconstruction is first evaluated.
The DMD methods based on the amplitude criterion, Tissot criterion, and the SP-DMD method achieve a reconstruction accuracy comparable to that of POD. In contrast, the DMD method based on the frequency criterion yields larger reconstruction error and is therefore unsuitable for the present case. The first seven modes are identified as dominant in the downstream stator. For both POD and DMD, the relative reconstruction error using these seven modes decreases progressively along the streamwise direction within the stator passage attributed to attenuation of upstream disturbances.

The spatial shapes of the first seven modes at 50\% span are compared between POD and DMD to assess the ability of each method to extract relevant flow structures. 
The first modes of both POD and DMD correspond to the time-averaged flow field. 
The second and third modes accurately capture the dominant pressure fluctuation structures in the stator passage. Both POD and DMD yield similar spatial distributions, with only minor differences in the locations of peaks and troughs.
The fourth and fifth POD modes, as well as the sixth and seventh DMD modes, capture higher-order structures of the pressure fluctuations, particularly in the region downstream of the trailing edge. 
The fourth and fifth DMD modes, along with the seventh POD mode, exhibit decaying behaviors along the streamwise direction.

The dynamic evolution of DMD modes is examined. Most DMD modes are neutrally stable, a small portion exhibits damping, and a few modes display growing. 
Low-frequency, high-amplitude, neutrally-stable modes dominate the downstream stator, consistent with the globally periodic nature of the unsteady flow driven by the rotating motion of the upstream rotor blades. For each complex-conjugate DMD mode pair, the real and imaginary parts of the time coefficient are phase-shifted by one quarter of the oscillation period. Each DMD mode is only associated with an individual harmonic of the rotor passing frequency. However, a POD mode may contain multiple frequencies beyond the primary harmonic.
POD analysis may misrepresent the fundamental frequencies of a dynamic system with multiple frequencies because its principle is based on a least-squares approximation of snapshots of the combined system response. Although POD may be equally accurate in terms of reconstructing the system response, it is not capable of identifying the fundamental dynamic components of the system.
Properly applied and interpreted, however, both POD and DMD can be used to uncover important flow physics and construct reduced-order models for complex unsteady flow in multi-stage turbomachines.

The correlations between the dominant modes in the downstream stator and the turbine adiabatic efficiency are explored across different stator clocking configurations.
At 50\% span, a clocking configuration with higher adiabatic efficiency exhibits a larger modulus and amplitude of the second and third DMD mode pair, as well as an increased strength and singular value of the second POD mode. This suggests that the dominant POD and DMD modes, characterized by high pressure fluctuation levels, are correlated with improved aerodynamic performance of the turbine. These findings provide a data-driven guideline for aerodynamic design or shape optimization of multi-stage turbines.


\bibliography{main}

\end{document}